\documentclass[12pt]{iopart}

\newcommand{\commentout}[1]{}

\usepackage{siunitx}
\usepackage{graphicx}
\usepackage{subcaption}
\usepackage[backend=biber, style=numeric-comp, sorting=none, maxcitenames=2, uniquelist=false]{biblatex}

\usepackage{perpage} 
\MakePerPage{footnote} 

\bibliography{Linear_GK}

\begin{document}

\title[Nucl. Fusion Linear GK STEP]{Linear gyrokinetic stability of a high $\beta$ non-inductive spherical tokamak}

\author{B.S. Patel$^{1,2}$, D. Dickinson$^1$, C.M. Roach$^2$ and H.R. Wilson$^1$}
\address{$^1$ University of York, Heslington, York, YO10 5DD, UK}

\address{$^2$ Culham Centre for Fusion Energy, Abingdon OX14 3DB, UK}
%

\ead{bhavin.s.patel@ukaea.uk}

\vspace{10pt}
\begin{indented}
\item[]March 2021
\end{indented}

\begin{abstract}
Spherical tokamaks (STs) have been shown to possess properties desirable for a fusion power plant such as achieving high plasma $\beta$ and having increased vertical stability. To understand the confinement properties that might be expected in the conceptual design for a high $\beta$ ST fusion reactor, a $1$\si{\giga\watt} ST plasma equilibrium was analysed using local linear gyrokinetics to determine the type of micro-instabilities that arise. Kinetic ballooning modes (KBMs) and micro-tearing modes (MTMs) are found to be the dominant instabilities. The parametric dependence of these linear modes was determined and from the insights gained, the equilibrium was tuned to find a regime marginally stable to all micro-instabilities at $\theta_0=0.0$. This work identifies the most important micro-instabilities expected to generate turbulent transport in high $\beta$ STs. The impact of such modes must be faithfully captured in first principles based reduced models of anomalous transport that are needed for predictive simulations.
\end{abstract}

%
%
%
%

\section{Introduction}

Understanding the confinement of tokamak plasmas in reactor relevant regimes is critical in the design of future fusion power plants. Scaling laws, such as the ITER98 IPB(y,2) \cite{iter1999plasmaconf} and the Petty08 \cite{petty2008sizing} are often used as a metric for the quality of confinement. Both of these laws describe the confinement of existing conventional tokamaks equally well but extrapolate quite differently to reactor relevant regimes. For example, for the ITER baseline scenario, the 98 scaling predicts $\tau_{98}=3.7$\si{\second}, but Petty predicts an increased value of $\tau_{\mathrm{Petty}}=4.6$\si{\second}. Global scalings are fitted to data from existing tokamaks not in the reactor regime, and there are inevitably unknown uncertainties in their extrapolation, especially if the turbulent regime is different. Global scalings can be used for ballpark estimates in confinement, but reliable predictions require high fidelity models for the neoclassical and turbulent transport and should be based on first principle models. Gyrokinetics has proven to be a useful tool in modelling turbulence and its associated transport and has been able to match experimental fluxes with reasonable success \cite{rhodes2011mode, van2017subcritical, howard2012quantitative}. Applying linear gyrokinetics to a high $\beta$ ST regime will provide an insight into the turbulent modes that may arise. This work aims to identify the linear micro-stability properties of a conceptual equilibrium for a high $\beta$ burning ST reactor and highlight the necessary physics that needs to be captured in a transport model. This will be critical for the development of STEP \cite{wilson2020step}.

A potential non-inductive operating point is identified in Section \ref{sec:scene_eq}, where a plasma equilibrium has been obtained using SCENE \cite{wilson1994scene}, a fixed boundary Grad-Shafranov solver. The reasoning behind this equilibrium design is briefly discussed but further details can be found in \cite{patel2021confinement}. In Section \ref{sec:important_modes}, the dominant instabilities of the $\rho_\psi \equiv \psi/\psi_{\mathrm{LCFS}}=0.5$ surface was determined using linear GS2 \cite{gs2code}. It was found that KBMs and MTMs were the prevalent modes at the ion scale illustrating the electromagnetic nature of the turbulence that needs to be captured in a transport model. At deep electron scales the equilibrium was found to be stable. Section \ref{sec:mode_dependancy} explores the impact of different equilibrium parameters, such as the kinetic gradients and plasma $\beta$, on these modes. With the insights gained here, Section \ref{sec:options_stable} perturbs the global plasma parameters and finds a more stable equilibrium. An equilibrium marginally stable to all the micro-instabilities is found, which suggests that operating at neoclassical levels of transport and thus high levels of confinement, may be possible. The feasibility of achieving such an equilibrium is briefly discussed. \ref{app:cross_code} compares the linear predictions made by CGYRO and GS2 for this equilibrium showing them to be in good agreement and \ref{app:high_ky_etg} the explores the cause of stability in the deep electron scale region.

\section{1\si{\giga\watt} ST operating point}
\label{sec:scene_eq}
This plasma equilibrium was chosen to produce a fusion power over $1$\si{\giga\watt} as this was assumed to be sufficient to produce net electricity. In SCENE several global plasma parameters are specified such as the current going through the central column (which sets the vacuum magnetic field), the total toroidal plasma current and the plasma boundary. Furthermore, the temperature and density profiles are specified, though no transport models inform these profiles. The feasibility of the profiles could be tested by calculating the heat and particle sources required for a given transport model.

\begin{figure}[!htb]
    \begin{minipage}{0.5\linewidth}
        \begin{subfigure}{\textwidth}
        \centering
        \includegraphics[width=75mm]{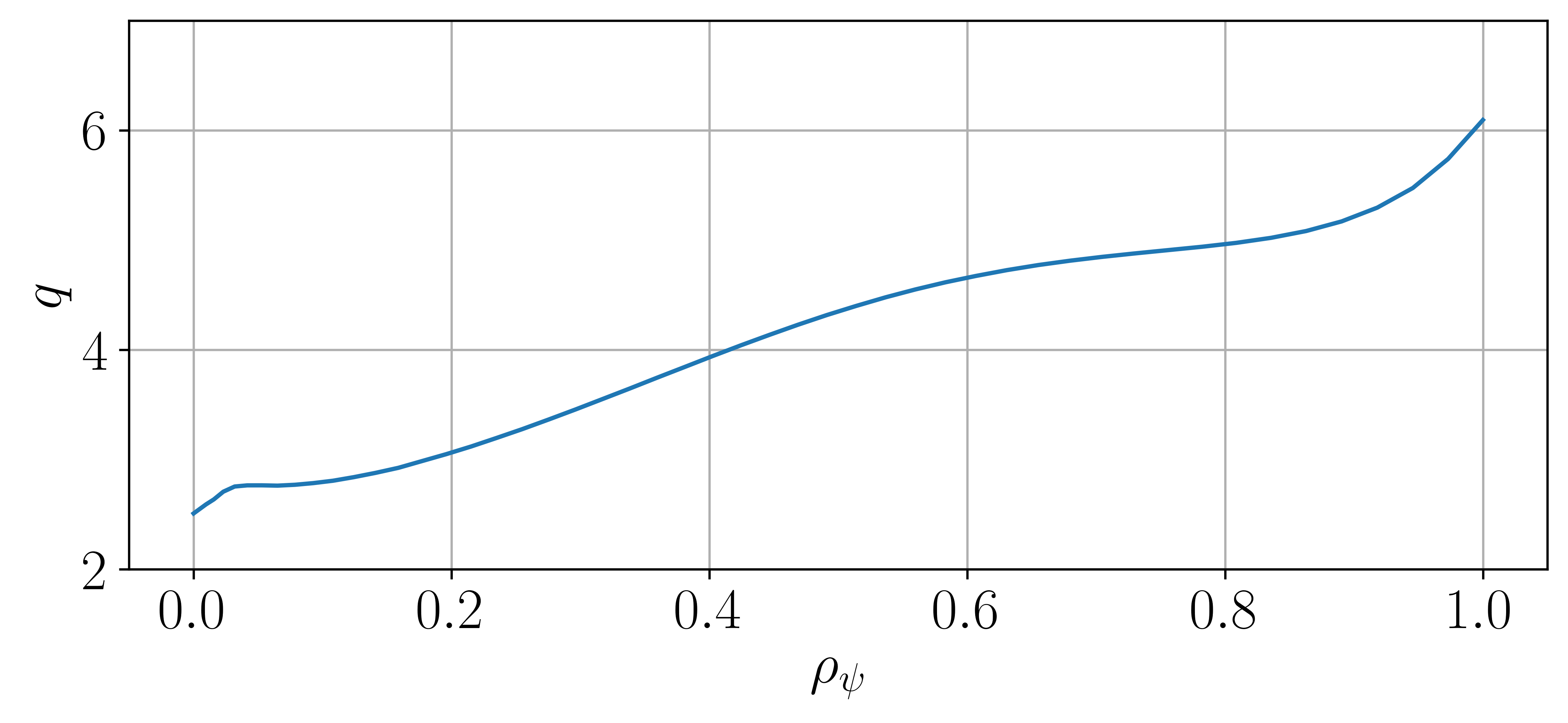}
        \caption{}
        \label{fig:baseline_q}   
    \end{subfigure}
        \begin{subfigure}{\textwidth}
        \centering
        \includegraphics[width=75mm]{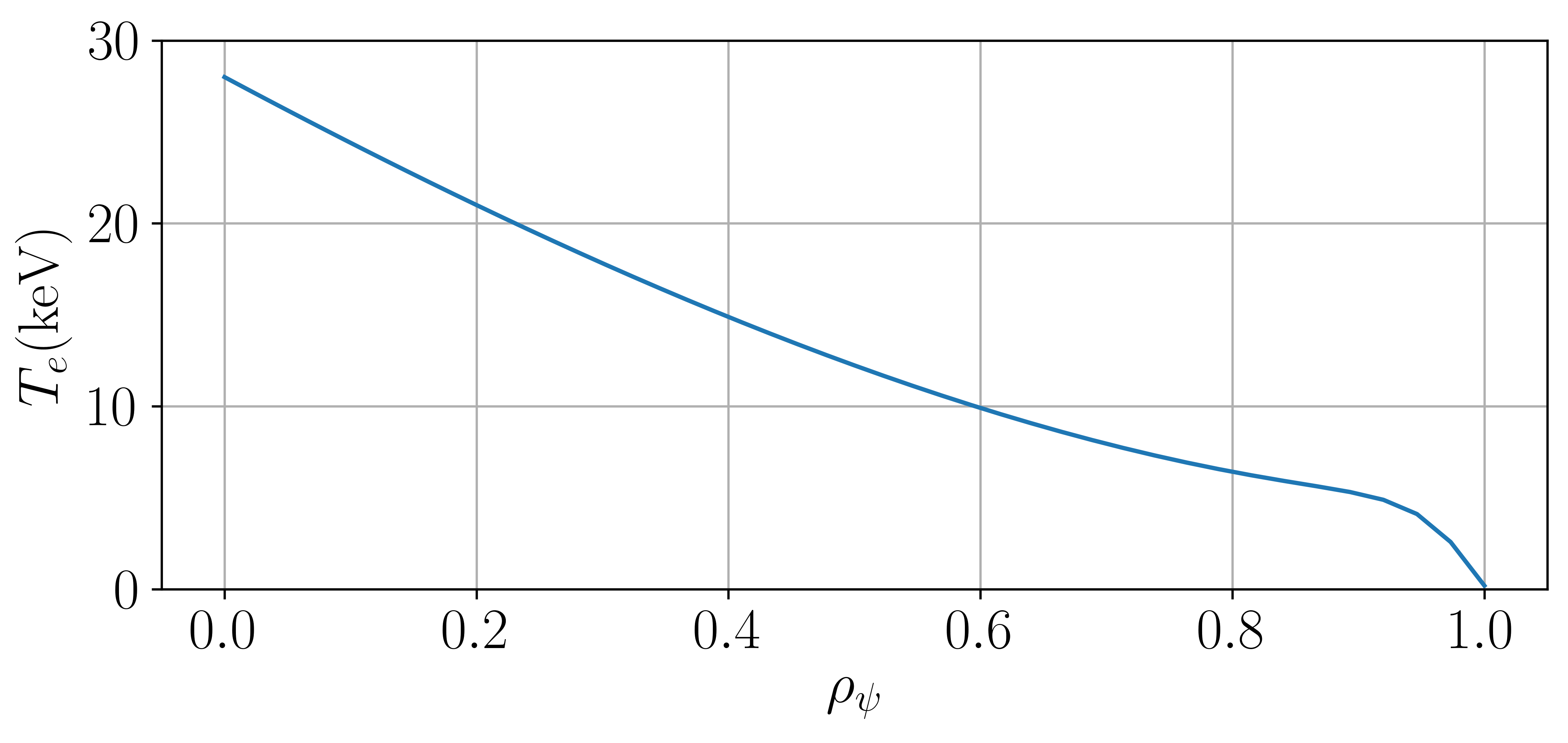}
        \caption{}
        \label{fig:baseline_te}   
    \end{subfigure}
    \begin{subfigure}{\textwidth}
        \centering
        \includegraphics[width=75mm]{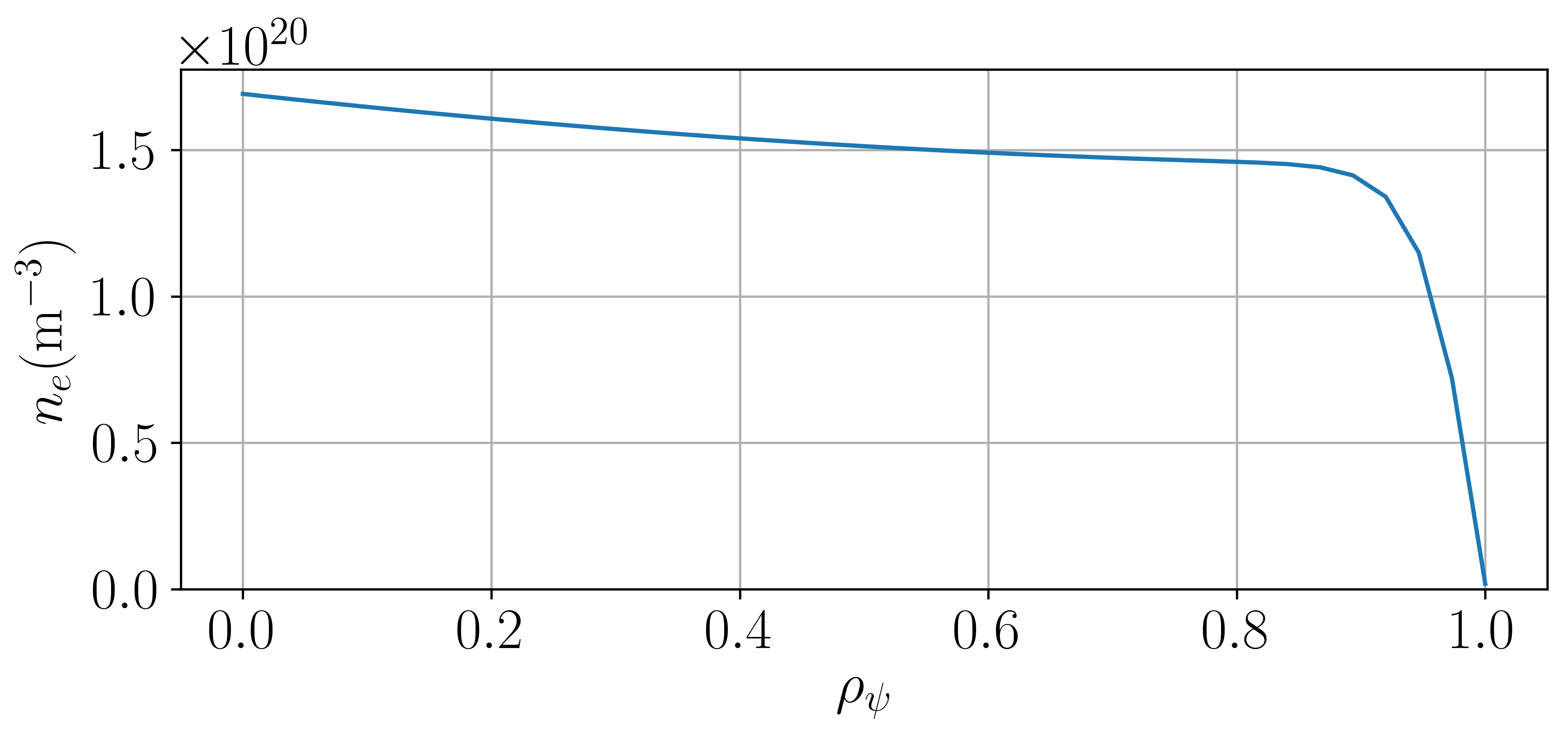}
        \caption{}
        \label{fig:baseline_ne}   
    \end{subfigure}
    \end{minipage}
    \begin{subfigure}{0.45\textwidth}
        \centering
        \includegraphics[width=67.5mm]{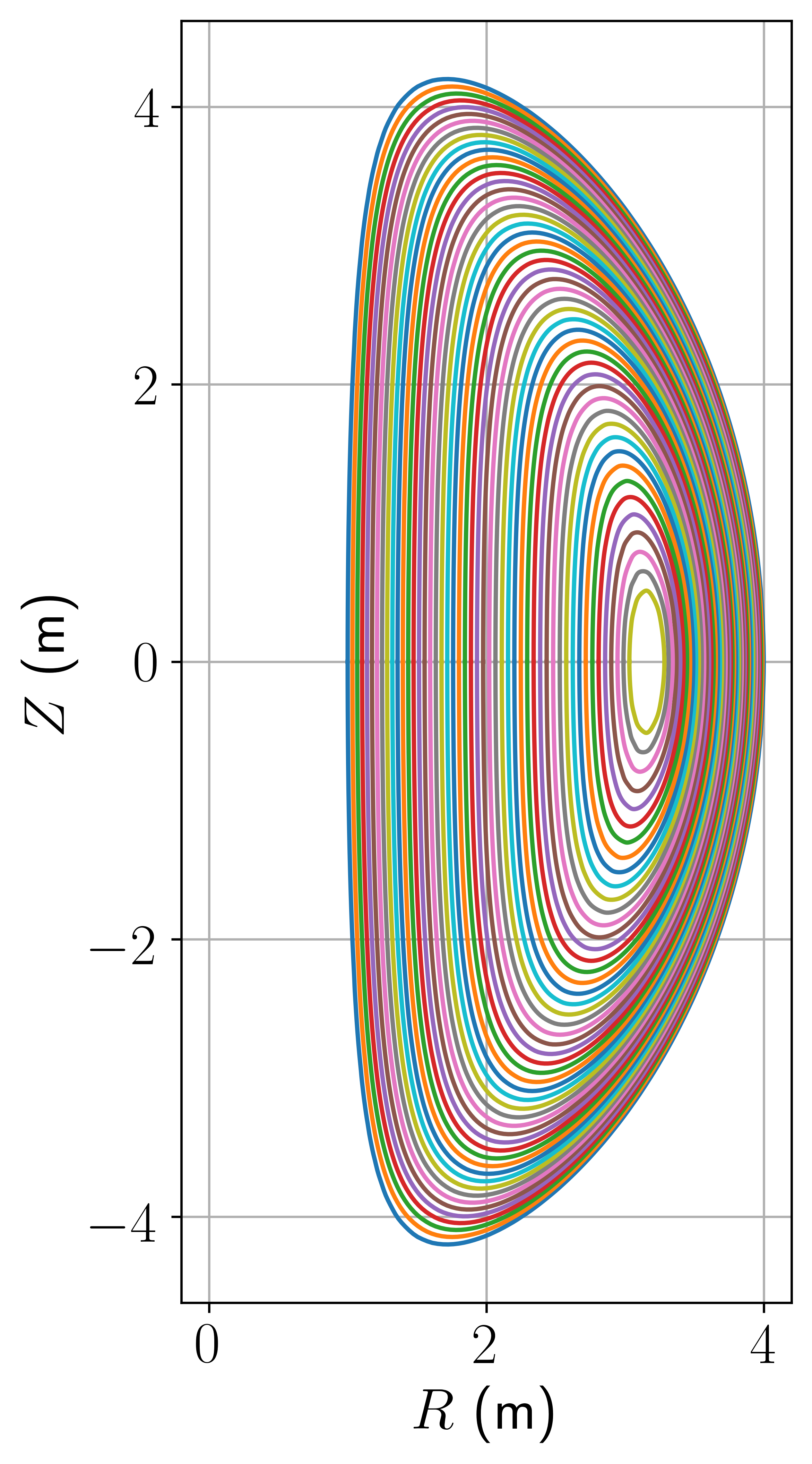}
        \caption{}
        \label{fig:baseline_flux_sur}   
    \end{subfigure}
    \caption{Profiles for the baseline equilibrium showing the a) safety factor, b) electron temperature where $T_e=T_i$  and c) electron density. d) The flux surface contours for the baseline equilibrium.}
    \label{fig:baseline}  
\end{figure}

\begin{table}[!ht]
    \centering
    \begin{tabular}{|c|c|c|}
        \hline
        Parameter & Value & Motivation\\
        \hline
        \hline 
        $R_{\mathrm{maj}}$ (m) & 2.5 & Power through the separatrix \\
        $a$ (m) & 1.5 & Allow space for centre column \\
        $R_0$ (m) & 3.15 & Magnetic axis position - output from SCENE \\
        $I_{\mathrm{rod}}$ (\si{\mega\ampere}) & 30.0 & STPP centre column \\
        $I_{\mathrm{p}}$ (\si{\mega\ampere}) & 21.0 & Trade-off between confinement/stability and current drive\\
        $I_{\mathrm{aux}}$ (\si{\mega\ampere}) & 8.2 & Output from SCENE\\
        $P_{\mathrm{fus}}$ (\si{\mega\watt}) & 1100 & Net electric device \\
        $P_{\mathrm{aux}}$ (\si{\mega\watt}) & 94 & NUBEAM\\
        $\kappa$ & 2.8 & Limits based off of NSTX data \\
        $\delta$ & 0.55 & Same as STPP \\
        $H_{\mathrm{98}}, H_{\mathrm{Petty}}$ & 1.35, 0.94 & Output given required $P_{\mathrm{aux}}$ \\
        $T_{e0},  \langle T_e\rangle$ (\si{\kilo\electronvolt}) & 28.0,  14.8 & Assumption\\
        $n_{e0}, \langle n_e\rangle (\times 10^{20}$\si{\per\cubic\metre}) & 1.72, 1.54 & Ensure $P_{\mathrm{fus}}=1.1$\si{\giga\watt} given $T_e$ assumption\\
        $f_{\mathrm{GW}}$ & 0.52 & Ensure $P_{\mathrm{fus}}=1.1$\si{\giga\watt} given $T_e$ assumption\\
        $l_i$ &  0.27 & Maximise elongation\\
        $\beta_{\mathrm{N}}$ & 5.5 & Output from SCENE - MHD stable\\
        $q_0$ & 2.51 & Avoid sawteeth/NTM/internal kink modes \\
        \hline
    \end{tabular}
    \caption{Basic global plasma parameters for this baseline operating point and the reasoning behind them.}
    \label{tab:baseline_parameters}
\end{table}

SCENE solves for the equilibrium whilst self-consistently calculating the neoclassical currents such as the bootstrap and diamagnetic current, returning the amount of auxiliary current needed to provide the total plasma current requested. The shape of the auxiliary current profile is prescribed and can be set to be consistent with the anticipated auxiliary current sources.

The major radius, $R_{\mathrm{maj}}$, was set a constraint on the power going through the separatrix $P_{\mathrm{sep}}$. Assuming that 80\% of the heating power is radiated away, similar to FNSF \cite{menard2016fusion} and requiring that $P_\mathrm{sep}/R_{\mathrm{maj}} < 20$\si{\mega\watt\per\metre}, similar to the ITER \cite{goldston2017new}, $R_{\mathrm{maj}} = 2.5$\si{\metre} was set. The spherical tokamak power plant (STPP) is an existing design of a high $\beta$ ST \cite{wilson2004STPP}. Using its centre column design \cite{voss2000toroidal} which had a rod current, $I_{\mathrm{rod}}=30$\si{\mega\ampere}, generates $2.4$\si{\tesla} on axis for this design. To allow for sufficient space for this centre column the minor radius $a$ was set to $1.5$\si{\metre}, resulting in an aspect ratio of $A=1.67$. 

With the plasma current set to $I_\mathrm{p}=21$\si{\mega\ampere}, the prescribed pressure leads to $\beta_\mathrm{N}=5.5$, which satisfies MHD stability constraints with the $q$ profile shown in Figure \ref{fig:baseline_q}. Note, that this was designed with $q_{\mathrm{min}}>2$ to avoid the 2/1 NTM and sawteeth modes. Increasing $I_\mathrm{p}$ further would increase the requirements on the auxiliary current drive system. The elongation was set to $\kappa=2.8$, inline with results from NSTX \cite{menard2016fusion} and the triangularity of $\delta=0.55$ to match STPP. Using the Europed model \cite{saarelma2017integrated}, a temperature pedestal height of $5.3$\si{\kilo\electronvolt} at $\rho_\psi=0.92$ was found for utilising the assumption that the width of the pedestal scales like $0.1 (\beta_\theta^{\textrm{ped}})^{1/2}$, consistent with MAST data \cite{diallo2011dynamical}. This work set the temperature pedestal height to $5$\si{\kilo\electronvolt} located at $\rho_\psi=0.9$, which happens to be similar to the ITER pedestal \cite{snyder2009pedestal}. The core temperature was assumed to be $28$\si{\kilo\electronvolt} with the $T_e=T_i$. A small amount of density peaking was specified such that $n_{e,\mathrm{ped}}/n_{e0}=0.9$, similar to ITER. The density was then scaled up such that $P_{\mathrm{fus}}>1$\si{\giga\watt}, leading to a Greenwald fraction of $f_{\mathrm{GW}}=0.52$. These profiles are shown in Figure \ref{fig:baseline_te} and \ref{fig:baseline_ne}. The resulting equilibrium is outlined in Table \ref{tab:baseline_parameters}, with Figure \ref{fig:baseline_flux_sur} illustrating the flux surfaces.

The auxiliary current profile was set to be primarily off axis which reduced the internal inductance $l_i$ to help with vertical stability, with a small amount necessary on axis to fill the hole in the bootstrap current. From Table \ref{tab:baseline_parameters}, it can be seen that 8.2\si{\mega\ampere} of auxiliary current was necessary for a non-inductive scenario. Using NUBEAM \cite{pankin2004tokamak} an NBI configuration was identified capable of driving the required current. For the on axis current, an 8\si{\mega\watt} 1\si{\mega\electronvolt} beam was necessary and 86\si{\mega\watt} 500\si{\kilo\electronvolt} for the off axis current. This totals to an auxiliary power of $P_{\mathrm{aux}} = 94$\si{\mega\watt}, which requires a $H_{98}=1.35$ and $H_{\mathrm{Petty}}=0.94$. This $H_{98}$ has been seen before in STs \cite{menard2010physics} and is very close to the Petty prediction suggesting that this is a reasonable starting point. This equilibrium is further described in \cite{patel2021confinement}.



\commentout{
\begin{figure}[!htb]
    \begin{subfigure}{0.5\textwidth}
        \centering
        \includegraphics[width=75mm]{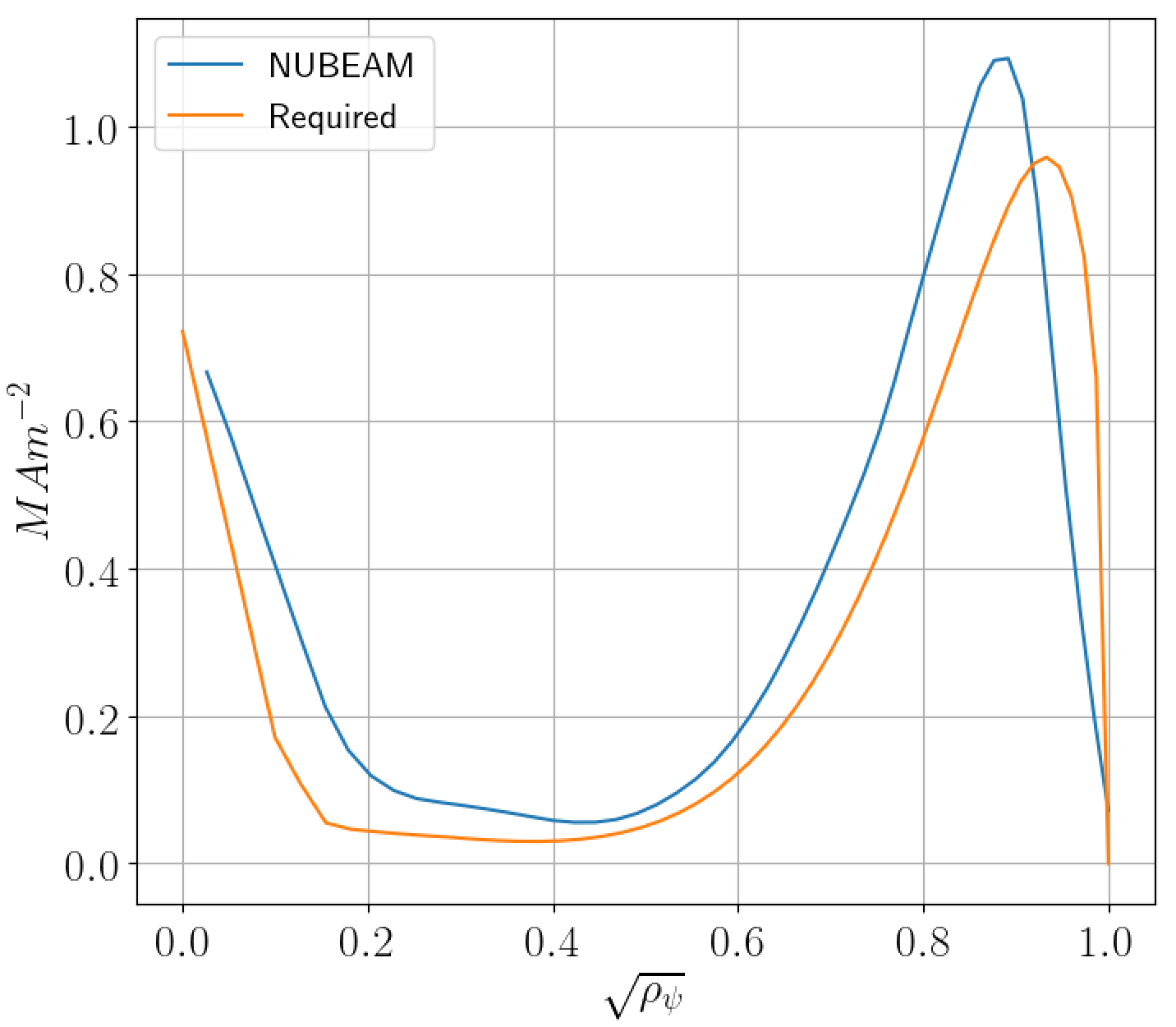}
        \caption{}
        \label{fig:nubeam_baseline}
    \end{subfigure}
    \begin{subfigure}{0.5\textwidth}
        \centering
        \includegraphics[width=75mm]{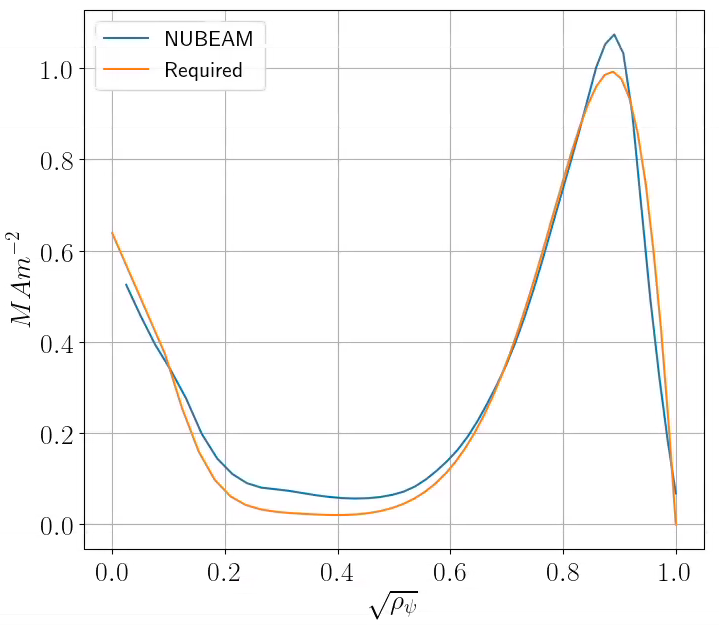}
        \caption{}
        \label{fig:modified_nubeam}
    \end{subfigure}
    \caption{Comparing the a) initial specified $J_{\mathrm{aux}}$ profile and the b) NUBEAM matched $J_{\mathrm{aux}}$ to the NUBEAM predictions of the final beam configuration.}
    \label{fig:final_config}
\end{figure}
}

\subsection{Local equilibrium for micro-stability analysis}

To begin with the dominant micro-instabilities will be examined using local linear gyrokinetics on a core flux surface where $\rho_\psi=0.5$ . Once the nature of the modes is identified, the parametric dependence of these modes will be determined which, in turn, will inform design choices that will help stabilise the equilibrium. 

\begin{table}[!htb]
    \begin{minipage}{0.4\linewidth}
    \centering
    \begin{tabular}{|c|c|}
        \hline
        Parameter & $\rho_\psi=0.5$ \\
        \hline\hline
        $r/a$ & 0.66\\
        $R_{\mathrm{maj}}/a$  & 1.79 \\
        $n_{e}$ ($\times 10^{20}$\si{\per\cubic\metre}) & 1.51 \\
        $T_e$ (\si{\kilo\electronvolt}) & 12.2 \\
        $B_0$ (\si{\tesla}) & 2.16 \\
        $B_\mathrm{unit}$ (\si{\tesla}) & 7.52 \\
        $\rho_s$ (\si{\metre}) & 0.0021\\
        $a/L_n$  & 0.43 \\
        $a/L_T$  & 2.77 \\
        $\Delta$ & -0.57\\
        $q$ & 4.30\\
        $\hat{s}$ & 0.78 \\
        $\kappa$ & 3.03 \\
        $s_\kappa$ & -0.14\\
        $\delta$ & 0.45 \\
        $s_\delta$ & 0.19\\
        $\beta_{e}$ & 0.15 \\
        $\beta_{e,\mathrm{unit}}$ & 0.012 \\
        $\nu_{ee} (c_s/a)$ & 0.017 \\
        $\gamma_{\mathrm{dia}} (c_s/a)$& 0.08  \\
        \hline
    \end{tabular}
    \caption{Local plasma and Miller parameters for the $\rho_\psi = 0.5$ surface of the equilibrium in Table \ref{tab:baseline_parameters}.}
    \label{tab:gk_flux_params}
    \end{minipage}
    \begin{minipage}{0.6\linewidth}
    \begin{subfigure}{\textwidth}
        \centering
        \includegraphics[width=60mm]{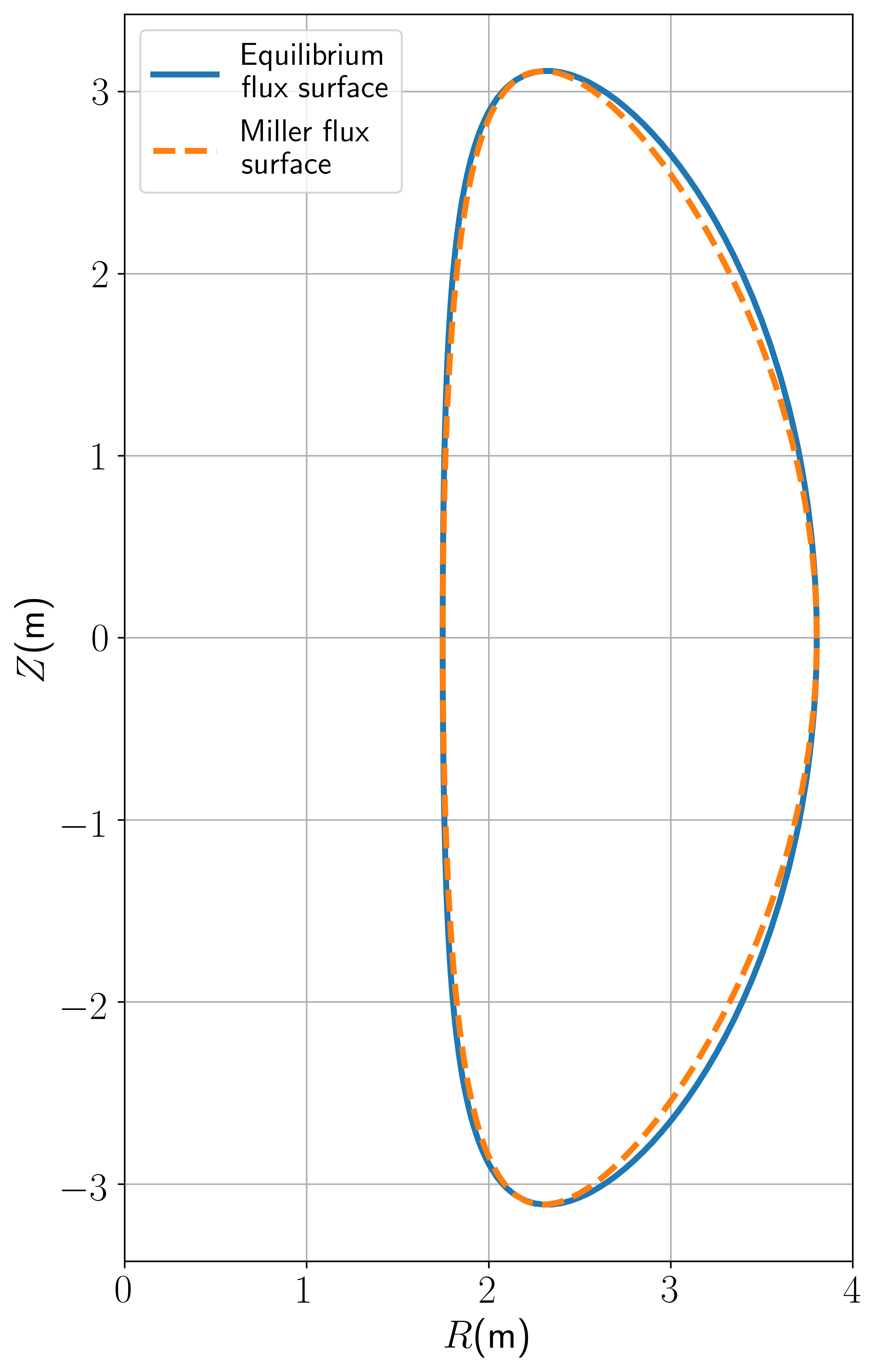}
        \caption{}
        \label{fig:miller_fit_bdy}
    \end{subfigure}
    \begin{subfigure}{\textwidth}
        \centering
        \includegraphics[width=80mm]{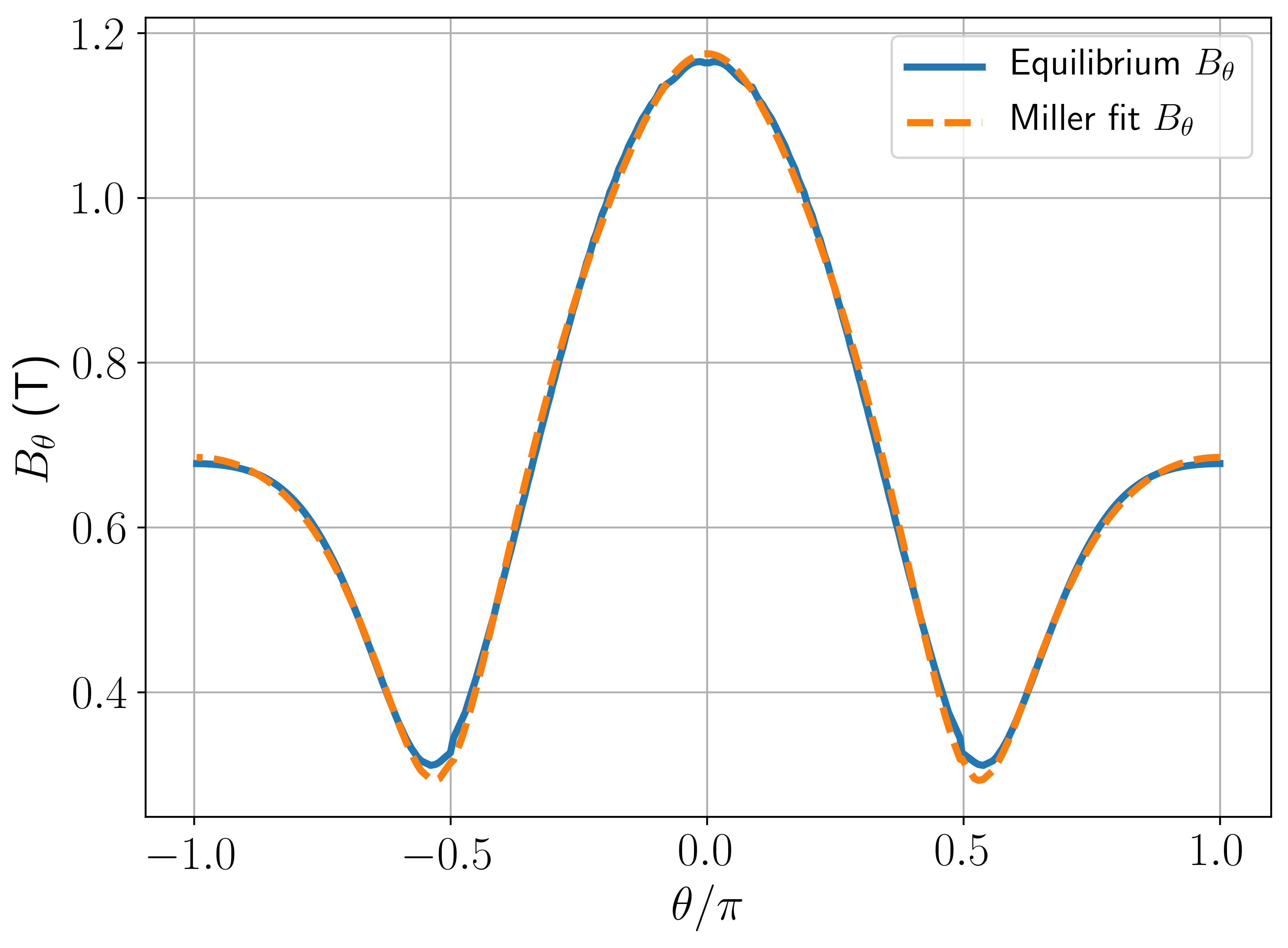}
        \caption{}
        \label{fig:miller_fit_bp}
    \end{subfigure}
    \captionof{figure}{Miller fit shown for $\rho_\psi=0.5$ surface showing the a) flux surface contour and b) the poloidal field}
    \label{fig:miller_fit}
    \end{minipage}
\end{table}

A Miller parameterisation \cite{miller1998noncircular} was used to model this equilibrium, with the parameters outlined in Table \ref{tab:gk_flux_params}. The fit of the flux surface and poloidal field are shown in Figures \ref{fig:miller_fit}a and \ref{fig:miller_fit}b respectively, showing that Miller fit to the local equilibrium is reasonable. Several simulations were conducted using the full numerical equilibrium and it was found to agree with the Miller simulations. 

\subsubsection{Numerical set-up}
The gyrokinetics code GS2 was used to examine the micro-stability of this plasma \cite{gs2code}. Convergence tests in GS2 indicated that 128 poloidal ($\theta$) grid points, 8 energy grid points and 16 un-trapped grid points in pitch angle \footnote{In GS2 the number of trapped grid points in pitch angle is given by $n_\theta/2 + 1$.} was sufficient to resolve the modes seen here. Two species were simulated, a thermal deuterium and electron species. It should be highlighted that no fast ions were included in this work, but must be examined in future work as they may have a significant impact on the micro-stability \cite{chowdhury2021low, kumar2021turbulent}. Furthermore, the impact of including tritium and impurities is also left as future work. 

Although there will be neutral beam injection, the driven rotation is expected to be low due to the high energy beams \cite{patel2021confinement}. However, there will be a contribution to the $E\times B$ shearing rate from the pressure gradient. This diamagnetic flow shear, $\gamma_{\mathrm{dia}}$, can be calculated from force balance. In this work Equation A10 from \cite{applegate2004microstability} is used to calculate $\gamma_{\mathrm{dia}}$.

The normalisations used here was that from CGYRO \cite{candy2016high} where the bi-normal wavenumber $k_y=\frac{nq}{r}$, where $n$ is the toroidal mode number and $r$ is the minor radius of the flux surface. $k_y$ is normalised to Larmor radius $\rho_s= \frac{c_s}{eB_\mathrm{unit}/{m_D c}}$. Here $c_s=\sqrt{\frac{T_e}{m_D}}$ and $B_\mathrm{unit}=\frac{q}{r} \frac{\partial \psi}{\partial r}$. This also leads to a different definition of normalised $\beta$ where $\beta_{e,\mathrm{unit}}= \frac{n_e T_e}{B_\mathrm{unit}^2/2\mu_0}$. It should be noted that this isn't the normalisation used in GS2, but was chosen to allow for easier comparison to CGYRO in \ref{app:cross_code}. The traditional normalising field is the $B_0 = f /R_{\mathrm{maj}}$ with $\beta_{e}= \frac{n_e T_e}{B_0^2/2\mu_0}$.

\section{Identifying the important instabilities}
\label{sec:important_modes}
\subsection{Dominant instabilities at $\rho_\psi=0.5$}

This section will examine the unstable modes at $\theta_0=0.0$. Due to the up-down symmetry of the equilibrium, it is possible to force an odd or even parity $\phi$ eigenfunction in GS2. This allows for usage as a pseudo-eigensolver to determine the dominant odd and even eigenmode. Conventionally, even parity $\phi$ modes are called ``twisting parity'' and odd parity are called ``tearing parity''. However, an odd parity $\phi$ eigenfunction doesn't guarantee that the mode will tear the field line, so in this work we will exclusively refer to the parity as even or odd. For all the figures in this work, a hollow data point will correspond to an even parity $\phi$ eigenfunctions and a filled data point to an odd parity mode. The figure markers for each type of mode will be kept consistent throughout this work, to easier allow for identification of the different modes.

\begin{figure}[!thb]

    \centering
    \includegraphics[width=120mm]{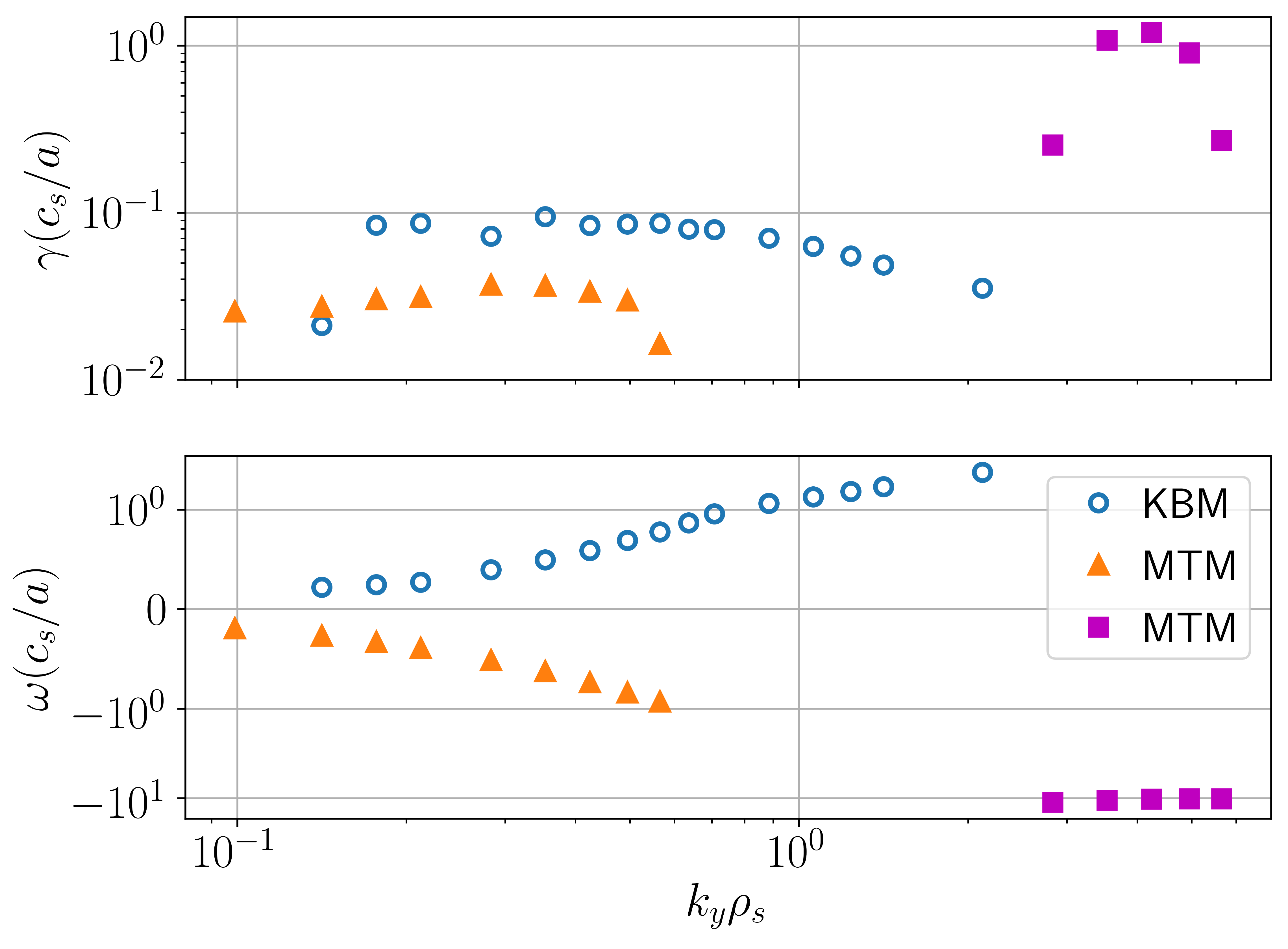}
    \caption{Growth rate and frequencies of the dominant odd and even parity micro-instabilities as a function of $k_y\rho_s$ with $\theta_0=0$ for the $\rho_\psi=0.5$ surface of the baseline equilibrium calculated by GS2. The hollow marker denotes an even parity eigenfunction and a filled marker an odd eigenfunction. The figure markers for each type of mode will be kept consistent throughout this work, to easier allow for identification of different modes. Note the log scale.}
    \label{fig:gs2_dom}
\end{figure}

The dominant odd and even parity instabilities of the $\rho_\psi=0.5$ surface at an initial ballooning angle of $\theta_0=0$ are shown in Figure \ref{fig:gs2_dom}. Three different types of modes were found; in the very long wavelength region when $k_y\rho_s<0.6$, micro-tearing modes (MTMs) were found which will be illustrated by filled orange triangles. At slightly shorter wavelengths, but still at the ion scale where $0.1<k_y\rho_s<2.0$, kinetic ballooning modes (KBMs) were found, denoted by hollow blue circles. Finally as the electron scale is approached between $3.0<k_y\rho_s<6.0$, another MTM is found shown by filled green squares, though this will be shown to be different in nature to the low $k_y$ MTMs. The eigenfunctions of these modes are shown in Figure \ref{fig:psin_0.5_eigfunc}.  

From Figure \ref{fig:gs2_dom}, it can be seen that dominant mode transitions from the MTMs to the KBMs, with the MTMs being subdominant above $k_y\rho_s=0.14$. Co-existing MTMs and KBMs have been seen before in simulations of JET, DIII-D, MAST and NSTX plasmas \cite{hatch2016microtearing, jian2019role, dickinson2012kinetic, guttenfelder2013progress, kumar2021turbulent}. Nonlinear simulations will be necessary to determine their relative contributions to the total turbulent transport.

\subsubsection{Low $k_y$ MTM}

In this work an MTM will refer to any mode with a group velocity in the electron diamagnetic flow direction that has field line tearing. MTMs are generally identified with an even parity $A_{||}$ eigenfunction which is symmetric about $\theta=0$ (when $\theta_0=0.0$) with $\phi$ being odd. However this does not guarantee field line tearing and a more precise definition can be used that quantifies this. A mode is tearing if the perturbation results in a field line that does not return to the equilibrium flux surface. This can be characterised using the following equation \cite{dickinson2011towards, hatch2010mode, ishizawa2015electromagnetic}

\begin{equation}
    \centering
    C_{\mathrm{tear}} = \frac{|\int A_{||} dl|}{\int |A_{||}| dl}
\end{equation}

which $\int dl$ corresponds to an integral along the field line. $C_{\mathrm{tear}}>0$ corresponds to some tearing of the field line. This low $k_y$ MTM has a $C_{\mathrm{tear}}=0.7$ indicating that this mode tears the equilibrium magnetic flux surfaces. 

There are several mechanisms that can drive an MTM. The first is from a parallel thermal force arising from the different frictional forces experienced by electrons travelling in opposite directions along a temperature gradient. This generates a parallel current which in turn generates a perturbation in $A_{||}$ \cite{hassam1980fluid}. It was found that this mechanism could be examined in different limits and in the collisionless regime $(\nu_{ee} < \omega)$ where this time dependent thermal force vanishes \cite{drake1977kinetic}. For this equilibrium, $\nu_{ee}=0.017 c_s/a$, which is well below the mode frequency where $\omega\sim0.5c_s/a$, suggesting this mechanism should not be relevant.

Another mechanism proposed by Catto \& Rosenbluth is where electrons close to the trapped-passing boundary can easily scatter across it, which increases the effective collisionality allowing for a destabilising current driving the tearing instability. This is valid when $\nu_{ee}<\varepsilon\omega$ \cite{catto1981trapped} where $\epsilon=a/R_{\mathrm{maj}}$ is the inverse aspect ratio. This condition is satisfied in this collisionality regime.

Both of these mechanisms require a finite collisionality. However, local gyrokinetic simulations have found MTMs not described by these two mechanisms \cite{applegate2007micro, dickinson2013microtearing}. 

In an MTM, the perturbed $A_{||}$ corresponds to a magnetic island forming at the rational surface. Particles stream freely along the perturbed field and undergo radial excursions about the rational surface. Radial transport is enhanced at large amplitudes when the islands on neighbouring rational surfaces overlap. The higher velocity of the electrons results in significant electron heat transport and is given as a potential reason for the electron heat transport often being found to dominate over ion heat transport in STs \cite{roach2001confinement, guttenfelder2011electromagnetic, kaye2007confinement}.

\begin{figure}
    \begin{subfigure}{0.5\textwidth}
        \centering
        \includegraphics[width=75mm]{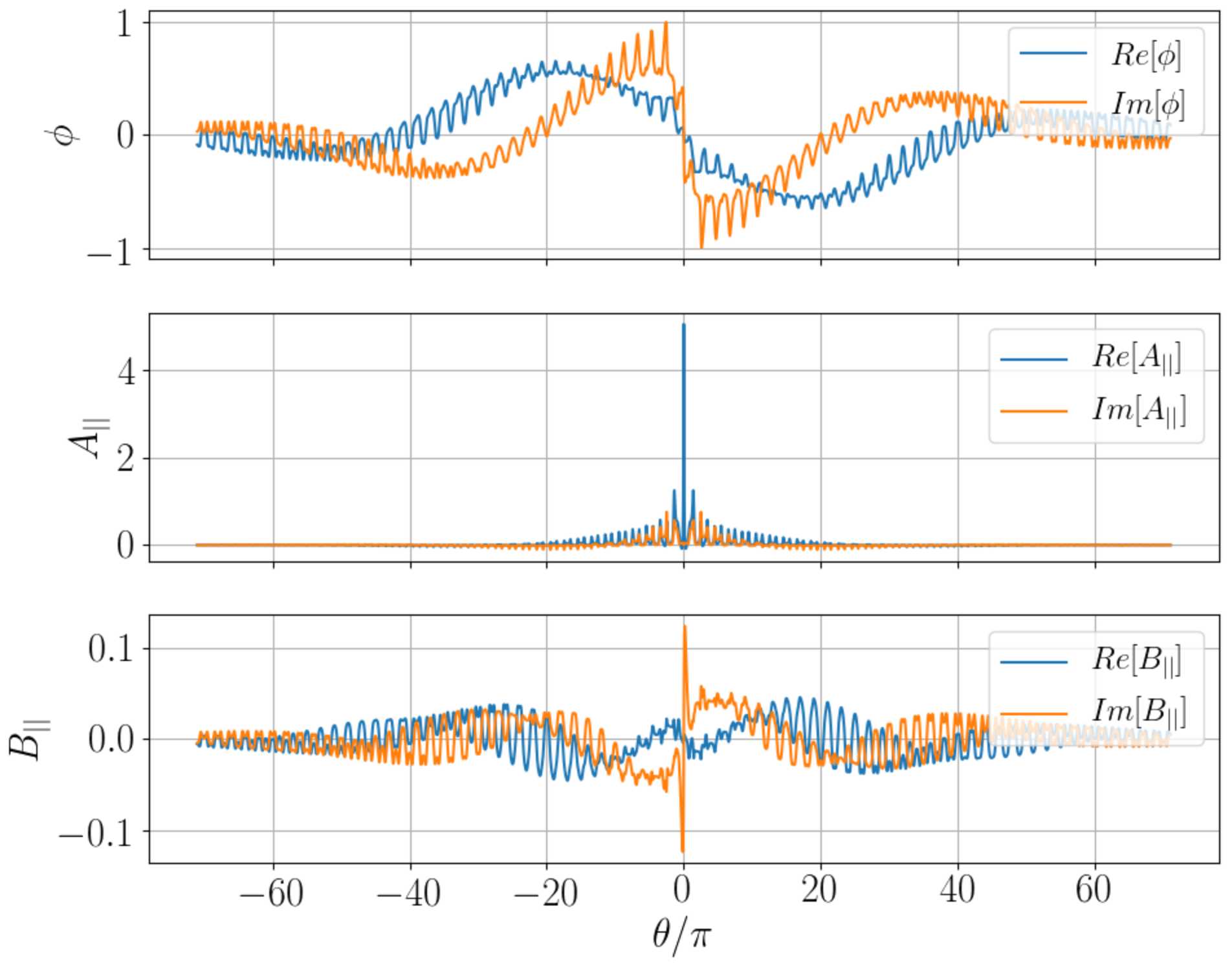}
        \caption{MTM at $k_y\rho_s=0.1$}
        \label{fig:ky_0.1_eigfunc}
    \end{subfigure}
    \begin{subfigure}{0.5\textwidth}
        \centering
        \includegraphics[width=75mm]{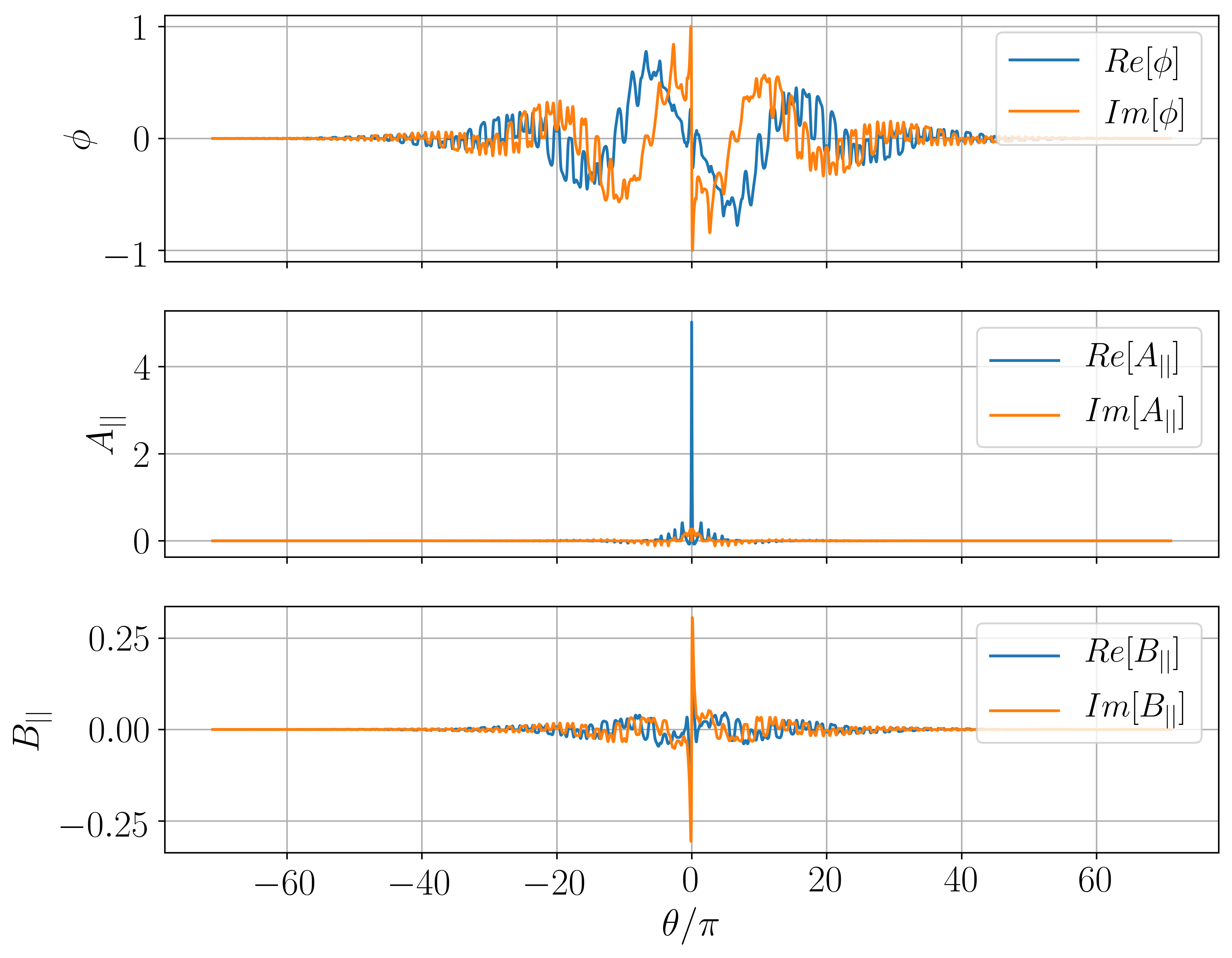}
        \caption{MTM at $k_y\rho_s=0.35$}
        \label{fig:ky_0.35_eigfunc_odd}
    \end{subfigure}
    \begin{subfigure}{=0.5\textwidth}
        \centering
        \includegraphics[width=75mm]{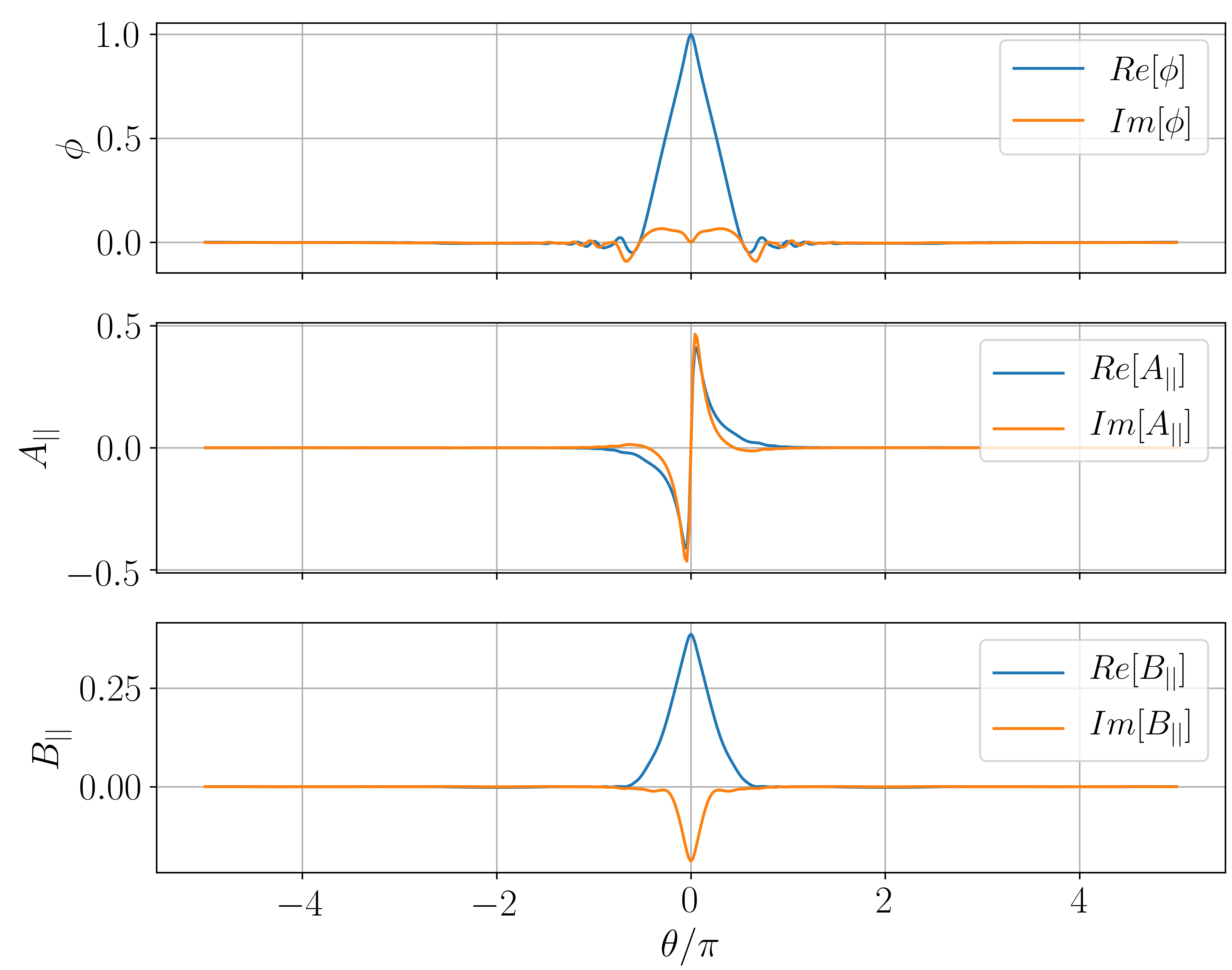}
        \caption{KBM at $k_y\rho_s=0.35$}
        \label{fig:ky_0.35_eigfunc}
    \end{subfigure}
    \begin{subfigure}{0.5\textwidth}
        \centering
        \includegraphics[width=75mm]{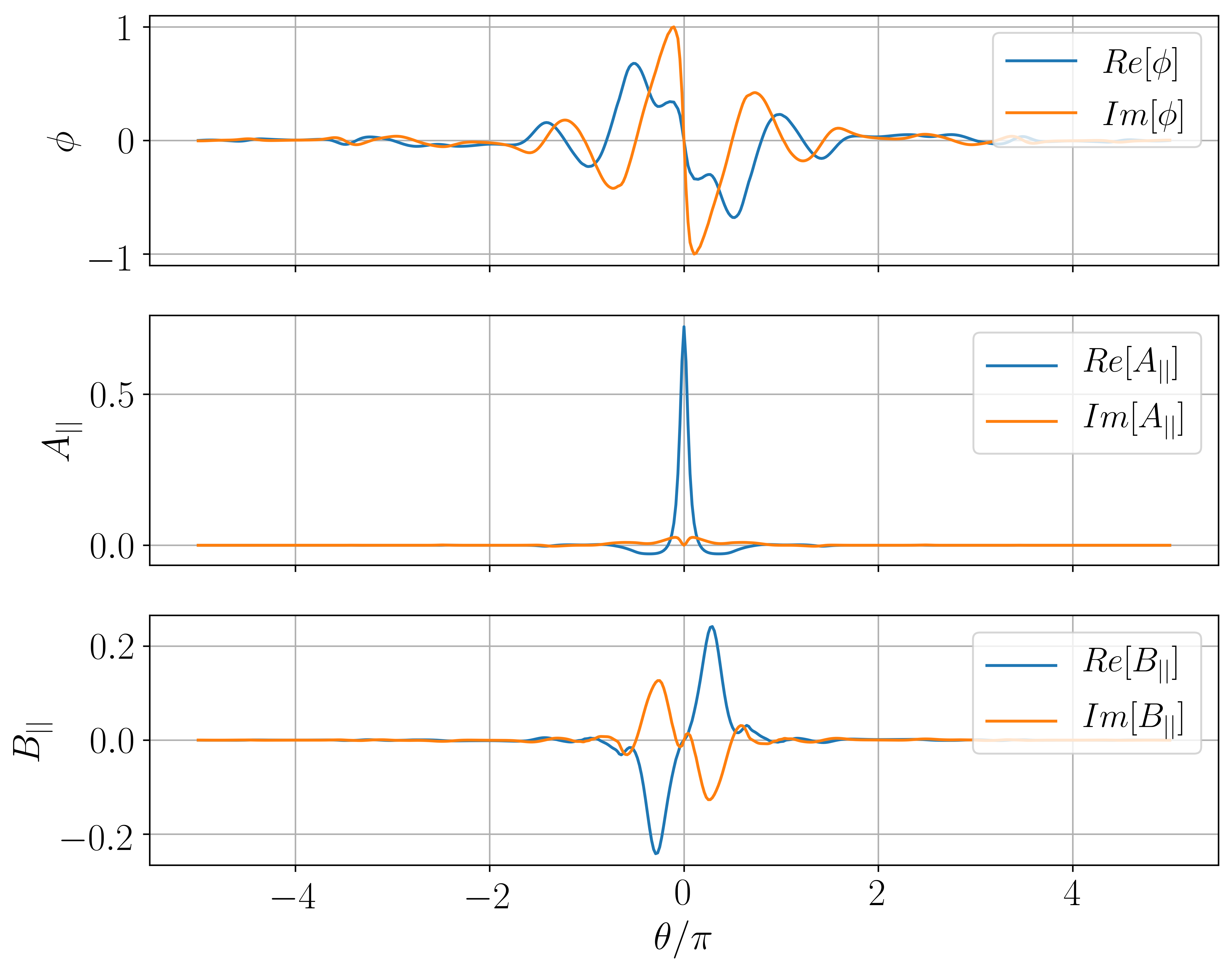}
        \caption{MTM at $k_y\rho_s=4.2$}
        \label{fig:ky_4_eigfunc}
    \end{subfigure}
    \caption{Eigenfunctions for the different modes seen across the $k_y$ spectrum for the $\rho_\psi=0.5$ surface. The amplitude of each mode have been normalised with respect to the maximum value of $\phi$}
    \label{fig:psin_0.5_eigfunc}
\end{figure}

The eigenfunctions of the MTM at $k_y\rho_s=0.1$ is shown in Figure \ref{fig:ky_0.1_eigfunc}. The eigenfunctions have been normalised to the maximum value of $\phi$, indicating that the $A_{||}$ fluctuation is significantly larger than the electrostatic fluctuations at $\theta=0$. Furthermore, these MTMs were found unstable when $B_{||}$ fluctuations were excluded which is typical for these modes. 
The eigenfunction at $k_y\rho_s=0.35$ is shown in Figure \ref{fig:ky_0.35_eigfunc_odd} and it is significantly less extended  in the parallel direction compared to the $k_y\rho_s=0.1$ mode shown in Figure \ref{fig:ky_0.1_eigfunc}, as would be expected if $k_{||} \propto k_y$.

The eigenfunctions exhibit two scales at work here, a broad oscillation in $\theta$, and a much narrower oscillation in $\theta$ which corresponds to a single poloidal revolution due to the equilibrium variation. In the $A_{||}$ eigenfunction, there is a central peak at $\theta=0.0$ and then the other peaks along the field line occur at same value of $\theta \bmod 2\pi$, at the top and bottom of the flux surface. This can be seen clearer in Figure \ref{fig:multi_theta0}. These extended modes are generated by the parallel electron dynamics as the ions would not be able to travel that far down the field line in the mode period due to their lower velocity. The extended nature of the mode in ballooning space required a parallel domain from $-71\pi \rightarrow 71\pi$, corresponding to $k_x= k_y\hat{s}\theta=26.1$. Here $\hat{s} = \frac{r}{q} \frac{\partial q}{\partial r}$. Even linearly resolving such modes becomes computationally expensive. Similar extended MTM eigenfunctions have been seen in simulations of MAST and NSTX discharges \cite{applegate2004microstability, guttenfelder2012scaling}.

Interestingly, these low $k_y$ MTMs were unstable with only $A_{||}$ (both $\phi$ and $B_{||}$ turned off), adiabatic ions, and without contributions from the trapped particles. This indicates that the important physics lies within the passing electrons. Moreover, these MTMs required collisions for instability. Furthermore, if the $\nabla B$ and curvature drifts were turned off the mode went stable even if the drive for these MTMs was increased, suggesting it wouldn't exist in a slab geometry. This differentiates it from the MTMs derived in Drake \textit{et al} \cite{drake1977kinetic} which assumed no toroidal effects, though Rafiq \textit{et al} \cite{rafiq2016microtearing} performed a similar derivation including these effects.

\subsubsection{Low $k_y$ KBM}

The KBM eigenfunction, shown in Figure \ref{fig:ky_0.35_eigfunc}, is significantly less extended compared to the low $k_y$ MTM, only extending a single poloidal revolution. The $B_{||}$ perturbation has a significant amplitude and without $B_{||}$, these KBMs are not driven unstable which has been seen before in high $\beta$ NSTX-like simulations \cite{belli2010fully}. The KBMs do not cause field line tearing as they had $C_{\mathrm{tear}}=0.0$. KBMs drive similar levels of electron and ion heat transport as well as significant particle transport differentiating them from MTMs \cite{hatch2017gyrokinetic}.

\subsubsection{High $k_y$ MTM}

From $3.0<k_y\rho_s<7.0$, a higher $k_y\rho_s$ MTMs is seen with the eigenfunction at $k_y\rho_s=4.2$ shown in Figure \ref{fig:ky_4_eigfunc}. This mode is also tearing as they had $C_{\mathrm{tear}}=0.5$. These are much less extended than the MTMs seen at lower $k_y$. Again, $B_{||}$ has little effect on these modes.

These MTMs are fundamentally different to the low $k_y$ MTMs as they will be shown to be unstable in the collisionless regime, which have been found previously \cite{swamy2015collisionless, dickinson2013microtearing, geng2020physics}, though the driving mechanisms are not completely understood.

Above $k_y\rho_s=7.0$ all the modes were found to be stable which suggests that ETG-like modes may not be significant for this equilibria. The cause of the stability is examined in more detail in \ref{app:high_ky_etg}.

\subsection{Impact of flow shear}
\label{sec:low_ky_kbm_mtm}

\subsubsection{Low $k_y$ modes}
Flow shear has been shown to stabilise turbulence by shearing the modes. Linearly this corresponds to making the radial wavenumber, $k_x$, of the modes time dependent. The effectiveness of flow shear stabilisation can thus be determined by examining how the stability varies with $k_x$ and in GS2 this is done with the parameter $\theta_0 \equiv \frac{k_x}{k_y \hat{s}}$ \cite{highcock2012zero, roach2009gyrokinetic}. If the mode is stable at different $\theta_0$, then when flow shear advects the mode it can be moved into a stabilising region, reducing its impact on the transport.

\begin{figure}[!htb]
    \begin{subfigure}{0.49\textwidth}
        \centering
        \includegraphics[width=75mm]{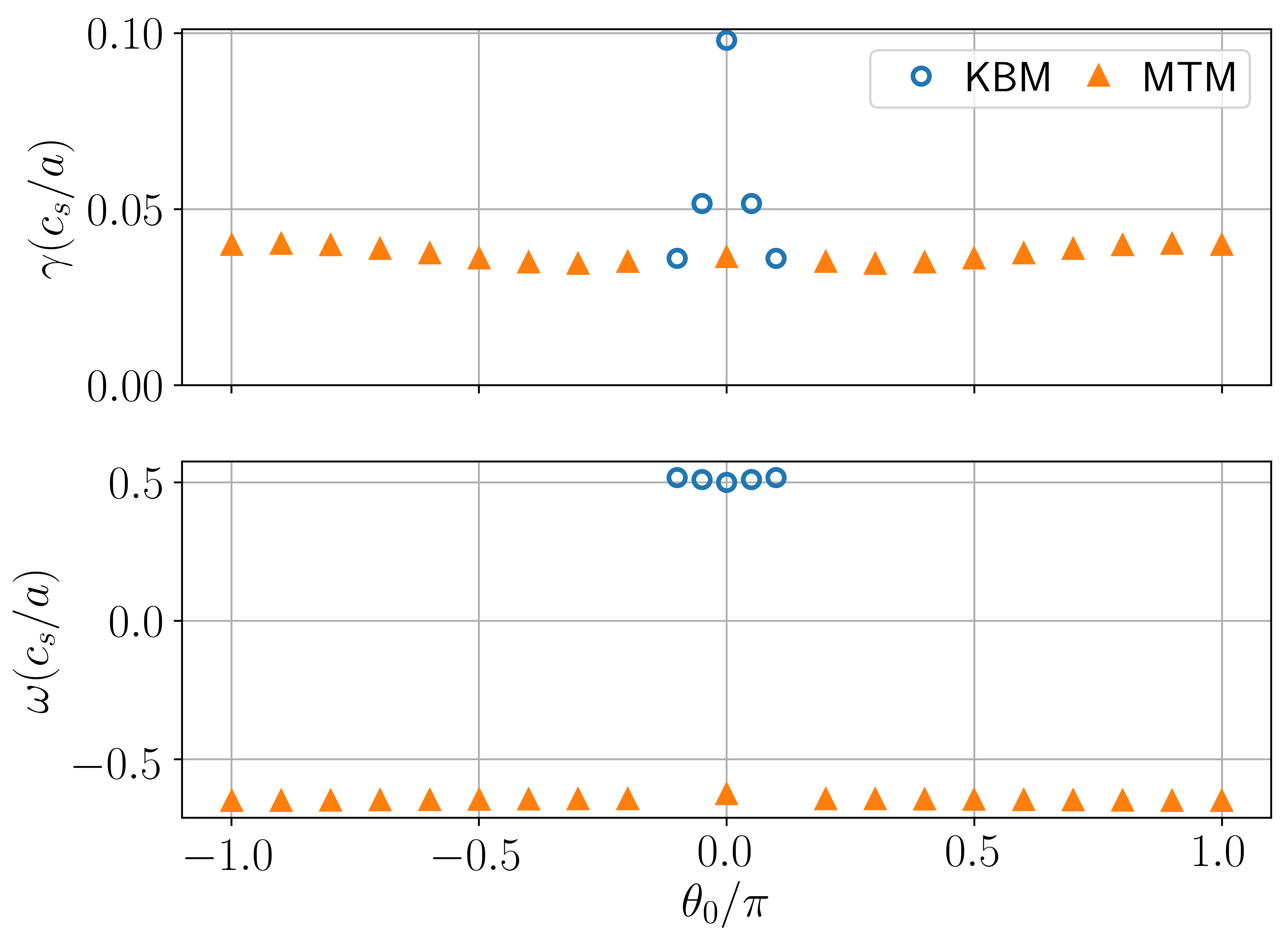}
        \caption{}
        \label{fig:ky_0.35_theta0_scan}
    \end{subfigure}
    \begin{subfigure}{0.5\textwidth}
        \centering
        \includegraphics[width=75mm]{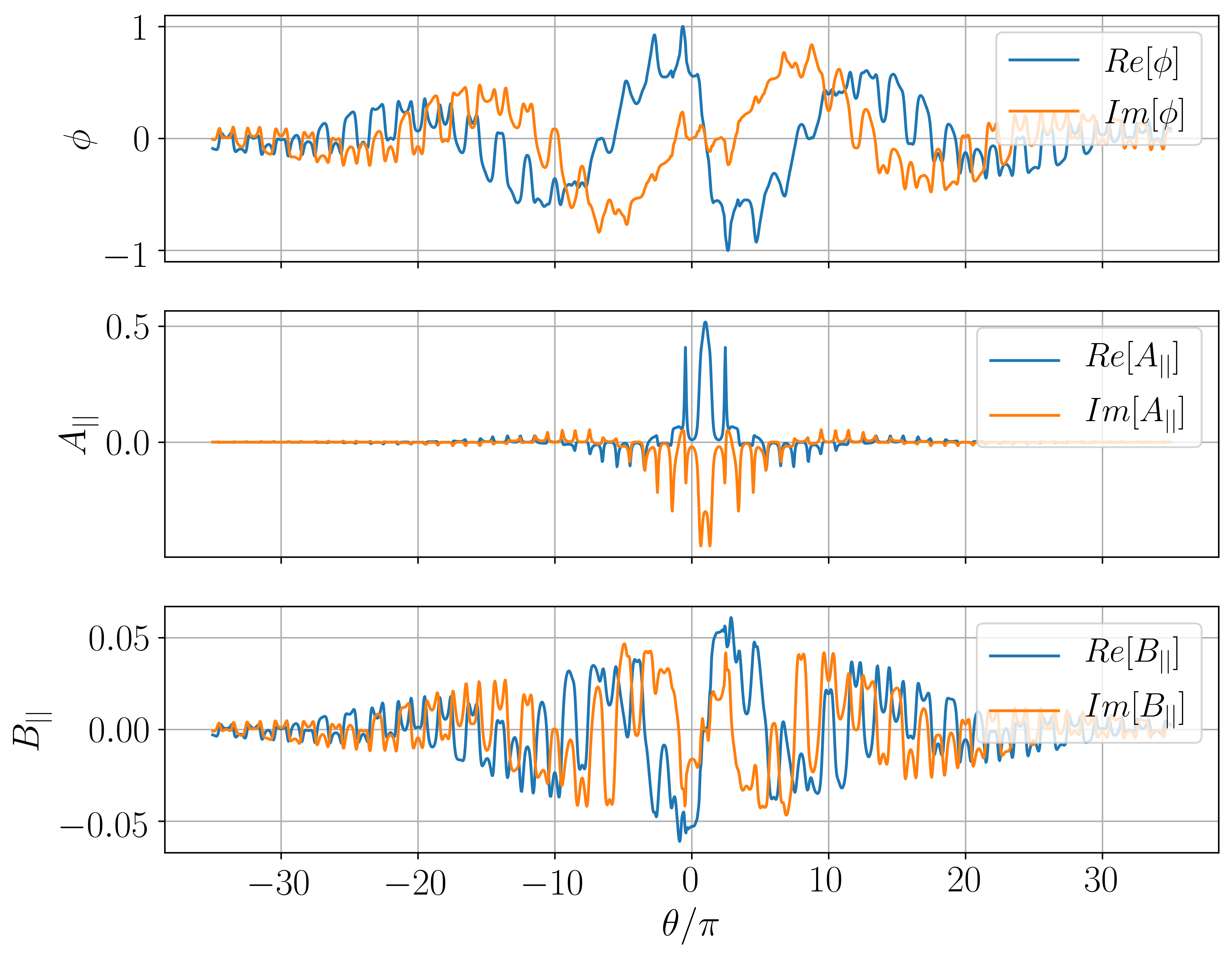}
        \caption{}
        \label{fig:ky_0.35_theta0_pi}
    \end{subfigure}
    \begin{subfigure}{0.5\textwidth}
        \centering
        \includegraphics[width=75mm]{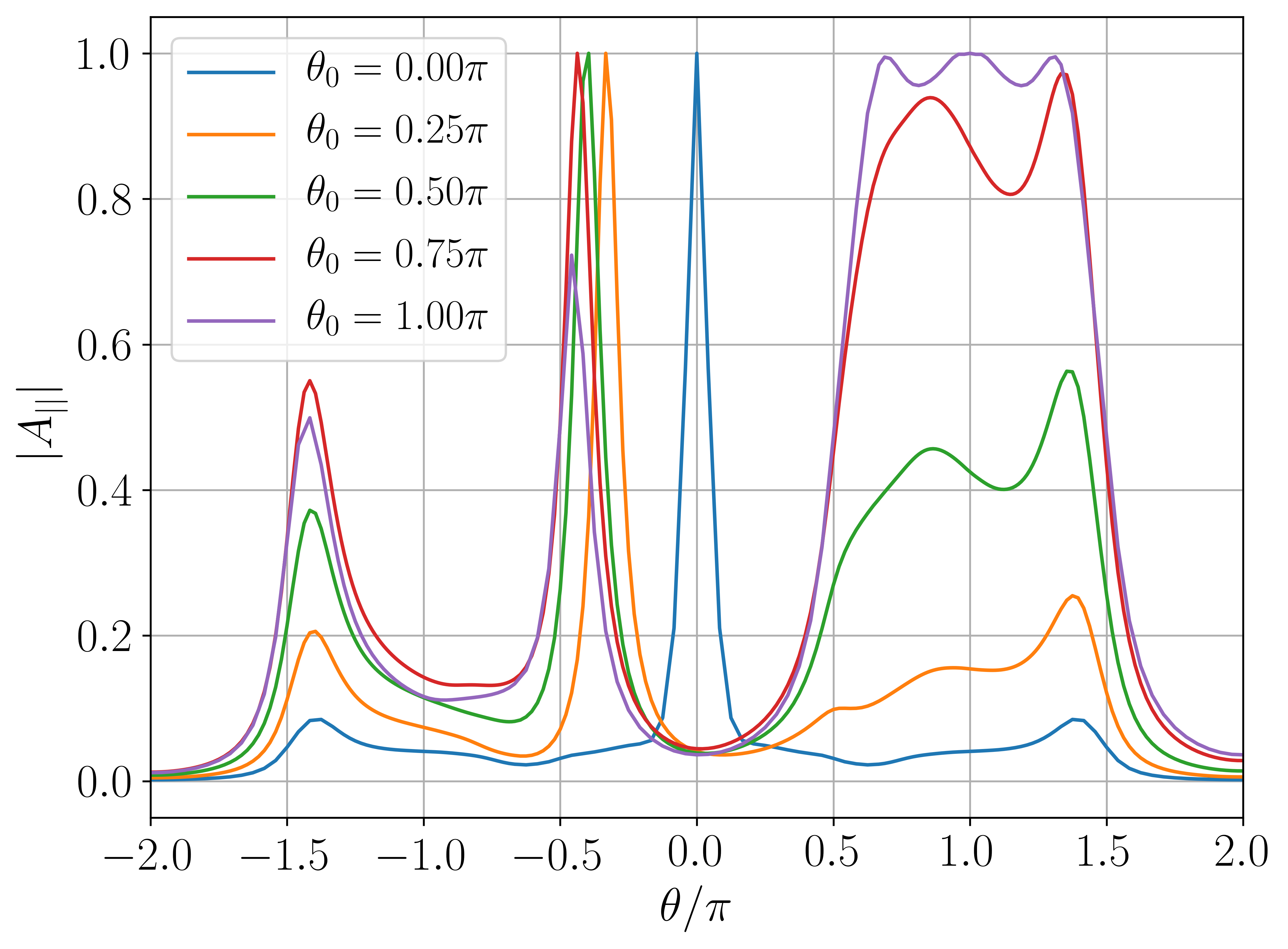}
        \caption{}
        \label{fig:multi_theta0_close}
    \end{subfigure}
    \begin{subfigure}{0.5\textwidth}
        \centering
        \includegraphics[width=75mm]{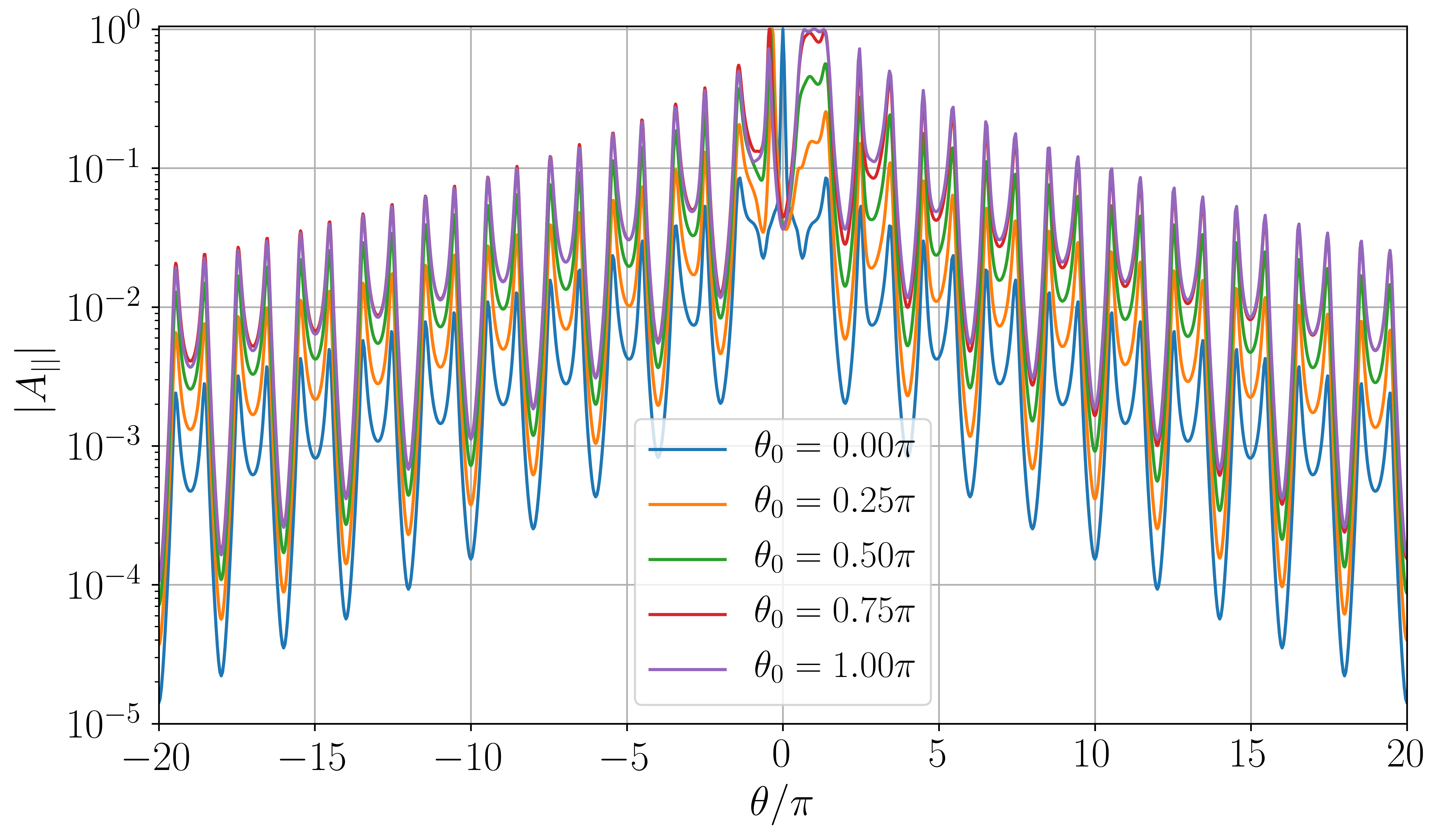}
        \caption{}
        \label{fig:multi_theta0}
    \end{subfigure}
    \caption{a) Growth rate and frequency of dominant mode at $k_y\rho_s=0.35$, over a scan in $\theta_0$. The KBM only dominates around $\theta_0=0$, while MTMs take over as the dominant mode for $|\theta_0|>0.15$ with eigenvalue that is independent of $\theta_0$. b) The MTM eigenfunction when $\theta_0=\pi$ and it can be seen that the peak in $A_{||}$ occurs at $\theta=\pi$. c) A close up of the eigenfunctions for the MTMs for a range of $\theta_0$. d) Full eigenfunction of the MTMs at various $\theta_0$ on a vertical log plot.}
    \label{fig:ky_0.35_theta0}
\end{figure}

A scan was done from $\theta_0=-\pi \rightarrow \pi$ at $k_y\rho_s=0.35$ to determine the impact of flow shear on the low $k_y$ modes. Even or odd solutions only exist when the system is up-down symmetric which is only true when $\theta_0=0$. Therefore, only the dominant mode can be examined with GS2 initial value simulations when $\theta_0\neq0$. 

Figure \ref{fig:ky_0.35_theta0_scan} illustrates the sensitivity of the growth rates and mode frequency to $\theta_0$ at $k_y\rho_s=0.35$. It is clear that the KBM is highly ballooning and only dominates in an extremely narrow range of $\theta_0 \sim 0$. This suggests that KBMs are highly susceptible to flow shear stabilisation and that flow shear will act to increase their critical gradient.

However, Figure \ref{fig:ky_0.35_theta0_scan} shows the MTM growth rate is largely unaffected by variations in $\theta_0$ suggesting that the MTMs will be more resilient against flow shear stabilisation. The eigenfunction for $\theta_0=\pi$ is shown in Figure \ref{fig:ky_0.35_theta0_pi}; this can be compared with Figure \ref{fig:ky_0.35_eigfunc_odd} showing the MTM at the same $k_y\rho_s$ with $\theta_0=0.0$ where the similar structure is evident. This MTM is clearly insensitive to curvature as the mode is equally unstable at the inboard side compared to the outboard side. This indicates that the drifts reversal, which can occur at the outboard side, may not impact these modes. The central peak in $A_{||}$ occurs at $\theta=\pi$ for $\theta_0=\pi$ in Figure \ref{fig:ky_0.35_theta0_pi}. Figure \ref{fig:multi_theta0_close} overlays the eigenfunctions at different $\theta_0$ around $\theta=0$ which shows that the central peak moves across $\theta$ as $\theta_0$ in increased. Figure \ref{fig:multi_theta0} shows the full eigenfunction, and the extended tail region all have a similar decaying envelope, indicating that the physics along these tails is not impacted by the central region, which is qualitatively similar to the extended electron tails seen recently in \cite{hardman2021extended}. It is noted that compared to the $\theta_0=0.0$ case the relative amplitude of the tails is smaller than for the cases with finite $\theta_0$.

\begin{figure}[!htb]
    \centering
    \includegraphics[width=75mm]{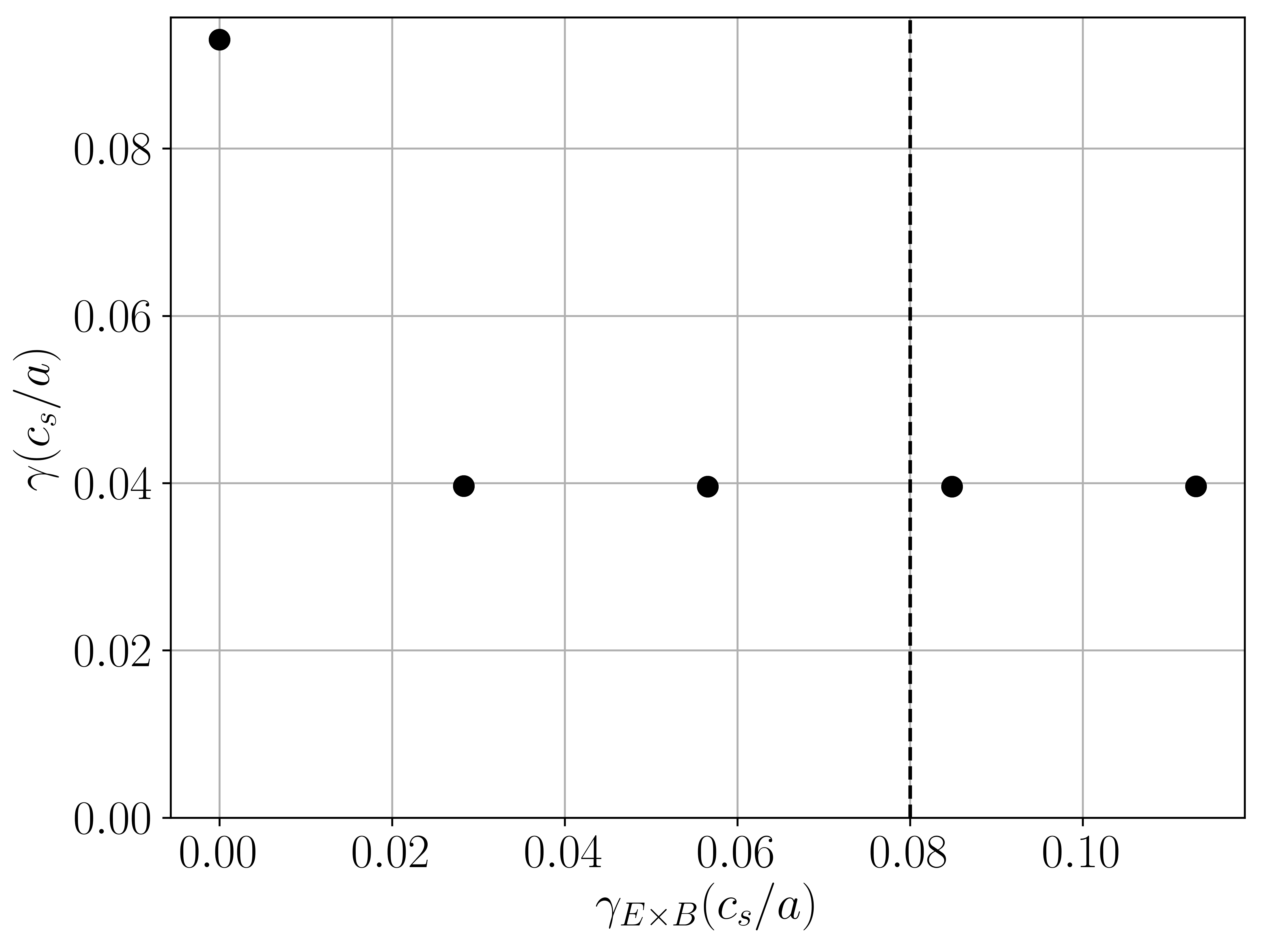}
    \caption{Effective growth rate when including flow shear in the baseline equilibrium where the vertical black line represents the diamagnetic flow shear $\gamma_{\mathrm{dia}}$.}
    \label{fig:ky_0.35_flowshear}
\end{figure}

These results demonstrate that there is a competition between KBMs and MTMs in this region. A small amount of flow shear will reduce the impact the KBM has on the transport as the mode will spend very little time in the $\theta_0$ region that is KBM unstable. However, the MTM which persists across $\theta_0$, will likely contribute significantly to the fluxes regardless of flow shear. This is illustrated in Figure \ref{fig:ky_0.35_flowshear}, where $\gamma_{E\times B}$ is included in a linear simulation by allowing $\theta_0$ to vary in time. The vertical dashed black line shows diamagnetic flow shear level $\gamma_{\mathrm{dia}}$ \cite{applegate2004microstability}. With no flow shear the effective growth rate of a $\theta_0=0.0$ mode is that of the KBM with $\gamma=0.093 c_s/a$. This is stabilised to some extent by a small amount of flow shear, but the effective growth rate remains at $\gamma\approx0.04 c_s/a$, which is due to the MTM. On inclusion of flow shear, the eigenmode becomes that of the MTM, which dominates over the KBM for most of the $\theta_0$ range apart from a narrow region around $\theta_0=0.0$.

\subsubsection{High $k_y$ modes}

\begin{figure}[!htb]
    \begin{subfigure}{0.49\textwidth}
        \centering
        \includegraphics[width=75mm]{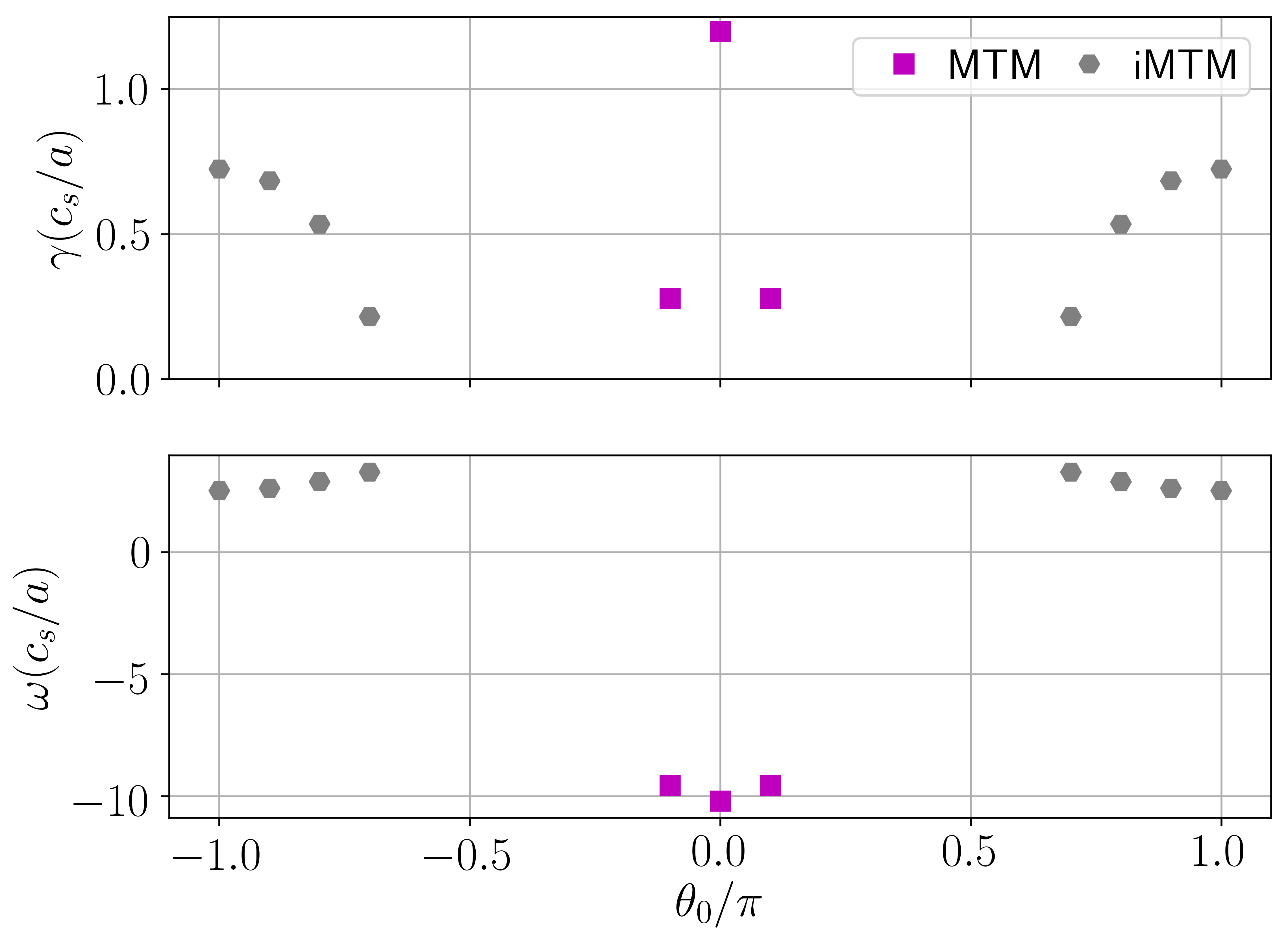}
        \caption{}
        \label{fig:ky_4_theta0_scan}
    \end{subfigure}
    \begin{subfigure}{0.5\textwidth}
        \centering
        \includegraphics[width=75mm]{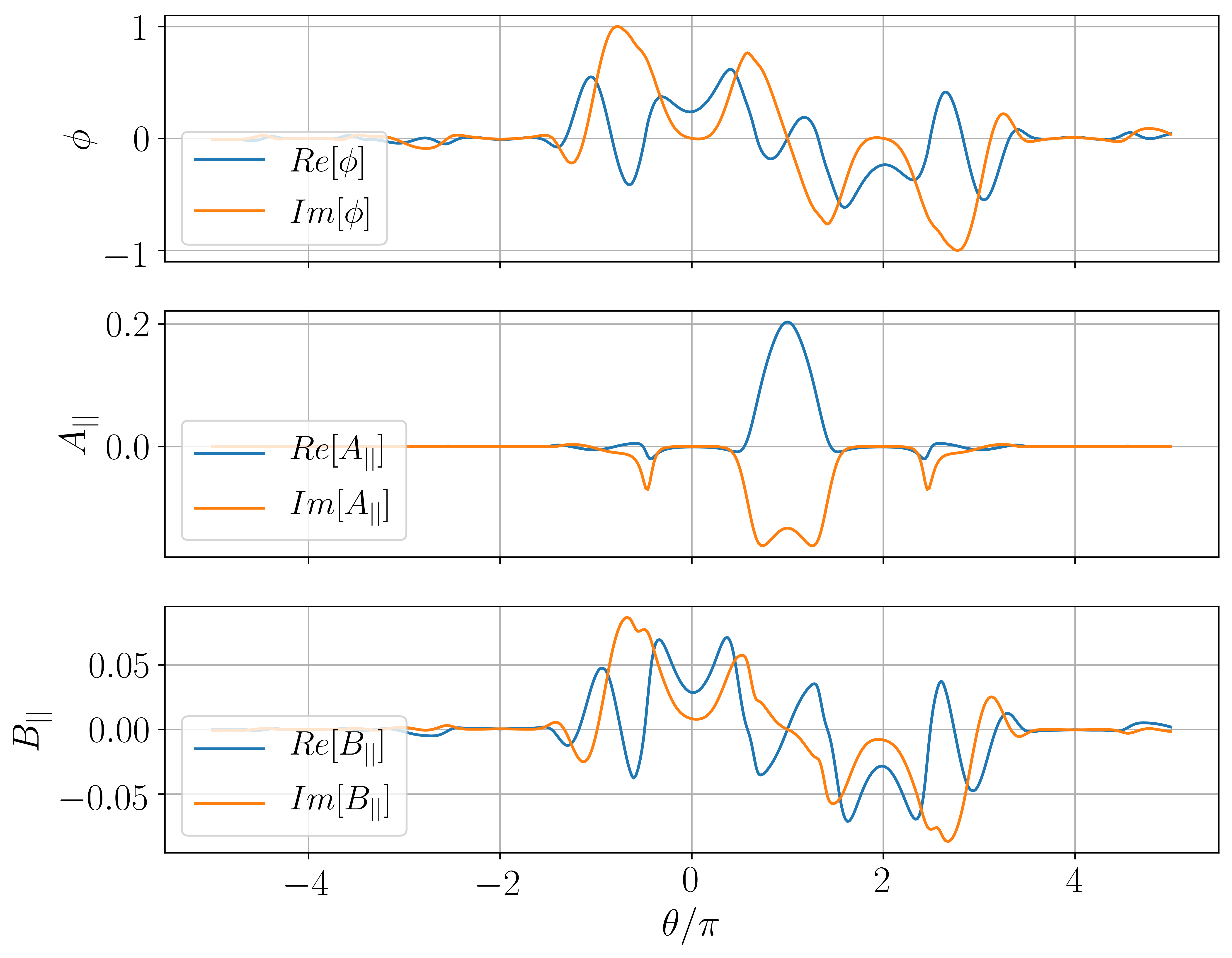}
        \caption{}
        \label{fig:ky_4_theta0_pi}
    \end{subfigure}
    \begin{subfigure}{\textwidth}
        \centering
        \includegraphics[width=75mm]{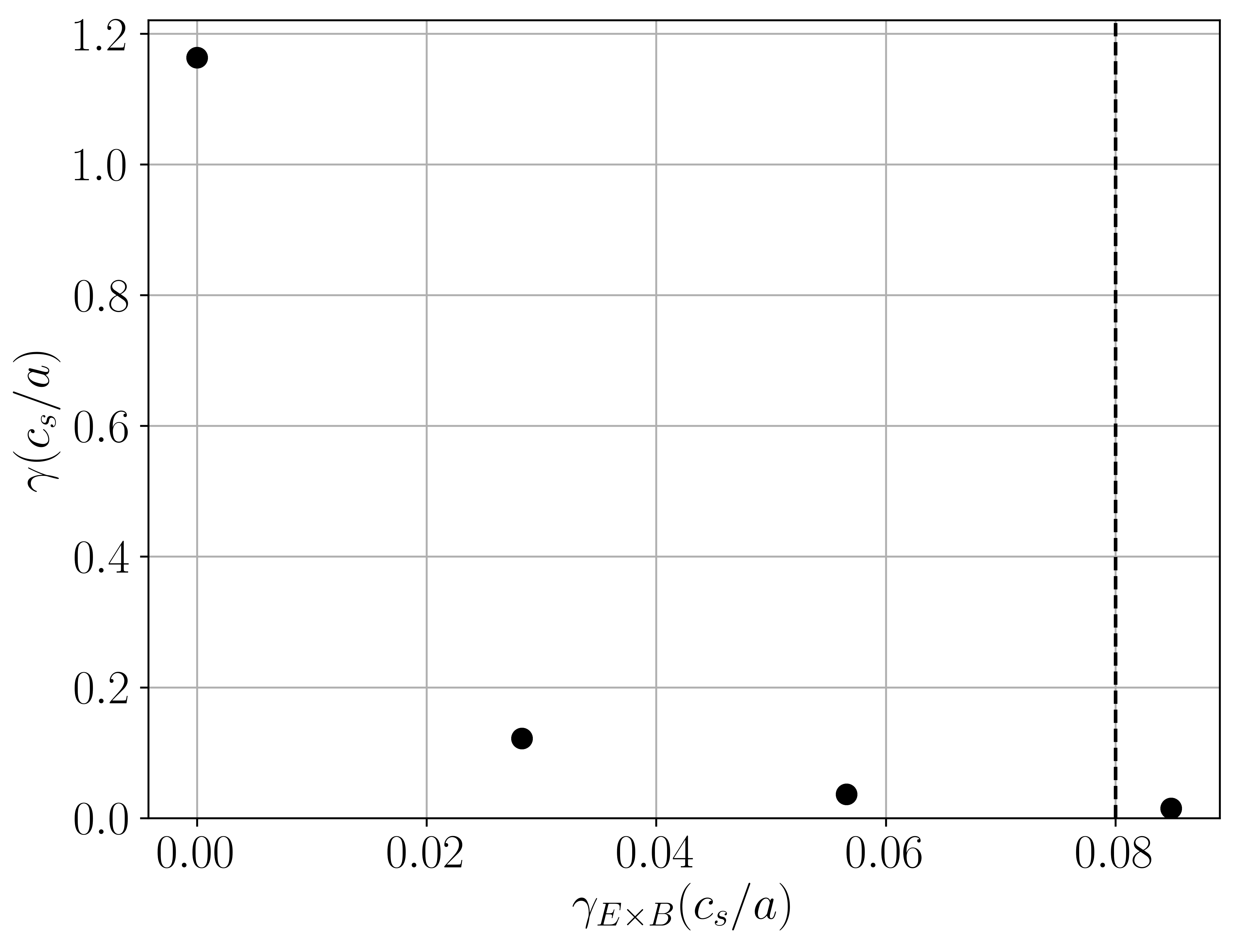}
        \caption{}
        \label{fig:ky_4_flowshear}
    \end{subfigure}
       \caption{a) $\theta_0$ scan of the dominant mode at $k_y\rho_s= 4.2$, b) eigenfunction for the inboard iMTM which should be compared with the eigenfunction of the outboard MTM at the same $k_y$ shown in Figure \ref{fig:ky_4_eigfunc}. c) effective growth rate with $E\times B$ shear. The vertical black line shows $\gamma_{\mathrm{dia}}$.}
    \label{fig:ky_4_theta0}
\end{figure}

This type of scan was then repeated for the high $k_y$ MTM. Figure \ref{fig:ky_4_theta0_scan} illustrates this MTM has a narrow peak in $\gamma$ around $\theta_0=0$, indicating that it is highly ballooning (like the KBM) and will also be stabilised by a small amount of flow shear.

However, a different tearing parity mode appears at $\theta_0=\pm \pi$ that rotates in the ion diamagnetic direction but has an odd parity $\phi$ about $\theta_0=\pi$. The eigenfunction is shown in Figure \ref{fig:ky_4_theta0_pi} and it has $C_{\mathrm{tear}}=0.9$, making it more tearing than the MTM situated around $\theta_0=0$. This is labelled as an iMTM (grey filled hexagons) and a tearing ion direction mode has not been reported in the literature before (to the best of the authors' knowledge), especially one that has a maximum growth rate on the inboard side. This highlights the exotic nature of this equilibrium.

Between $0.1\pi<|\theta_0|<0.7\pi$, both modes are stable. Including flow shear will move the modes through the stable region, resulting in a lower overall growth rate. Figure \ref{fig:ky_4_flowshear} shows that the inclusion of flow shear reduces the effective growth, even for very low values of $\gamma_{E\times B}$, such that at diamagnetic levels of flow shear the mode is close to stable. A $\gamma/k_\perp^2$ argument would suggest that the impact of these MTMs on the total transport may not be significant.

Reduced levels of transport might be anticipated in a local equilibrium where the low $k_y$ MTMs could be stabilised given that the KBMs and high $k_y$ MTMs will likely be stabilised by flow shear.

\section{Parametric dependence of micro-instabilities}
\label{sec:mode_dependancy}

The rest of this work aims to understand what drives these modes such that the equilibrium can be re-designed to help to stabilise them. The parametric dependence of the linear modes will be determined which will help to identify which plasma parameters should be the focus of optimisation. The primary focus here will be to find actuators to stabilise the low $k_y$ MTMs which appear to be the most dangerous from the transport point of view. The KBMs and high $k_y$ MTMs can be largely ignored when optimising the equilibrium, so long as they aren't driven significantly more unstable. 

This section will examine the impact of the impact of certain local parameters which were found to be the most significant in the stability properties of this equilibrium. The parameters examined were $a/L_{Te}$, $a/L_{Ti}$, $a/L_n$, $\nu_{ee}$, $Z_{\textrm{eff}}$, $\beta_{e, \mathrm{unit}}$, $\beta'_{e, \mathrm{unit}}$, $q$ and $\hat{s}$. This will then guide global changes that can be made to the equilibrium

\subsection{Impact of kinetic profiles}

In SCENE, the density and temperature profiles are prescribed, meaning the main transport assumption is quantified via the $H_{98}$ and $H_{\mathrm{Petty}}$ scaling laws. This sub-section will investigate the impact of the kinetic profiles on the micro-stability. Changing the kinetic profiles will change the kinetic gradients $a/L_n$ and $a/L_T$, so the impact of these needs to be quantified to identify desirable operating scenarios. In this section the electron and ion temperature gradient will be independently changed to explore their impact on the KBMs and MTMs. Next the density gradient of the electrons and ions will be scanned together such that quasi-neutrality is maintained. 

In the gyrokinetic equation the pressure gradient influences the local equilibrium in two ways: firstly the impact on local magnetic equilibrium including magnetic drifts and local shear via the current profile; secondly its impacts on the profile gradients providing the linear drive terms. Usually in gyrokinetic codes, it is possible to define these terms independently allowing for the impact of those terms to be isolated. In GS2 the pressure gradient parameter, $\beta'$, affecting the magnetic geometry is prescribed independently of the kinetic profile gradients. A self-consistent pressure gradient would depend on the profile gradients, and satisfy: $\beta_{e}'\equiv -\beta_{e} a/L_{p}$ where $a/L_p = \sum_s\frac{ n_sT_s}{n_{ref}T_{\mathrm{ref}}} (a/L_{Ts} + a/L_{ns})$\footnote{This work is missing the contribution from the fast ions which may be significant at high $\beta$}. In the scans in this section we vary the kinetic profile gradients (inconsistently) at fixed magnetic geometry, to allow for the impact of the kinetic gradients to be isolated\footnote{$\beta'$ is related to the parameter $\alpha$ often used to define the pressure gradient with $\alpha = Rq^2 \beta'$ in the infinite aspect ratio shifted circular geometry limit.}.

Changing the density and temperature profiles will also impact the collisionality as $\nu_* \propto n_s/T_s^{2}$, so a higher density, lower temperature scenario will have a higher collisionality. Furthermore, the impact of impurities and fast ions will be important. This work won't include an impurity species, but will investigate the impact $Z_{\textrm{eff}}$ has on these modes. Examining the impact of fast ions such as fusion $\alpha$'s is left as future work.

\subsubsection{Electron Temperature gradient}

\begin{figure}[!bth]
    \begin{subfigure}{0.5\textwidth}
        \centering
        \includegraphics[width=75mm]{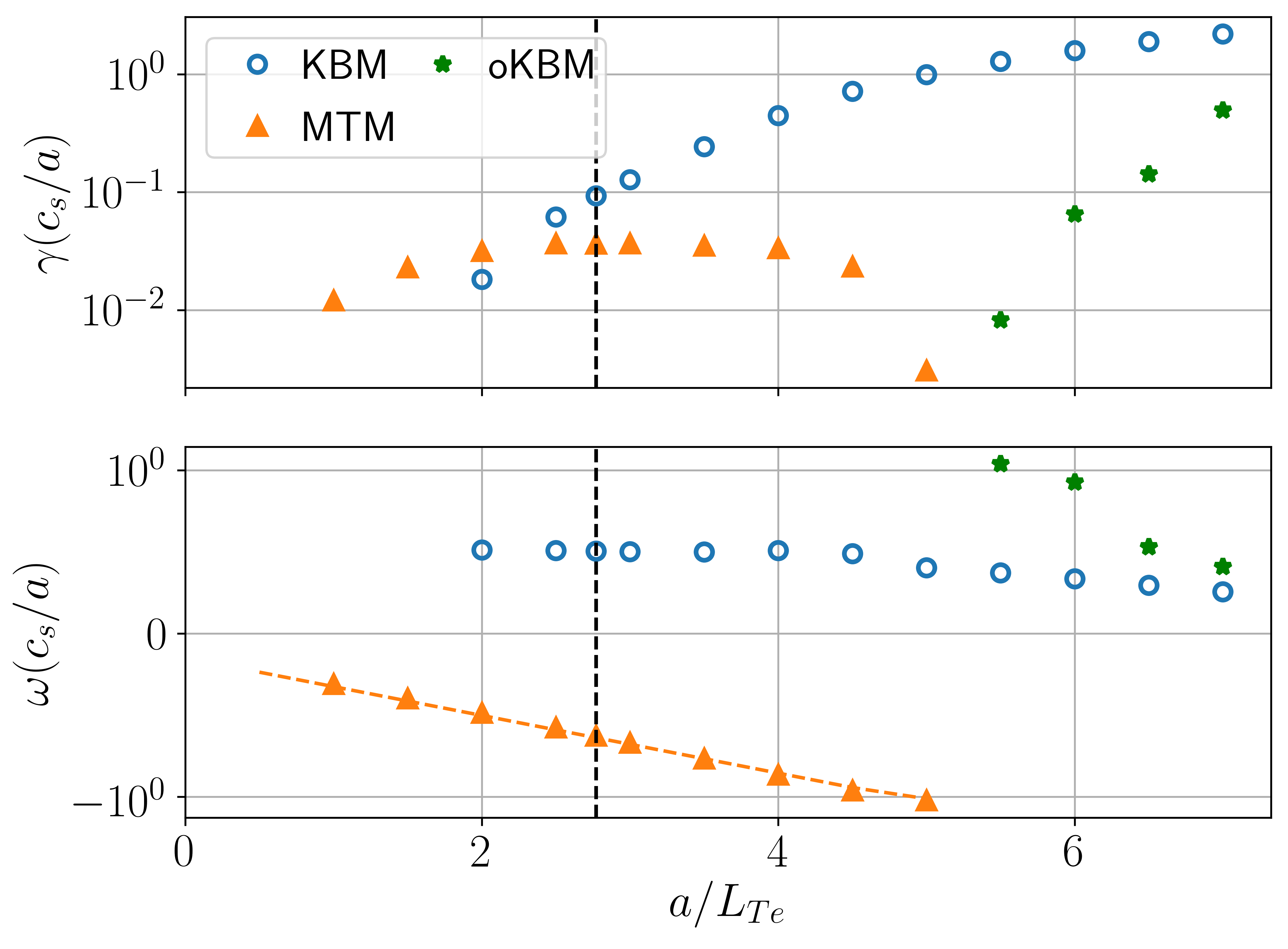}
        \caption{}
        \label{fig:ky_0.35_lte_scan}
    \end{subfigure}
    \begin{subfigure}{0.5\textwidth}
        \centering
        \includegraphics[width=75mm]{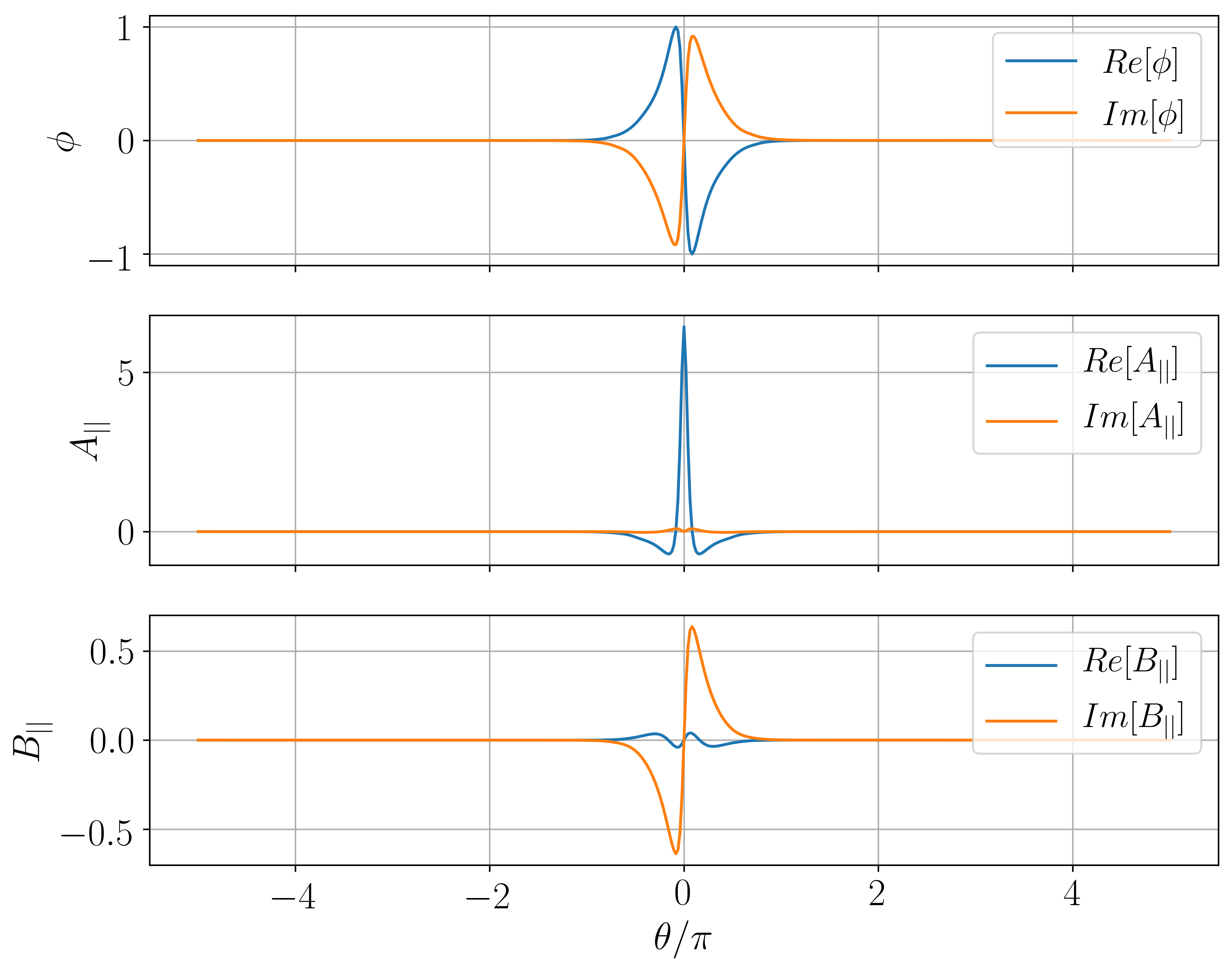}
        \caption{}
        \label{fig:ky_0.35_lte_7}
    \end{subfigure}
       \caption{a) Examining the impact of plasma $a/L_{Te}$ when $k_y\rho_s=0.35$. The dominant even and odd instabilities are shown, with the KBM demonstrating much stiffer behaviour. Note the log scale in $\gamma$. The dashed orange line shows the analytic prediction $\omega^{CR}_{\mathrm{MTM}}$ defined in the text. The vertical dashed black line shows the equilibrium value of $a/L_{Te}$. b) Eigenfunction of the ion direction odd parity mode (filled green stars) for $a/L_{Te}=7.0$; despite having odd parity, the mode is not tearing as $C_{\mathrm{tear}}=0.0$.}
    \label{fig:ky_0.35_lte}
\end{figure}

The impact of the electron temperature gradient was examined by scanning from $a/L_{Te}=0\rightarrow 7.0$ at $k_y\rho_s=0.35$. It is expected that this will have a significant impact on all the modes seen thus far, as MTMs are driven unstable by the electron temperature gradient and KBMs are driven unstable by the total pressure gradient. The eigenvalues are shown in Figure \ref{fig:ky_0.35_lte_scan}, with the reference equilibrium value shown with the vertical black dashed line.

There appears to be a critical gradient where the KBM is completely stable at $(a/L_{T_e})^{\mathrm{KBM}}_{\mathrm{crit}}=2.0$, which would correspond to $a/L_p = 5.63$. Its growth rate increases exponentially with $a/L_{Te}$ which may lead to stiff transport and if $\gamma$ becomes sufficiently large then the flow shear stabilisation may not be sufficient in suppressing the transport\footnote{Of course as $a/L_{Te}$ increases, so would $\beta'_{e,\mathrm{unit}}$, which will be shown in Section \ref{sec:beta_betaprime} to be stabilising so the combined impact is not clear.}. The mode frequency is dropping as the temperature gradient increases. A small drop in $a/L_{T_e}$ would allow for the MTM to become the dominant instability.

Looking at the dominant odd eigenmode, it can be seen that there's a critical gradient $(a/L_{Te})^{\mathrm{MTM}}_{\mathrm{crit}}=1.0$. This critical gradient may be the limiting factor on the electron temperature profile as MTMs can drive significant electron heat flux. The mode frequency scales with the temperature gradient and this follows predictions made by Catto and Rosenbluth \cite{catto1981trapped}, that the mode frequency of an MTM is given by $\omega_{\mathrm{MTM}}^{CR} = \omega_e^*[1 + \eta_e/2]$ where $\omega_e^*$ is the electron diamagnetic frequency defined as $\omega_e^* = k_y (a/L_{ne})$ and $\eta_e = (L_{ne}/L_{Te})$. The orange dashed line shows $\omega_{\mathrm{MTM}}^{CR}$, indicating that this scaling fits well.

\begin{figure}[!bth]
    \begin{subfigure}{0.5\textwidth}
        \centering
        \includegraphics[width=75mm]{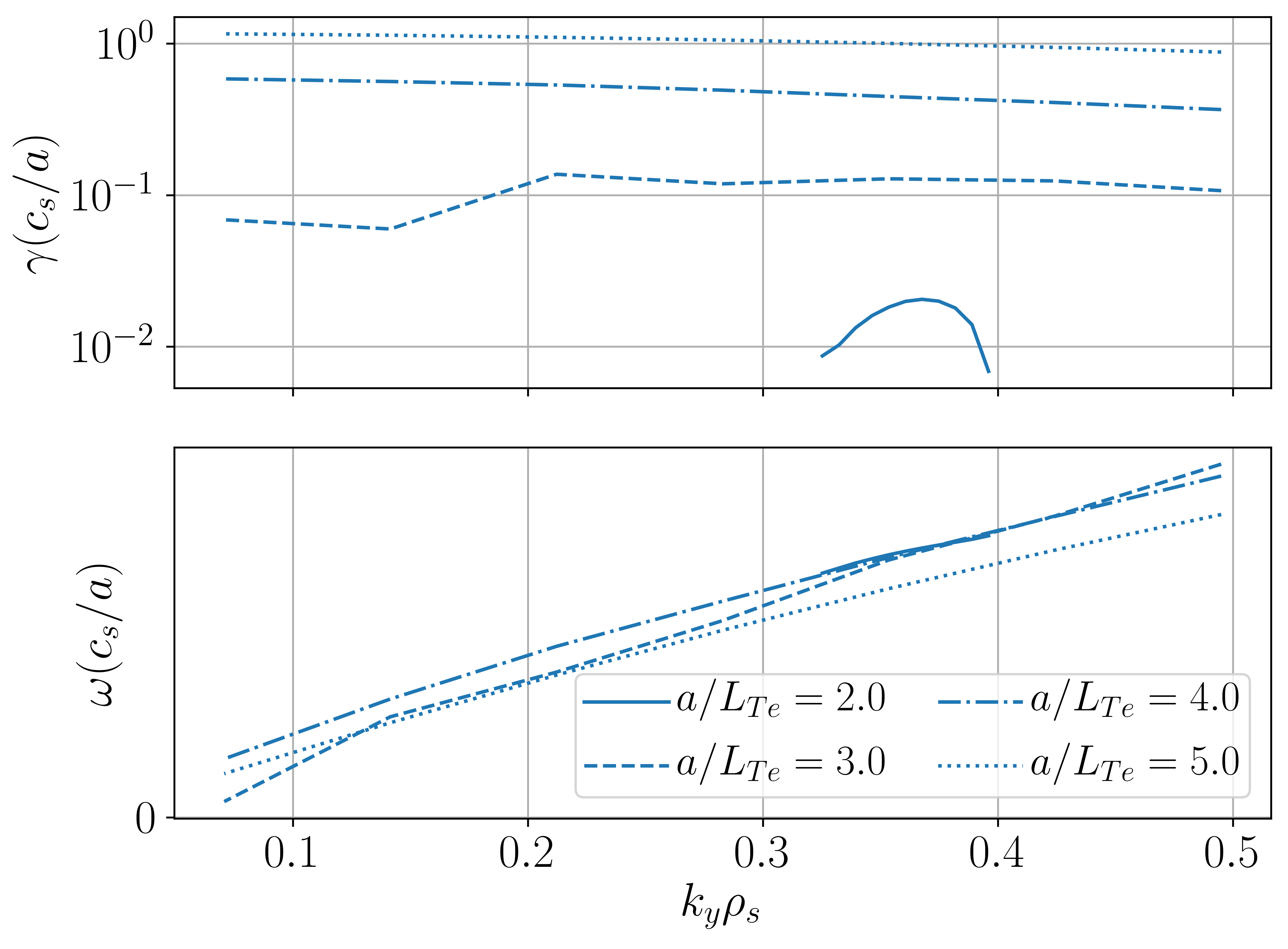}
        \caption{}
        \label{fig:ky_lte_scan_even}
    \end{subfigure}
    \begin{subfigure}{0.5\textwidth}
        \centering
        \includegraphics[width=75mm]{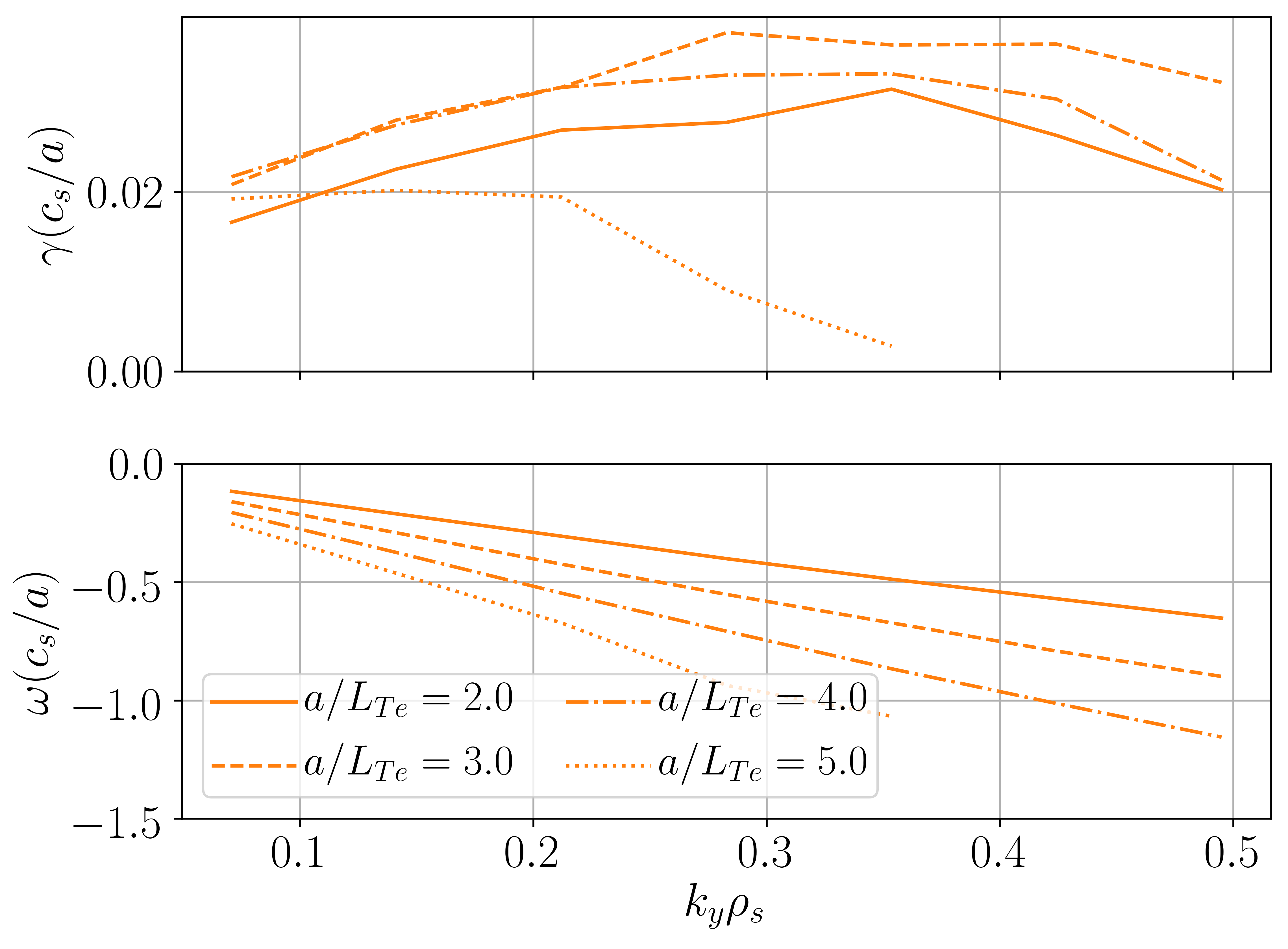}
        \caption{}
        \label{fig:ky_lte_scan_odd}
    \end{subfigure}
    \caption{A scan in $k_y$ at several different values of $a/L_{Te}$ showing the dominant a) even and b) odd mode. The colour denotes the mode type (KBM - blue, MTM - orange) and the line-style shows the electron temperature gradient. Note the log scale for the KBMs.}
    \label{fig:ky_lte_scan}
\end{figure}

The MTM growth rate has a much weaker dependence on the $a/L_{Te}$ compared to the KBM and actually appears to level out, suggesting that small changes made to the electron temperature gradient may not have an impact on the transport. At sufficiently high gradient the MTM gets stabilised and this has been seen before in MAST simulations \cite{applegate2007micro} where the MTMs also had a mode frequency like $\omega_{\mathrm{MTM}}^{CR}$. This was thought to be related to a resonance with a drift frequency. If $|\omega|$ is increased sufficiently then this resonance is disturbed and the mode becomes damped. As observed previously for MTMs, instability only arises over a finite range in $a/L_{Te}$, which interestingly, is consistent with recent analytic calculations demonstrating that toroidal ETG modes can only be unstable over a finite range in $(\eta_e \omega_e^*)/\omega_{\kappa e}$\cite{parisi2020toroidal}. Increasing $a/L_{Te}$ reduces the growth rate at a given $k_y$, but pushes the MTM spectrum to longer wavelength as seen in Figure \ref{fig:ky_lte_scan_odd}.

However, when $a/L_{Te}>5.0$, an ion direction mode emerges that has an odd $\phi$ eigenfunction, shown in Figure \ref{fig:ky_0.35_lte_7}. Its frequency is tending towards the KBM frequency. This however, is not a tearing mode as it has $C_{\mathrm{tear}}=0.0$. This rather appears to be an odd parity KBM, which will be labelled as an oKBM and will be represented with the filled green stars as seen in Figure \ref{fig:ky_0.35_lte_scan}. This oKBM is a higher order eigenstate of the KBM and has been seen before in steep-gradient simulations \cite{xie2018kinetic}. Any quasi-linear model may need to account for this extra source of transport. Access to these temperature gradients may be possible if these oKBMs are also stabilised by flow shear, but the level of transport driven by the MTM must first be quantified. If this is an oKBM, it should be seen when scanning through $a/L_{Ti}$.

A scan was also done across $k_y\rho_s$ to see how the spectra changed with $a/L_{Te}$ and the eigenvalues for the KBMs and MTMs are shown in \ref{fig:ky_lte_scan_even} and \ref{fig:ky_lte_scan_odd} respectively. It can be seen that as $a/L_{Te}$ increases the KBM remains unstable at very low values of $k_y\rho_s$, suggesting a local model may fail at high gradients\footnote{Once again, this would also increase $\beta'_{e\mathrm{unit}}$ which is stabilising so a consistent study should is required for each equilibrium.}. For the MTMs the non-monotonic behaviour persists across the $k_y$ spectrum, with the $a/L_{Te}=5.0$ having the lower growth rates than the $a/L_{Te}=4.0$ scan.

\begin{figure}[!htb]
    \centering
    \includegraphics[width=75mm]{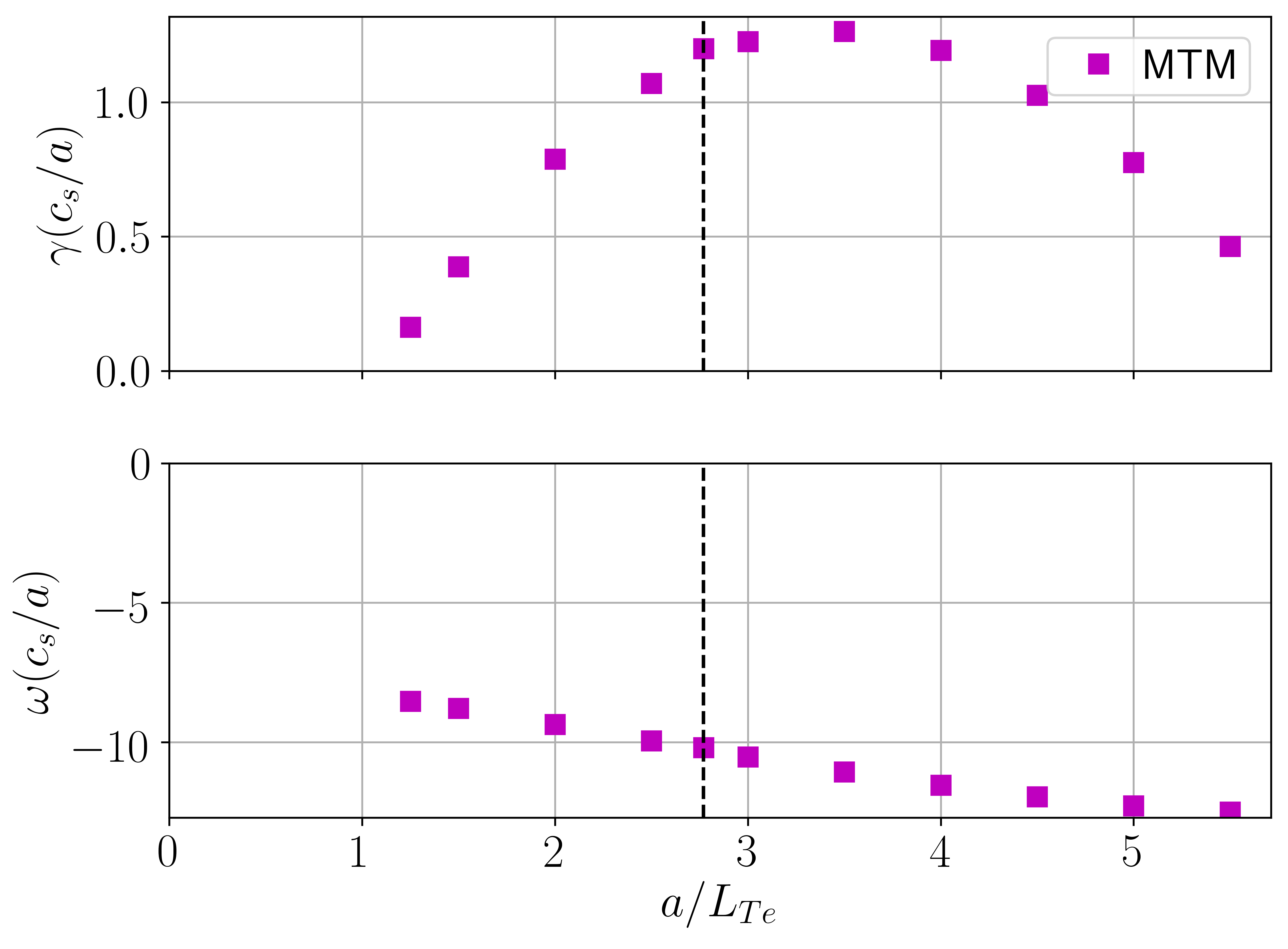}
    \caption{Examining the impact of $a/L_{Te}$ on the MTM seen at $k_y\rho_s=4.2$. The vertical dashed black line shows the equilibrium value of $a/L_{Te}$. }
    \label{fig:ky_4_lte}
\end{figure}

It is expected that the high $k_y$ MTMs will be impacted by $a/L_{Te}$ so a similar scan is shown in Figure \ref{fig:ky_4_lte} for $k_y\rho_s=4.2$. Once again non-monotonic behaviour is seen. The critical electron temperature gradient occurs at $a/L_{Te}=1.0$, similar to the lower $k_y\rho_s$ MTM seen earlier. It peaks at $a/L_{Te}=3.5$ and then begins to drop off. Significant changes in $a/L_{Te}$ would be required to stabilise this MTM, though given the sensitivity of the growth rate to $\theta_0$, the transport from these modes will be mitigated by flow shear.

\subsubsection{Ion temperature gradient}
\begin{figure}[!htb]
    \centering
    \includegraphics[width=100mm]{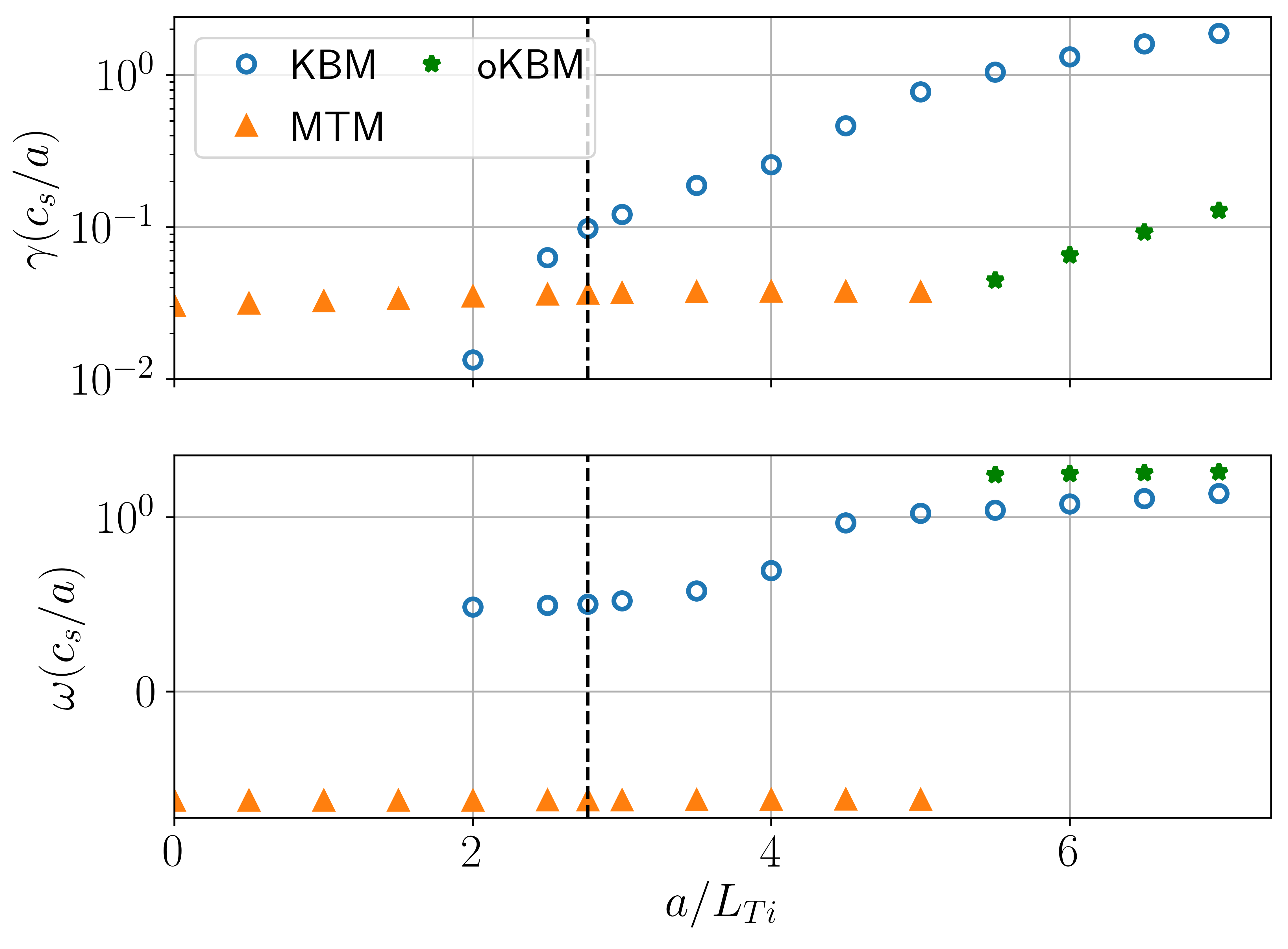}
    \caption{Examining the impact of plasma $a/L_{Ti}$ when $k_y\rho_s=0.35$. The dominant even and odd instabilities are shown, with the KBM, once again, demonstrating much stiffer behaviour. The oKBM is again found at high $a/L_{Ti}$. The vertical dashed black line shows the equilibrium value of $a/L_{Ti}$.}
    \label{fig:ky_0.35_lti_scan}
\end{figure}

A similar scan was performed for the ion temperature gradient by scanning from $a/L_{Ti}=0\rightarrow 7$ whilst keeping the other kinetic gradients fixed. For $k_y\rho_s=0.35$, the KBM and MTM eigenvalues are shown in Figure \ref{fig:ky_0.35_lti_scan}. The KBM has a similar critical gradient to the previous scan with $(a/L_{Ti})^{\mathrm{KBM}}_{\mathrm{crit}}=2.0$, as expected if the relevant parameter is $a/L_p$. Once again the KBM is strongly destabilised by $a/L_{Ti}$ and the mode frequency in this case actually increases with $a/L_{Ti}$, which suggests that the KBM frequency scales like $\eta_i/\eta_e$.

Examining the low $k_y$ MTM, its growth rate and mode frequency are largely unaffected by $a/L_{Ti}$ as expected. Once again if the ion temperature gradient is pushed high enough then when $a/L_{Ti}>5.5$ an oKBM appears at a similar threshold as the $a/L_{Te}$ scan. The high $k_y$ MTM is also unaffected by the ion temperature gradient.

It has been found in MTM driven transport, that $98\%$ of the heat transport can occur in the electron channel \cite{guttenfelder2013progress}. This suggests that $a/L_{Ti}$ will predominantly be determined by the balance between the neoclassical transport and the electron-ion exchange power, assuming the KBMs is suppressed by flow shear.

\subsubsection{Density gradient}
\begin{figure}[!bth]
    \begin{subfigure}{0.5\textwidth}
        \centering
        \includegraphics[width=75mm]{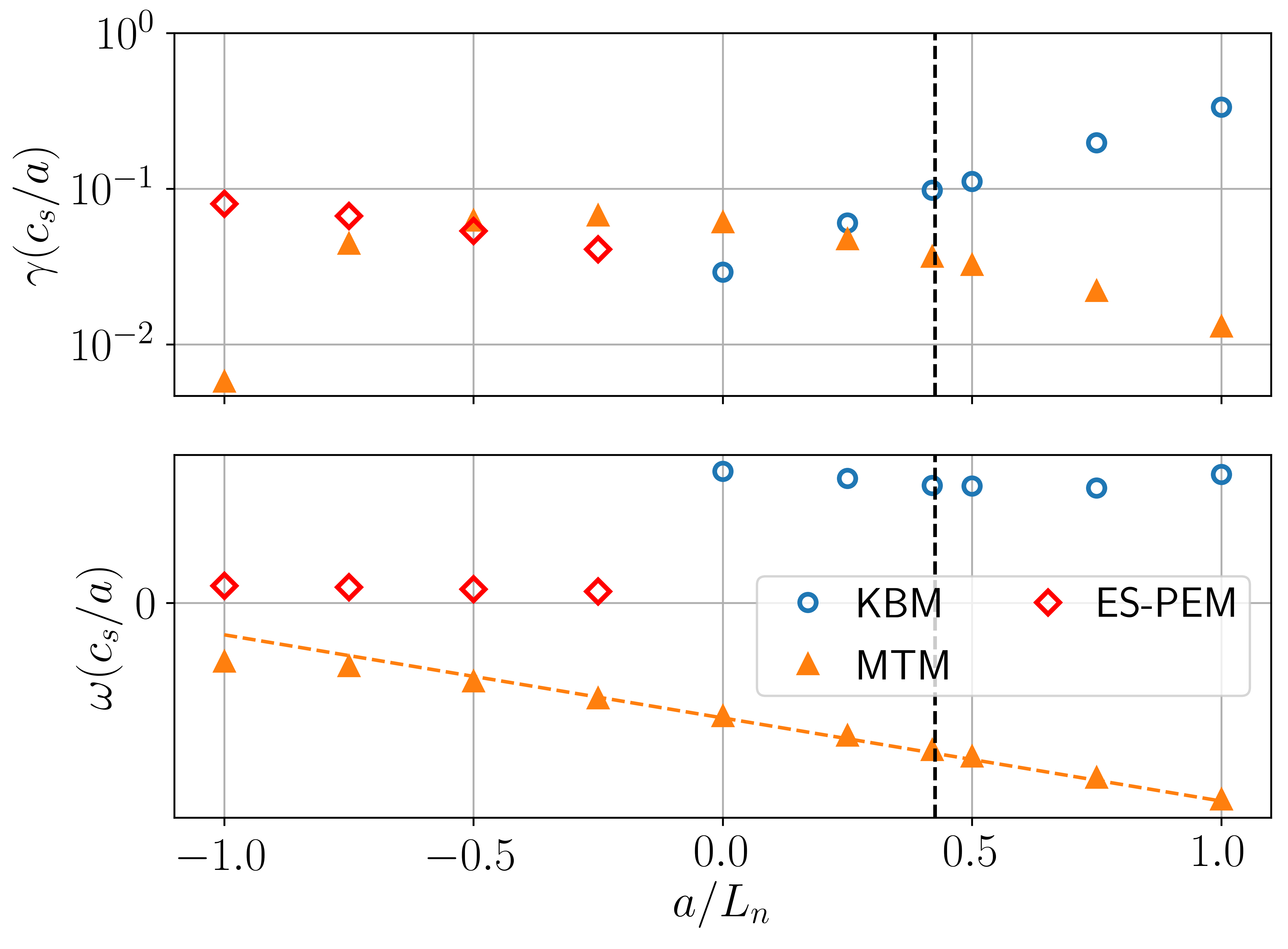}
        \caption{}
        \label{fig:ky_0.35_ln}
    \end{subfigure}
    \begin{subfigure}{0.5\textwidth}
        \centering
        \includegraphics[width=75mm]{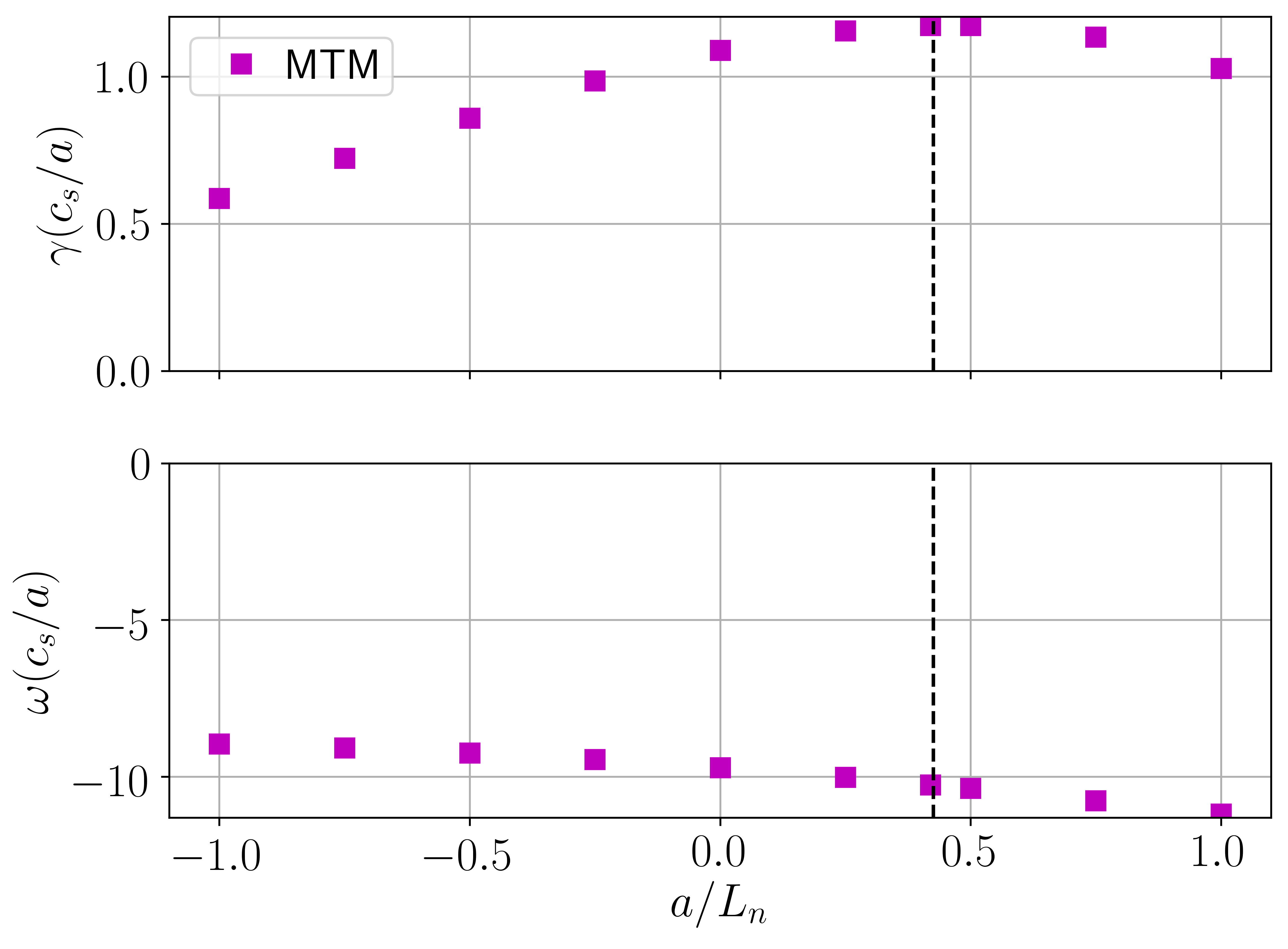}
        \caption{}
        \label{fig:ky_4_ln}
    \end{subfigure}
    \caption{Examining the impact of $a/L_{n}$ when a) $k_y\rho_s=0.35$ and b) $k_y\rho_s=4.2$. The dominant even and odd instabilities are shown. The vertical dashed black lines shows the equilibrium value of $a/L_{n}$.}
    \label{fig:ky_ln}
\end{figure}

A density gradient scan was performed from $a/L_n=-1 \rightarrow 1$, corresponding to $a/L_p=3.54 \rightarrow7.54$, with the negative density gradient allowing for a similar lower value of $a/L_p$ as the temperature gradient scans. It can be seen from Figure \ref{fig:ky_0.35_ln} that when $a/L_n$ is increased the KBM is destabilised, which further supports that this is a pressure gradient driven mode. The KBM is stable when $a/L_n<0$ corresponding to $a/L_p=5.54$ which is similar to the critical value in the temperature gradient scan. The mode frequency is unaffected which supports the idea that $\omega_{\mathrm{KBM}}$ scales with $\eta_i/\eta_e$ which remains unchanged in this scan. When the density gradient is negative the KBM is stabilised and an electrostatic passing electron mode (ES-PEM shown by hollow red diamonds) appears; though this is a scenario that should be avoided for a reactor, though may occur locally and transiently if pellet fuelling occurs at the edge. This mode was found to only require passing electrons and depended on both the electron and ion temperature gradient.

The low $k_y$ MTM seems to be stabilised by a large $|a/L_n|$, which has been seen before on NSTX \cite{guttenfelder2012scaling} and on MAST \cite{dickinson2012kinetic}. This again was thought to be due to the mode frequency changing and disrupting a resonance, though the peak growth rate here occurs at $\omega=0.4 c_s/a$ and in the $a/L_{Te}$ scan it occurred at $\omega=0.6 c_s/a$.

For the high $k_y$ MTMs, the equilibrium happens to lie at the peak of the growth rate spectrum so increasing the density gradient would help to stabilise the mode, but the effect is not as large compared to the impact on the low $k_y$ MTMs. Given the similar behaviour as the low $k_y$ modes, similar measures can be taken to stabilise these modes, noting that their impact on transport will not be as significant given the flow shear stabilisation.

\subsubsection{Scans in $a/L_T$ and $a/L_n$ at fixed $a/L_p$}

\begin{figure}[!htb]
    \centering
    \includegraphics[width=100mm]{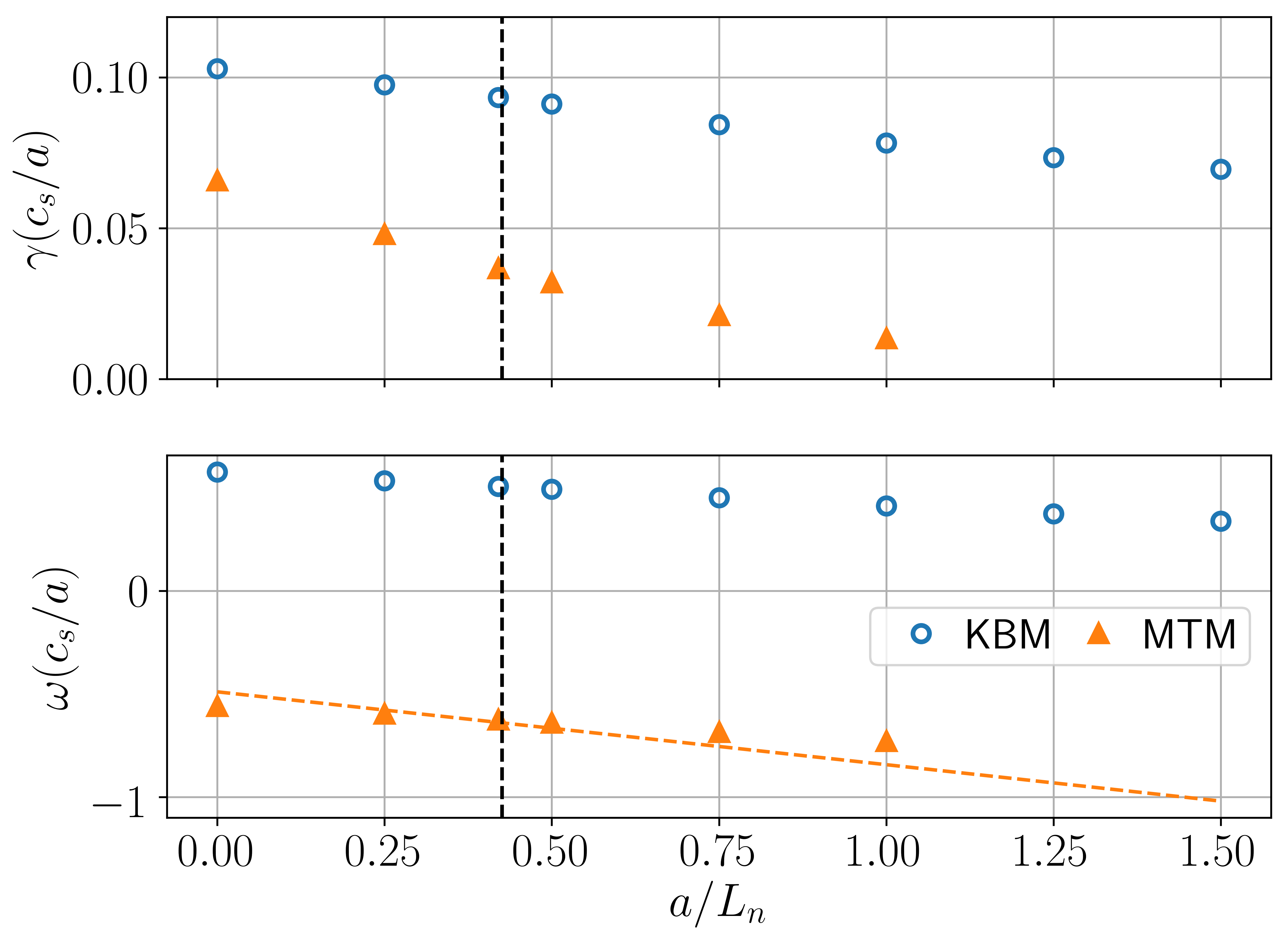}
    \caption{a) A scan was performed at fixed $a/L_p$ where the contribution of $a/L_n$ was varied for $k_y\rho_s= 0.35$. $a/L_{Te}=a/L_{Ti}$ was enforced. Note the linear scale. The vertical dashed black line shows the equilibrium value of $a/L_{n}$.}
    \label{fig:ky_0.35_fix_alp}
\end{figure}

It has been shown that increasing any of the kinetic gradients drives the KBM unstable. Assuming the KBM is driven by the total pressure gradient, if the total pressure gradient is kept fixed a similar growth rate should be seen. However, if the temperature gradient is exchanged for density gradient then it should be expected that the MTM will be stabilised. This should be doubly beneficial for the stability as the drive from $a/L_{Te}$ is reduced and the stabilisation from $a/L_n$ is being increased. A scan was performed at fixed $a/L_p$ where the density gradient was changed from $a/L_n=0 \rightarrow 1.5$ and $a/L_{Te}=a/L_{Ti}$ was set. The eigenvalues are shown in Figure \ref{fig:ky_0.35_fix_alp}. Note the linear scale here. 

It can be seen that the MTM is indeed stabilised, and when $a/L_n > 1.0$, the MTM is completely stable. The critical gradient occurs at the same value as the pure density gradient scan, indicating that reducing the electron temperature gradient is not having a significant impact on the MTM growth rate. This is explained by the ``levelling" out of the MTM growth rate seen in Figure \ref{fig:ky_0.35_lte_scan} during the $a/L_{Te}$ scan, indicating that small changes in $a/L_{Te}$ will not impact these MTMs significantly.

Overall, this implies that a peaked density profile could be beneficial in reducing MTM-based transport. The KBM is also stabilised slightly indicating the growth rate is slightly sensitive to $L_n/L_T$.


\subsubsection{Collision Frequency}

The kinetic profiles were specified to generate $P_{\mathrm{fus}}=1.1GW$, but it would be possible to operate at a higher density and lower temperature at the same fusion power. This would directly impact the collisionality of the plasma given that $\nu_* \propto n_e/T_e^{2}$.

MTMs are sometimes reported to be highly sensitive to the electron collision frequency, so a scan was conducted from $\nu_{ee}=0 \rightarrow 0.14 c_s/a$ and $\nu_{ei}$ was consistently changed assuming $T_e=T_i$. An electron collision frequency of $0.14 c_s/a$ (which corresponds to $\nu_*=0.05$) is rather high for reactor relevant conditions and would be approximately at the Greenwald limit for this device. Given the high temperatures of a reactor, it will be difficult to operate at a significantly higher collision frequency.

\begin{figure}[!htb]
    \begin{subfigure}{0.5\textwidth}
        \centering
        \includegraphics[width=75mm]{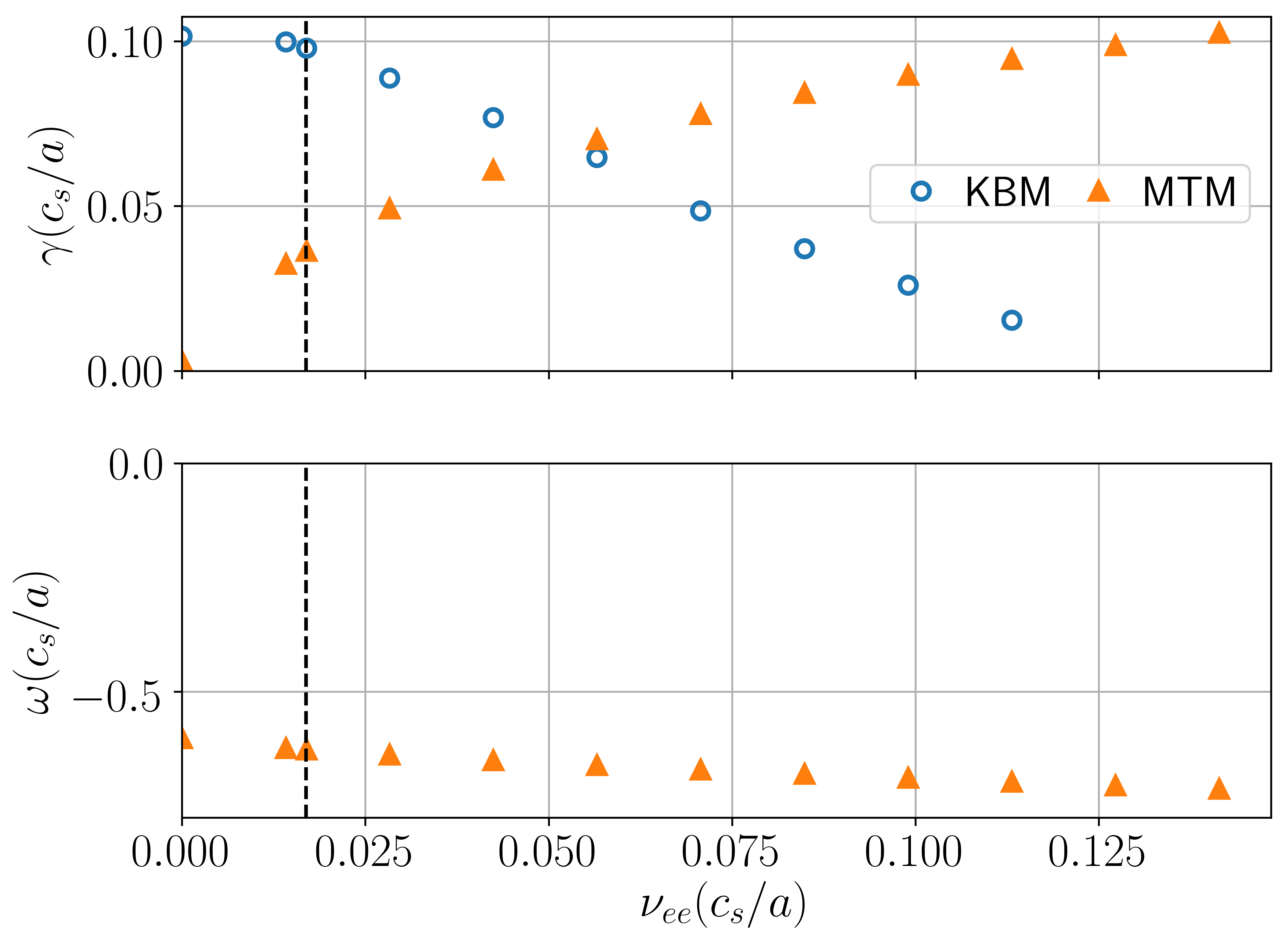}
        \caption{}
        \label{fig:ky_0.35_vnewk_scan}
    \end{subfigure}
    \begin{subfigure}{0.5\textwidth}
        \centering
        \includegraphics[width=75mm]{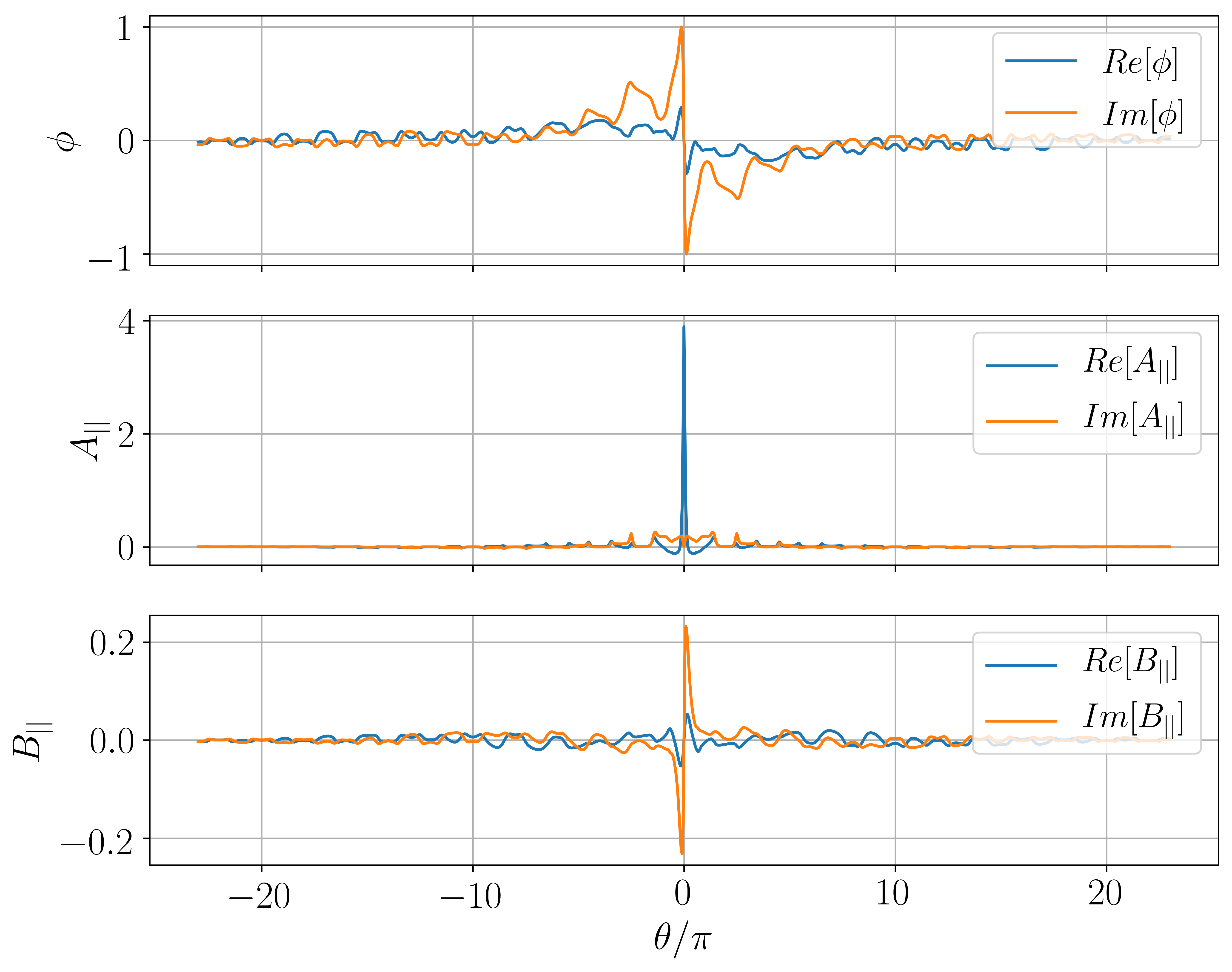}
        \caption{}
        \label{fig:ky_0.35_vnewk_0.1}
    \end{subfigure}
    \begin{subfigure}{\textwidth}
        \centering
        \includegraphics[width=75mm]{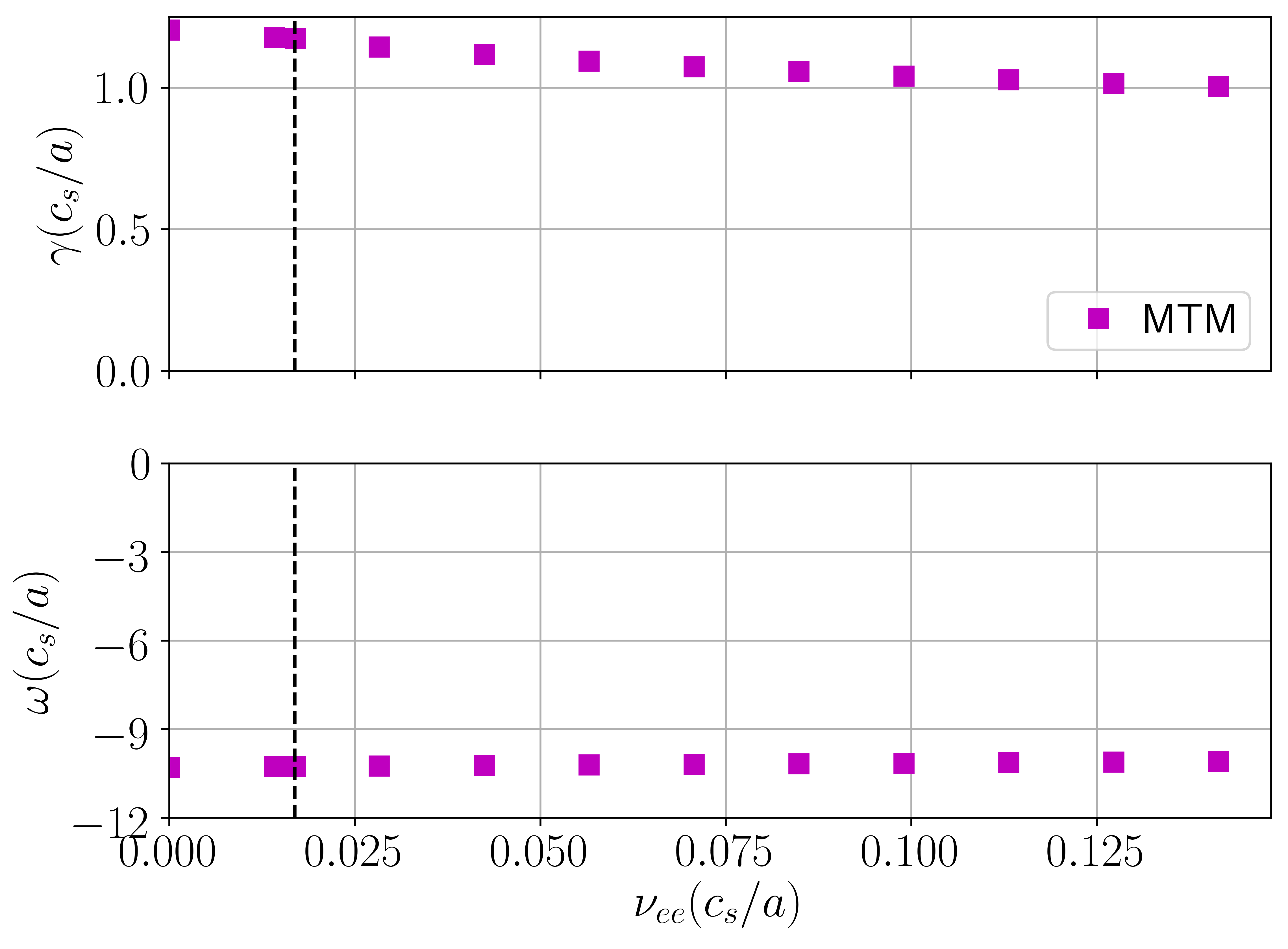}
        \caption{}
        \label{fig:ky_4_coll}
    \end{subfigure}
       \caption{ a) Impact of collision frequency showing the dominant odd and even parity mode for different values of $\nu_{ee}$ when $k_y\rho_s=0.35$. b) Eigenfunction of the low $k_y$ MTM when $\nu_{ee}=0.14c_s/a$ showing the reduced extent in ballooning space. c) Growth rate and frequency for the $k_y\rho_s=4.2$ MTM as a function of $\nu_{ee}$. The vertical dashed black line shows the equilibrium value of $\nu_{ee}$.}
    \label{fig:ky_vnewk}
\end{figure}

When examining the low $k_y$ MTM, as shown in Figure \ref{fig:ky_0.35_vnewk_scan}, it appears that $\gamma^{\mathrm{MTM}}$ increases with $\nu_{ee}$, and is only unstable with a finite collision frequency. Additionally, the collisions reduce the extent of the mode in ballooning space as shown in Figure \ref{fig:ky_0.35_vnewk_0.1} where $\nu_{ee} = 0.14 c_s/a$, compared to the nominal collisionality ($\nu_{ee} = 0.017 c_s/a$) eigenfunction shown in Figure \ref{fig:ky_0.35_eigfunc_odd}. This is due to the passing electrons undergoing a collision before they can propagate further along the field line. The reduced extent of the mode would make nonlinear simulations easier to resolve. 

As the collision frequency is dropped towards 0, the MTM growth rate tends to 0, which is also consistent with the collisionality scaling recovered from MTM simulations for NSTX by Guttenfelder \textit{et al} \cite{guttenfelder2012scaling}.

This suggests that the confinement will scale favourably as collisionality is reduced, aligned with previous confinement scaling laws in STs where $B\tau_e \propto \nu_{*e}^{-0.82}$ \cite{valovivc2011collisionality}. One conclusion is that to reduce the electron transport a low collisionality regime is favourable.

However, Figure \ref{fig:ky_0.35_vnewk_scan} also shows that as the collision frequency is increased the KBM is stabilised and a similar feature has been seen in a hybrid TEM/KBM mode in simulations of NSTX \cite{guttenfelder2013progress}. There is a critical collision frequency at which the dominant mode switches from a KBM to an MTM and the KBM becomes stable for $\nu_{ee}>0.12c_s/a$. Furthermore, this is consistent with NSTX observations of increasing anomalous ion transport at low $\nu^*$, attributed to KBM-TEM hybrid modes \cite{kaye2013dependence}.

Examining the impact of collisions on the high $k_y$ MTM, it can be seen from Figure \ref{fig:ky_4_coll} that these are collisionless MTMs, highlighting the completely different nature of these modes compared to the longer wavelength MTMs. Collisionless MTMs have been seen before \cite{predebon2013linear, swamy2015collisionless, dickinson2013microtearing, geng2020physics}, but their mechanism is not fully understood. These are weakly stabilised by collisions over the range of the scan, but it appears that the impact is not significant.



A scenario can be imagined where due to the large electron transport, the temperature gradient will drop, stabilising the MTM. However, the lower electron gradient will also lower the temperature and increase the collisionality which could drive the MTM even more unstable, further dropping the temperature. This bootstrapping process could result in very low plasma temperatures. This can then in turn impact the ion temperature through reductions in collisional exchange heating which is a significant source of ion heating in a reactor. Finding an equilibrium point between the lowered $a/L_{Te}$ and higher $\nu_{ee}$ will be critical in determining the electron temperature. Conversely, if electron heating is sufficient to get over the peak growth rate for the MTM such that increasing $a/L_{Te}$ is stabilising for these MTM, then this feedback loop is positive. This could lead to a bifurcation to a high $a/L_{Te}$, high confinement regime.

\begin{figure}[!htbp]
    \centering
    \includegraphics[width=100mm]{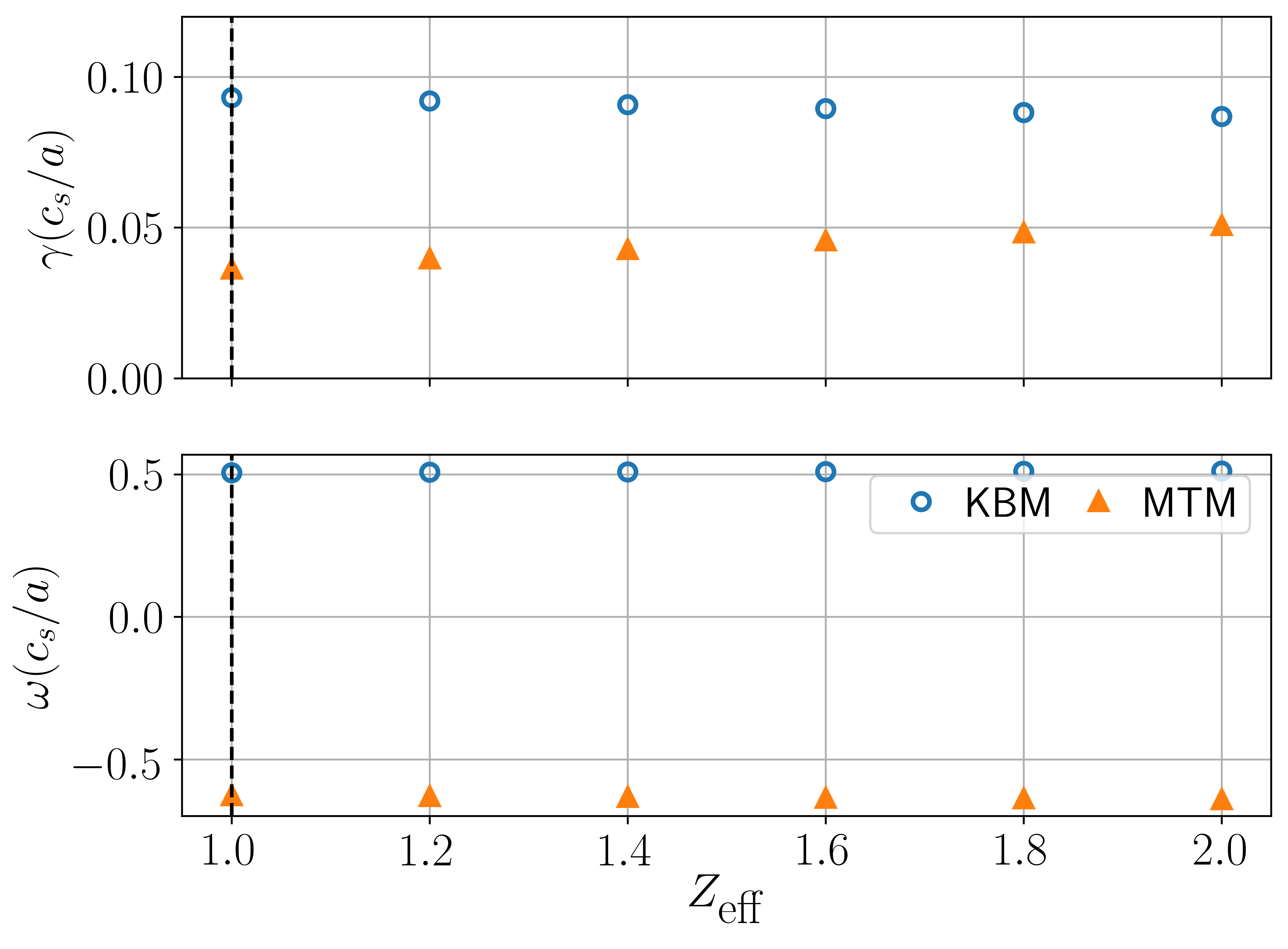}
    \caption{Examining the impact of $Z_{\textrm{eff}}$ on the KBMs and MTMs for $k_y\rho_s=0.35$ and $\nu_{ee}=0.14 c_s/a$. The vertical dashed black line shows the equilibrium value of $Z_{\textrm{eff}}$.}
    \label{fig:ky_0.35_zeff}
\end{figure}

Previous work has also found $Z_{\textrm{eff}}$ to be destabilising for MTMs, via its impact on the collision frequency as $\nu_{ei} \rightarrow Z_{\textrm{eff}}\nu_{ei}$ in the Lorentz collision operator \cite{guttenfelder2012scaling, guttenfelder2013progress}. The simulations conducted thus far had $Z_{\textrm{eff}}=1.0$, but in reality there will be impurities and helium in the plasma, raising the $Z_{\textrm{eff}}$. In the conceptual ST reactor design STPP a $Z_{\textrm{eff}}\approx1.6$ was assumed \cite{wilson2004STPP}, so it is crucial to quantify the impact it may have. A scan was performed from $Z_{\textrm{eff}}=1.0 \rightarrow 2.0$. It should be noted that impurities were not included in the simulation and $Z_{\textrm{eff}}$ only enters the gyrokinetic equations through its influence on the electron-ion collision frequency. Figure \ref{fig:ky_0.35_zeff} illustrates the dependency on $Z_{\textrm{eff}}$, with a doubling of $Z_{\textrm{eff}}$ causing the MTM growth rate to increase from $0.36 \rightarrow 0.51 c_s/a$. A similar growth is seen when the collision frequency is doubled instead where $\gamma =0.54 c_s/a$, which further confirms how $Z_{\textrm{eff}}$ acts to increase the effective collision frequency between the ions and electrons, which drives the MTM. Moreover, when removing electron-electron collisions the mode was unaffected, further highlighting that electron-ion collisions are the relevant drive for this MTM. Figure \ref{fig:ky_0.35_zeff} indicates the KBM growth rate is weakly stabilised, similar to the $\nu_{ee}$ scan. In summary, adding impurities in the plasma would cause a slight downshift in the MTM critical gradient.

\subsection{Magnetic equilibria}

Both MTMs and KBMs are inherently electromagnetic modes, so the total magnetic field will have a significant impact on the modes via $\beta_e$. Furthermore, there is evidence that both of these modes are impacted by both $q$ and $\hat{s}$, which are set by the toroidal field and plasma current profile. Both of these were also inputs to SCENE, so it is possible to vary the assumptions about the magnetic equilibrium to see how to further stabilise these modes. This section will perform local equilibrium scans to assess the impact of $\beta_e$, $\beta'_e$, $q$ and $\hat{s}$, which might provide a guide to the impact of changing the toroidal field and plasma current profile in a global Grad-Shafranov equilibrium from SCENE.

\subsubsection{Impact of $\beta_{e,\mathrm{unit}}$ and $\beta'_{e,\mathrm{unit}}$}
\label{sec:beta_betaprime}

\begin{figure}[!htb]
    \begin{subfigure}{0.5\textwidth}
        \centering
        \includegraphics[width=75mm]{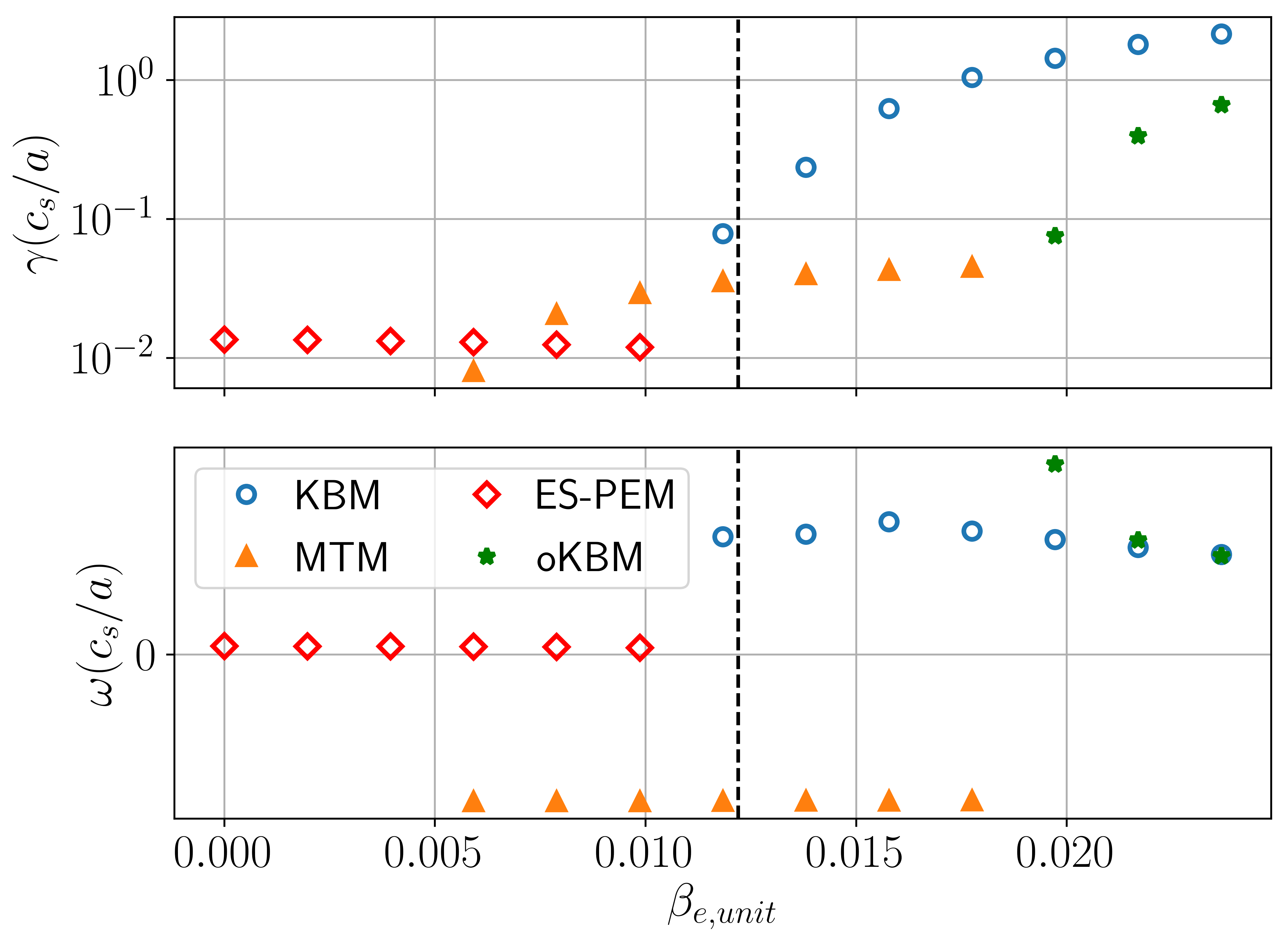}
        \caption{}
        \label{fig:ky_0.35_betascan}
    \end{subfigure}
    \begin{subfigure}{0.5\textwidth}
        \centering
        \includegraphics[width=75mm]{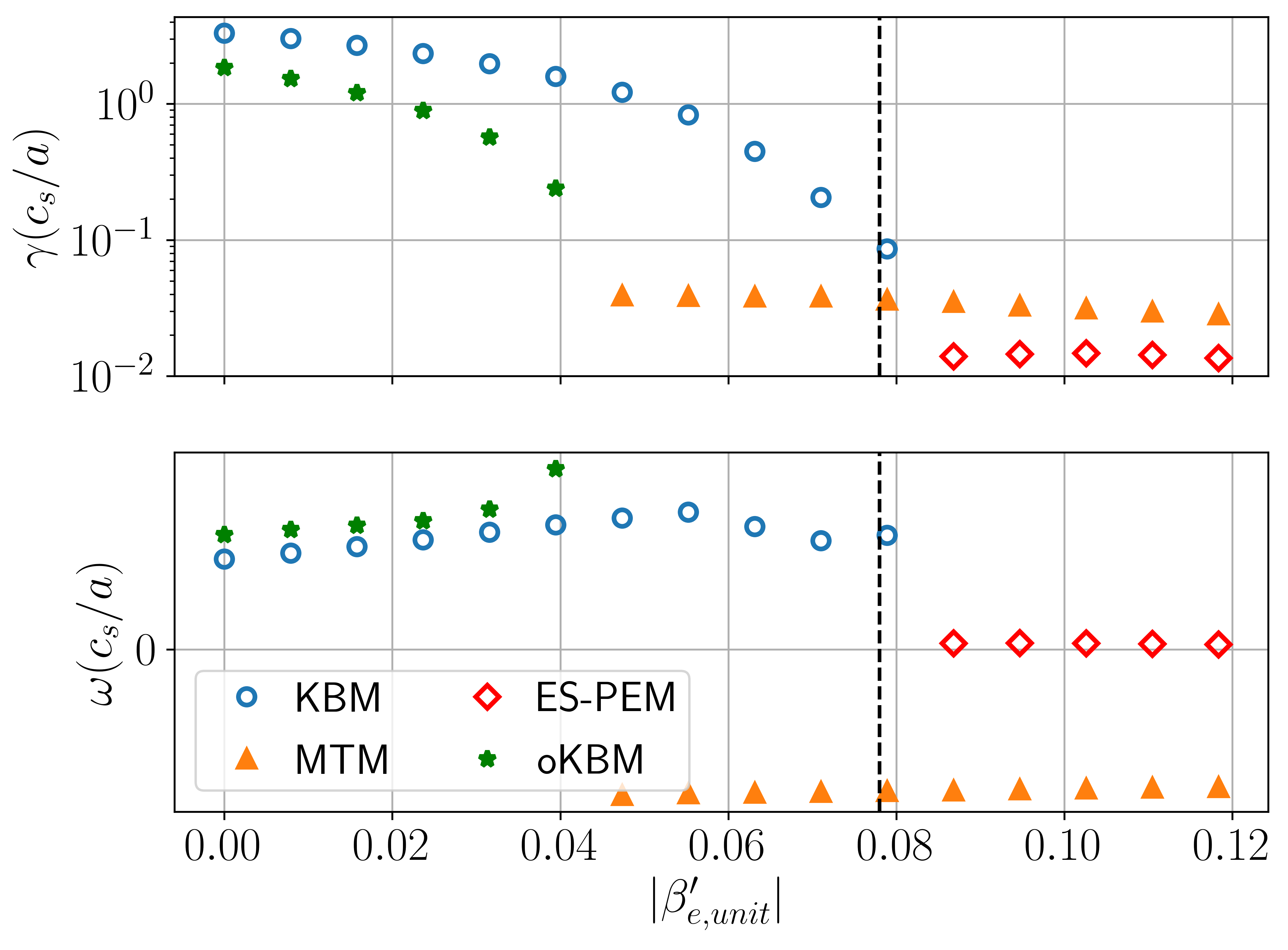}
        \caption{}
        \label{fig:ky_0.35_betaprime}
    \end{subfigure}
       \caption{Examining the impact of a) $\beta_{e,\mathrm{unit}}$ at fixed $\beta'_{e,\mathrm{unit}}=0.078$ and b) $\beta'_{e,\mathrm{unit}}$ at fixed $\beta_{e,\mathrm{unit}}=0.012$ on the dominant even and odd parity micro-instabilities at $k_y\rho_s=0.35$. The vertical dashed black lines shows the equilibrium values.}
    \label{fig:ky_0.35_beta}
\end{figure}

A scan was conducted in $\beta_{e,\mathrm{unit}}=0.0 \rightarrow 0.024$\footnote{The subscript $\mathrm{unit}$ corresponds to the normalising magnetic field being $B_{\mathrm{unit}} = \frac{q}{r} \frac{\partial \psi}{\partial r}$.} at $k_y\rho_s=0.35$, where $\beta_{e,\mathrm{unit}}=0.024$ at $\rho_\psi=0.5$ approximately corresponds to a 40\% drop in the field compared to the reference case in Table \ref{tab:gk_flux_params}. In a self-consistent scan both $\beta_{e,\mathrm{unit}}$ and $\beta'_{e,\mathrm{unit}} $ would be scanned together, but in the first scan shown in Figure \ref{fig:ky_0.35_betascan} $\beta_{e,\mathrm{unit}}$ alone is scanned with with $\beta'_{e,\mathrm{unit}}$ fixed. This effectively scans in the strength of the magnetic perturbations.

Looking at the even modes, when $\beta_{e,\mathrm{unit}}=0$, a weakly unstable electrostatic passing electron mode (ES-PEM) appears, which was seen in the negative density gradient scan in \ref{fig:ky_ln}. Above a critical value of $\beta_{e,\mathrm{unit}}=0.01$, the KBM mode becomes unstable, before it begins to saturate around $\beta_{e,\mathrm{unit}} = 0.02$. 
Looking at the MTMs, again there is a critical $\beta_{e,\mathrm{unit}}=0.006$, below which the MTM is stable. This is driven unstable by $\beta_{e,\mathrm{unit}}$, but when $\beta_{e,\mathrm{unit}}$ increases sufficiently then the oKBM is seen, which is very sensitive to $\beta_{e,\mathrm{unit}}$ and quickly begins to approach the KBM growth rate.

$\beta'_{e,\mathrm{unit}}$ is used to calculate the equilibrium and has large impacts on the magnetic drifts and local magnetic shear. A scan was performed in $\beta'_{e,\mathrm{unit}}$ at the fixed equilibrium $\beta_{e,\mathrm{unit}}$ to isolate its impact. This is shown in Figure \ref{fig:ky_0.35_betaprime} where it can be seen that at low $\beta'_{e,\mathrm{unit}}$, the KBM/oKBM are the dominant instabilities. These are quickly stabilised and the dominant odd mode switches from an oKBM into the MTM and the dominant even mode switches from a KBM to a weakly unstable ES-PEM. The MTM is not significantly affected by $\beta'_{e,\mathrm{unit}}$ and the reason can be determined by examining the drifts.

\begin{figure}[!htb]
    \begin{subfigure}{0.5\textwidth}
        \centering
        \includegraphics[width=75mm]{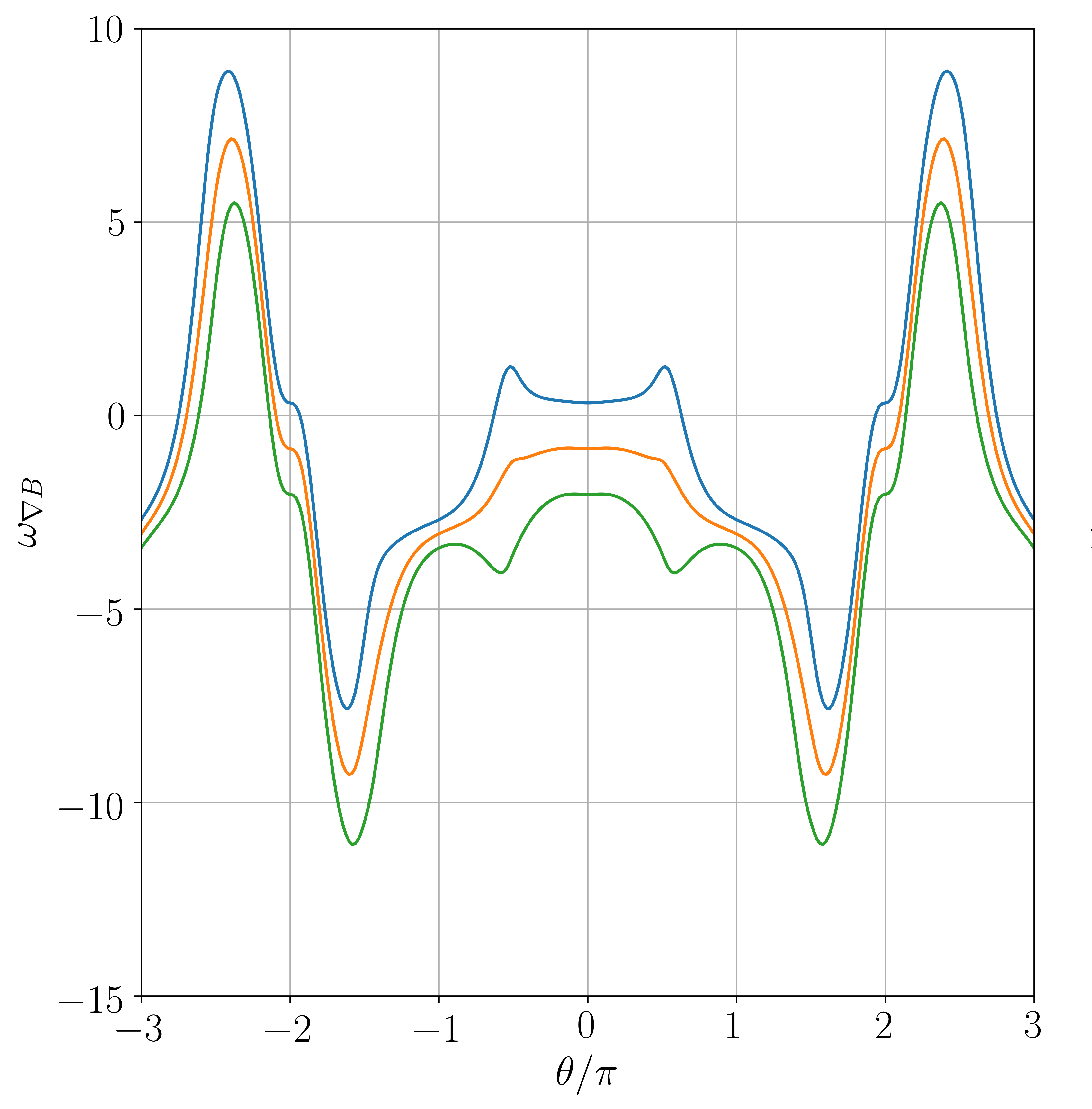}
        \caption{}
        \label{fig:gs2_gradb_bprime}
    \end{subfigure}
    \begin{subfigure}{0.5\textwidth}
        \centering
        \includegraphics[width=75mm]{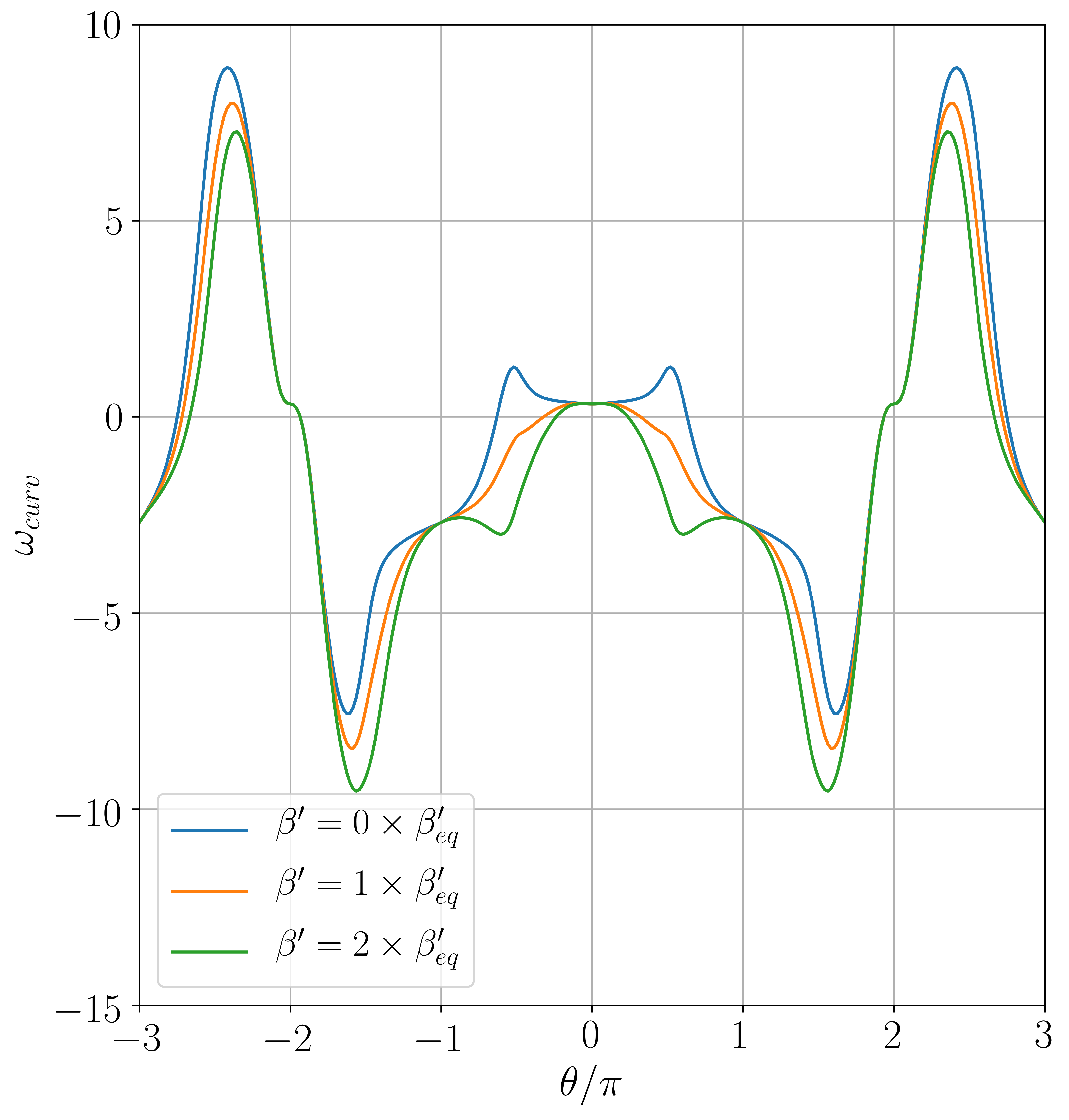}
        \caption{}
        \label{fig:gs2_curv_bprime}
    \end{subfigure}
    \caption{The impact $\beta'_{e,\mathrm{unit}}$ has on $\omega_{\nabla B}$ and $\omega_{curv}$. The $\nabla B$ drift is negative for the high $\beta'$ cases when $|\theta|<\pi$, which is stabilising for ballooning modes.}
    \label{fig:gs2_drifts}
\end{figure}

Figure \ref{fig:gs2_drifts} illustrates how $\omega_{\nabla B}$ and $\omega_{curv}$ are modified by changing $\beta'_{e,\mathrm{unit}}$. The total magnetic drift is approximately the sum of $\omega_{\nabla B}$ and $\omega_{curv}$. Here a negative drift frequency corresponds to ``good curvature" and positive is ``bad curvature". Increasing $\beta'_{e,\mathrm{unit}}$ makes $\omega_{\nabla B}$ negative such that the combination of the two drifts becomes stabilising on the outboard side, resulting in ``good curvature". This effect is highly stabilising for highly ballooning KBMs that are driven by bad curvature on the outboard side. The MTMs at low $k_y$ are considerably more extended in the parallel direction, and much less sensitive to drift reversal at the outboard mid-plane, with growth rates that are relatively insensitive to $\beta'_{e, \mathrm{unit}}$.

\begin{figure}[!thb]
    \begin{subfigure}{0.5\textwidth}
        \centering
        \includegraphics[width=75mm]{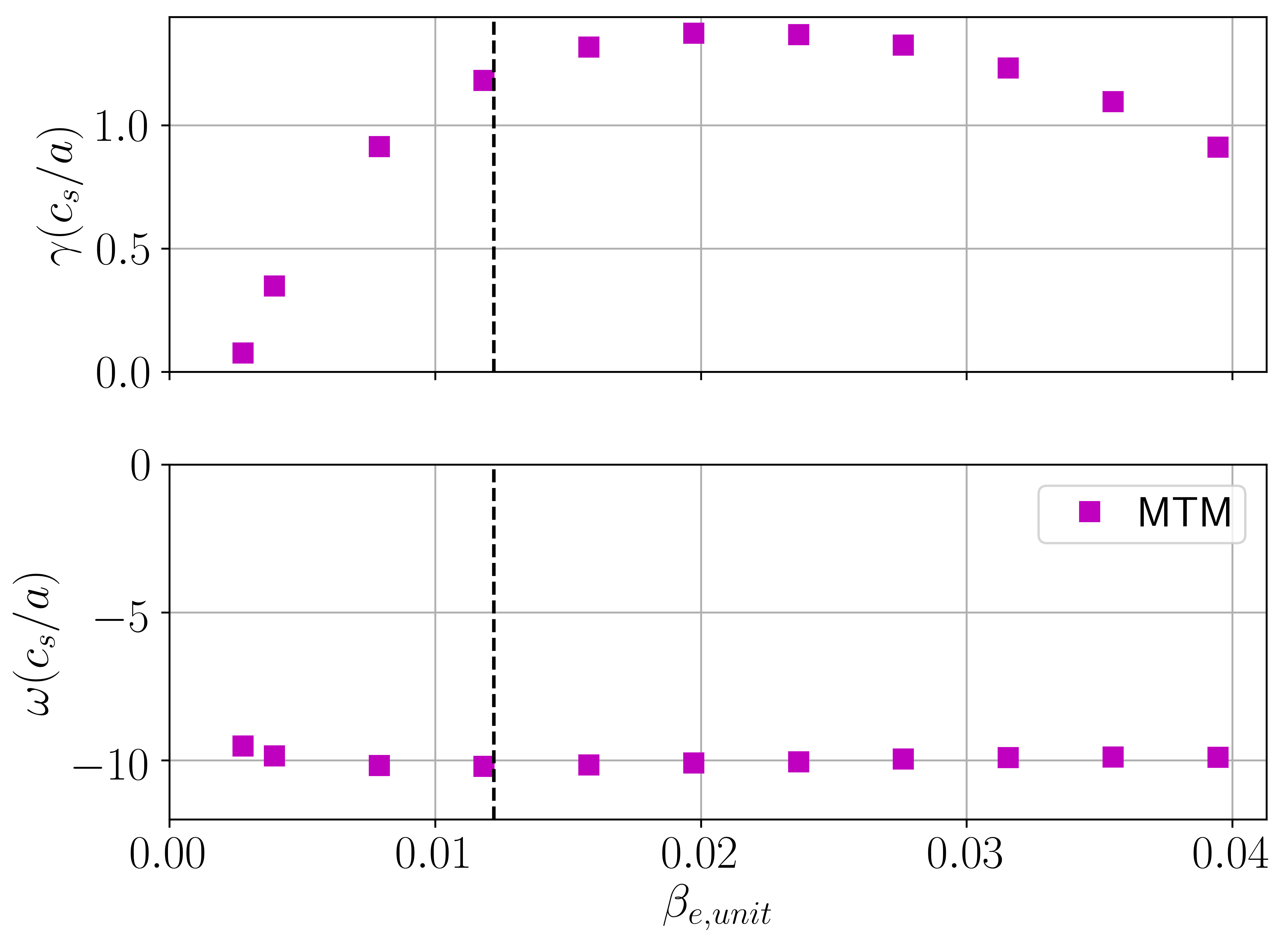}
        \caption{}
        \label{fig:ky_4_mtm_beta}
    \end{subfigure}
    \begin{subfigure}{0.5\textwidth}
        \centering
        \includegraphics[width=75mm]{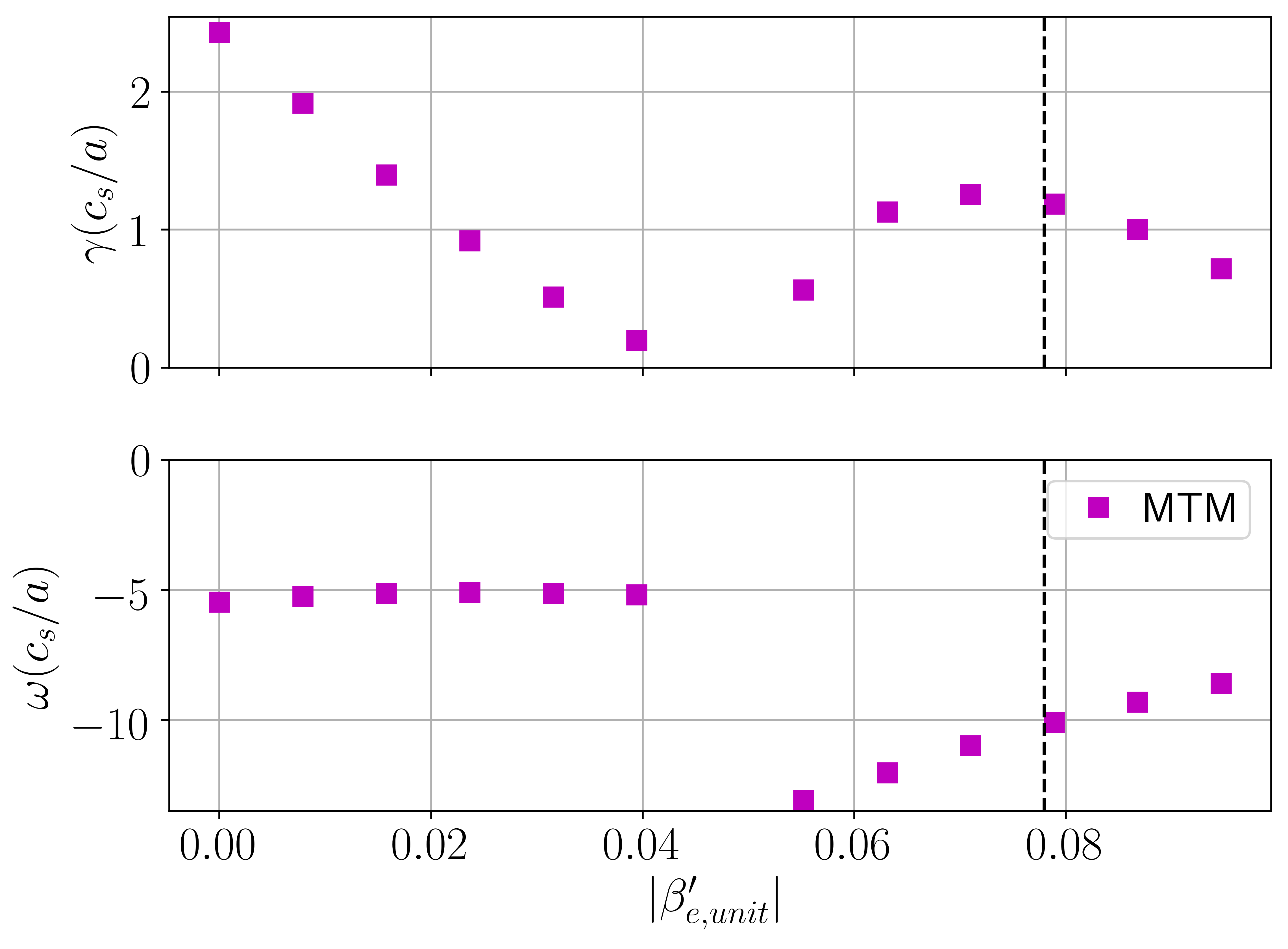}
        \caption{}
        \label{fig:ky_4_mtm_bprime}
    \end{subfigure}
    \begin{subfigure}{\textwidth}
        \centering
        \includegraphics[width=75mm]{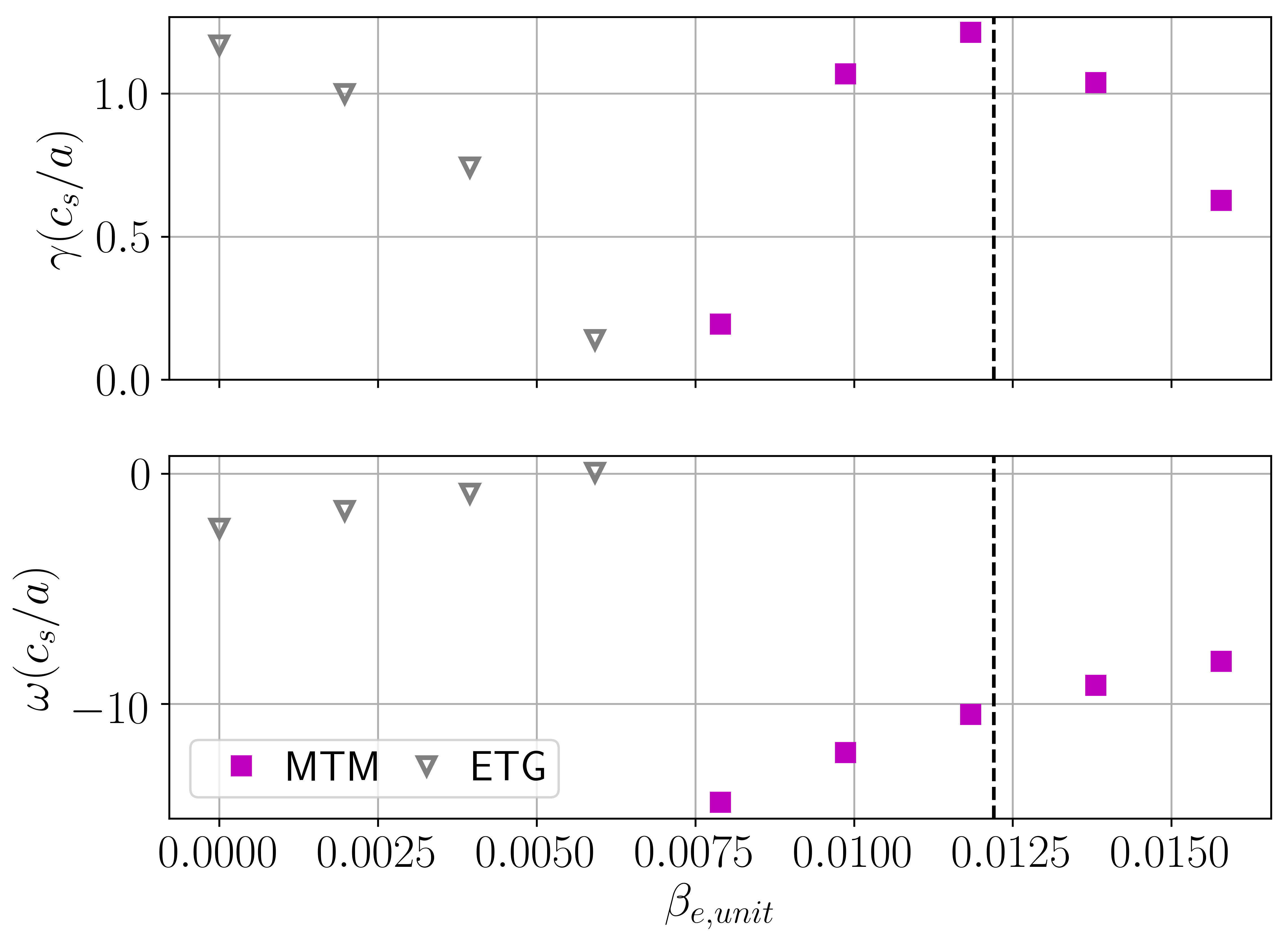}
        \caption{}
        \label{fig:ky_4_mtm_beta_bprime}
    \end{subfigure}
       \caption{Examining the impact of changing a) $\beta_{e,\mathrm{unit}}$ at fixed $\beta_{e,\mathrm{unit}}'=0.078$ and b) $\beta_{e,\mathrm{unit}}'$ at fixed $\beta_{e,\mathrm{unit}}=0.012$. c) $\beta_{e,\mathrm{unit}}$ and $\beta_{e,\mathrm{unit}}'$ are changed together for the MTMs at $k_y\rho_s=4.2$. Above $\beta_{e,\mathrm{unit}}=0.016$, the equilibrium was completely stable. The vertical dashed black lines shows the equilibrium values.}
    \label{fig:ky_4_mtm_em}
\end{figure}

For the high $k_y$ MTM, Figure \ref{fig:ky_4_mtm_beta} shows a scan of changing $\beta_{e,\mathrm{unit}}$ at fixed $\beta'_{e,\mathrm{unit}}$. Exceeding a critical $\beta_{e,\mathrm{unit}}=0.003$ was found to destabilise this mode, corresponding to a $70\%$ increase in the field. Figure \ref{fig:ky_4_mtm_bprime} shows how changing $\beta'_{e,\mathrm{unit}}$ impacts the high $k_y$ MTM which is very different to the low $k_y$ MTMs which were largely unaffected by $\beta'_{e,\mathrm{unit}}$, which is related to the MTMs respective sensitivity to $\theta_0$. At lower $\beta'_{e,\mathrm{unit}}$ this MTM is stabilised, but another MTM branch appears for $\beta'_{e,\mathrm{unit}}<0.04$, likely also driven unstable by $\beta_{e,\mathrm{unit}}$. At $\beta'_{e,\mathrm{unit}}=0.0$, this MTM has $C_{\mathrm{tear}}=0.3$ which is lower than that for the equilibrium $\beta'_{e,\mathrm{unit}}$ at $\theta_0=0$ which had $C_{\mathrm{tear}}=0.5$. However, a scenario with a very low $\beta'_{e,\mathrm{unit}}$ and high $\beta_{e,\mathrm{unit}}$ requires a very low pressure gradient which is not desirable for a reactor as this would result in a lower bootstrap fraction.

To examine the relevance of this low $\beta'_{e,\mathrm{unit}}$ MTM, a scan was done where $\beta'_{e,\mathrm{unit}}$ and $\beta_{e,\mathrm{unit}}$ were changed together i.e. $\beta'_{e,\mathrm{unit}}=\beta_{e,\mathrm{unit}}a/L_p$ was maintained throughout this scan. Figure \ref{fig:ky_4_mtm_beta_bprime} shows that at sufficiently low $\beta_{e,\mathrm{unit}}$, the original MTM is stabilised and the new MTM does not appear. However, an ETG mode (hollow grey upside-down triangle) appears when operating at lower $\beta_{e,\mathrm{unit}}$ and $\beta'_{e,\mathrm{unit}}$. This ETG is explored in more detail in \ref{app:high_ky_etg}. But this suggests that this low $\beta'_{e,\mathrm{unit}}$ MTM will not be relevant.

\subsubsection{Safety factor profile}
\label{sec:q_shat}

The $q$ profile has a large impact on both the MHD and the micro-stability of a reactor. Tailoring the $q$ profile requires careful control of the auxiliary current profile so it may be difficult to optimise it for turbulence but this section highlights how it will impact these modes.

\begin{figure}[!htb]
    \centering
    \includegraphics[width=75mm]{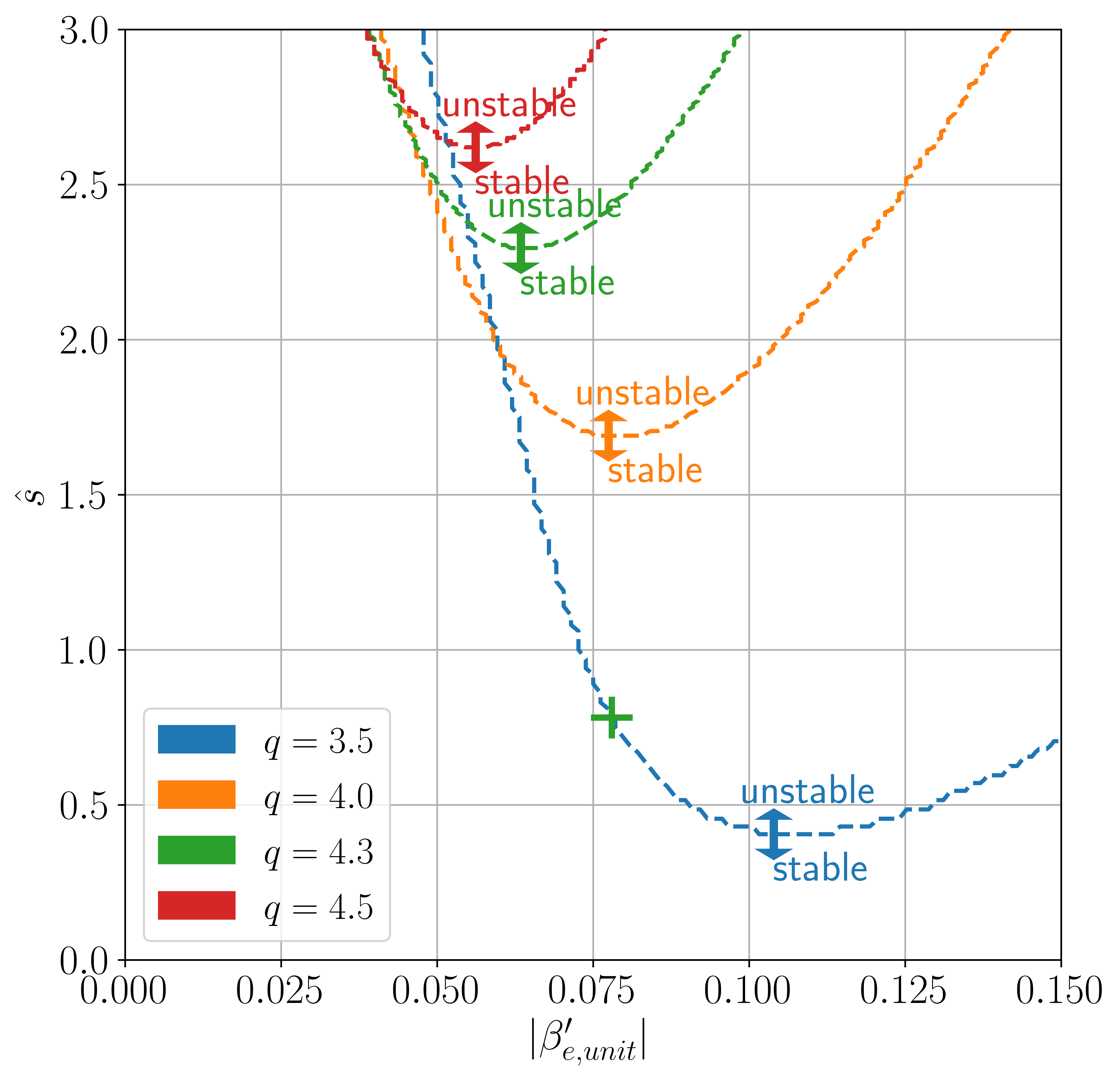}
    \caption{$\hat{s}-|\beta'|$ diagram showing how the ideal ballooning stability boundary moves with $q$. The reference equilibrium value of $\hat{s}$ and $|\beta'_{e,\mathrm{unit}}|$ is shown by the green cross and has $q=4.3$ (green curve). }
    \label{fig:gs2_ideal_ball}
\end{figure}

The $n=\infty$ ballooning stability boundary for ideal ballooning modes is often used as an initial indicator for the onset of KBMs \cite{groebner2010limits} and in GS2 there is a module that can calculate this boundary \cite{connor1979high}. Figure \ref{fig:gs2_ideal_ball} shows how this ideal stability boundary changes with $q$, illustrating a somewhat complicated relation between the ideal ballooning mode and $q$. As $q$ increases the stability boundary moves to high $\hat{s}$ and lower $|\beta'|$. Therefore pushing to a higher $q$ will make access to the second stability region easier. At sufficiently low $q$, the stability boundary gets pushed to higher $|\beta'|$ enabling the equilibrium to lie in the first stability region.

\begin{figure}[!thb]
    \begin{subfigure}{0.5\textwidth}
        \centering
        \includegraphics[width=75mm]{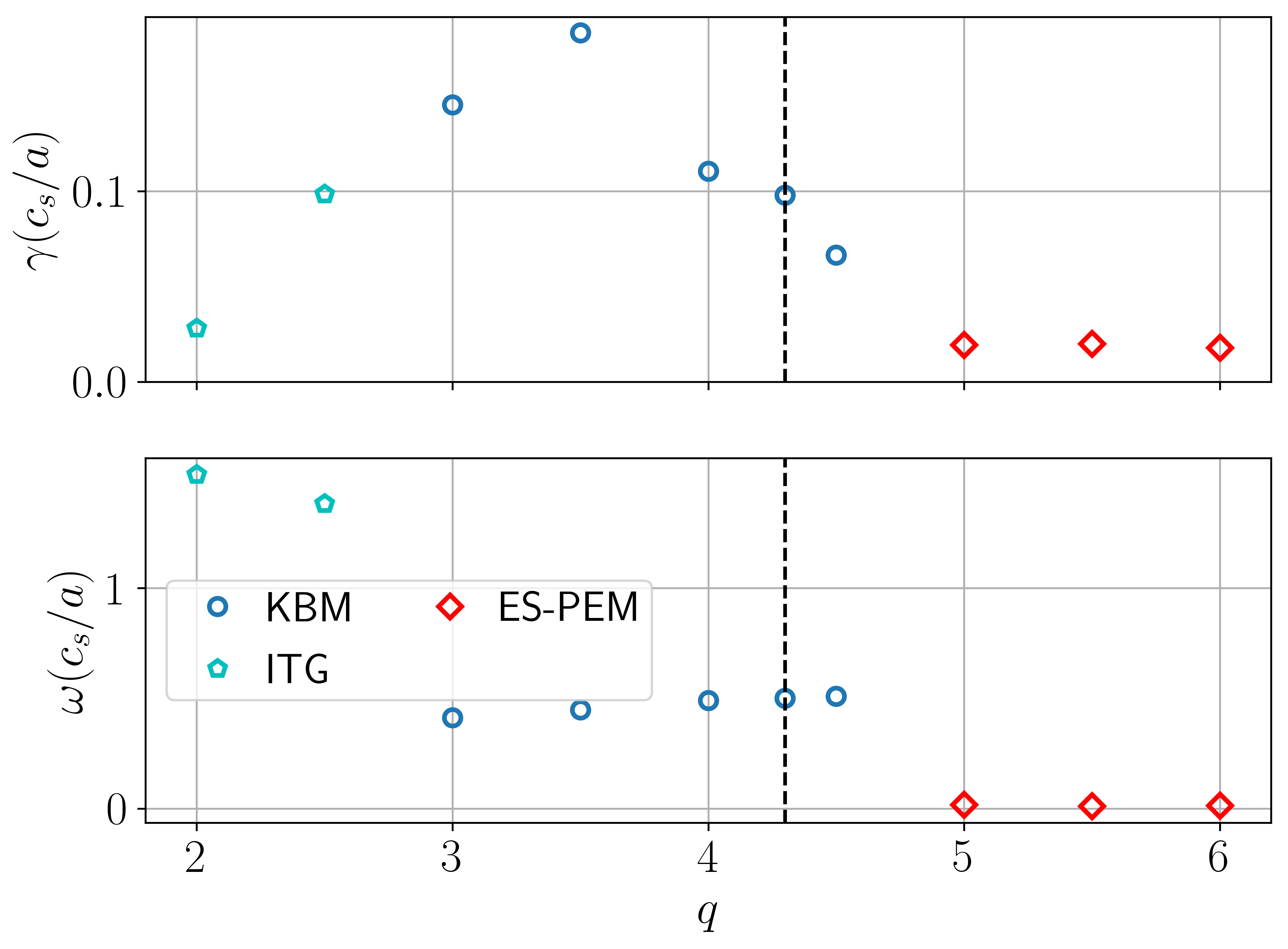}
        \caption{}
        \label{fig:ky_0.35_q_kbm}
    \end{subfigure}
    \begin{subfigure}{0.5\textwidth}
        \centering
        \includegraphics[width=75mm]{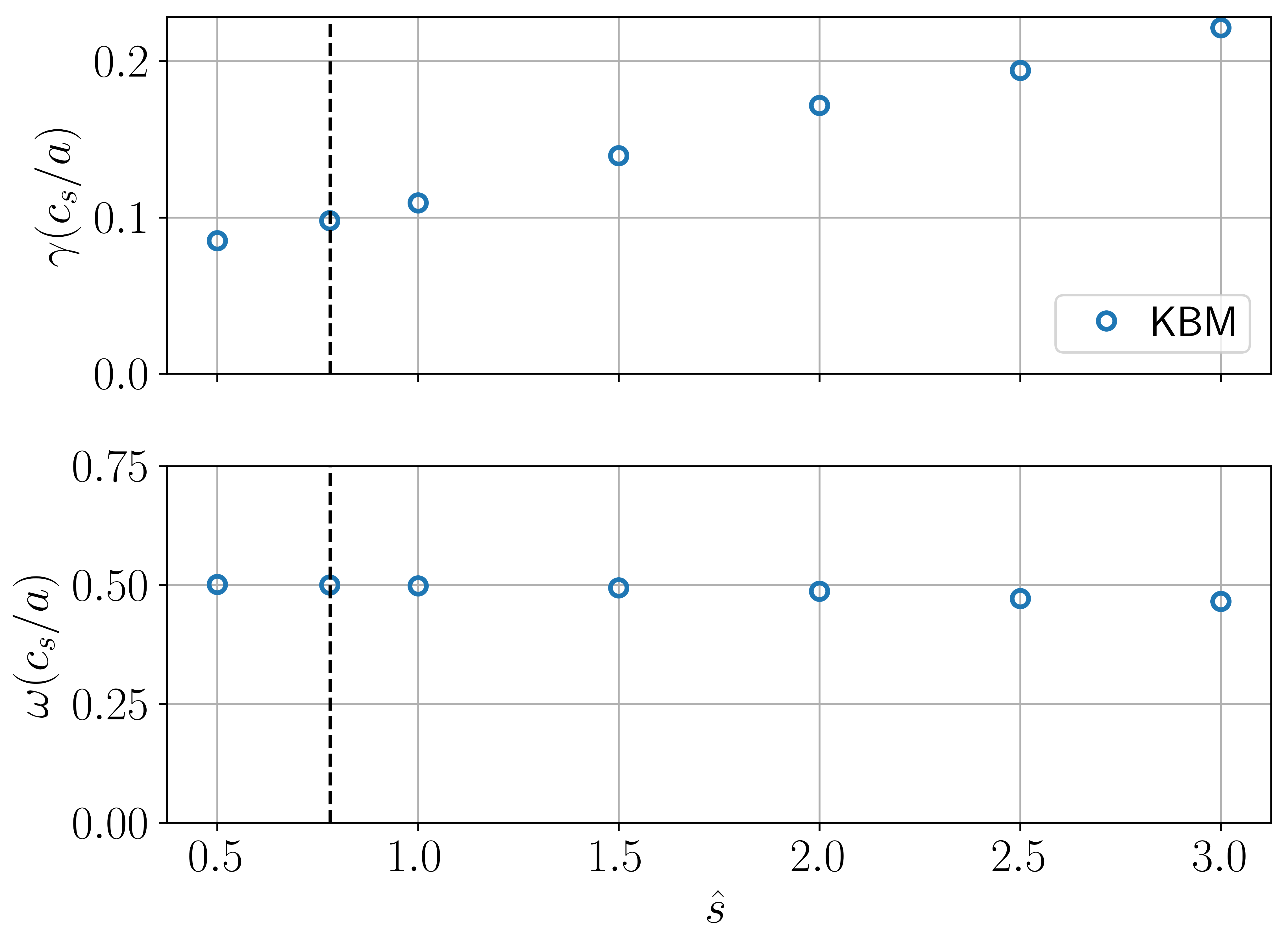}
        \caption{}
        \label{fig:ky_0.35_shear_kbm}
    \end{subfigure}
       \caption{Dominant even parity mode eigenvalues when changing a) $q$ and b) $\hat{s}$ for $k_y\rho_s=0.35$. The vertical dashed black lines shows the equilibrium values.}
    \label{fig:ky_0.35_kbm_magnetic}
\end{figure}

Evidently, the ideal ballooning mode is not sufficient in predicting the KBM threshold as it predicts this equilibrium to be stable, but it can give an idea of how the KBM will behave. Figure \ref{fig:ky_0.35_q_kbm} shows a scan in $q$ for the kinetic modes. For $q<3.0$, when the equilibrium is in the first stability region, an ITG (hollow cyan pentagons) mode is dominant but as the equilibrium enter the second stability region, the KBM becomes the dominant instability. As $q$ is increased the KBM peaks and then becomes stabilised as the ideal boundary gets pushed further away from the equilibrium value. When the equilibrium is in the second stability region then increasing $q$ stabilises the KBM and another ES-PEM appears, indicating that operating at high $q$ may help to increase the KBM critical gradient. This will be useful if the MTMs can be suppressed.

A scan was also conducted with $\hat{s}$ and Figure \ref{fig:ky_0.35_shear_kbm} shows how the KBM is destabilised by increasing $\hat{s}$, consistent with the ideal ballooning mode behaviour.

\begin{figure}[!bth]
    \begin{subfigure}{0.5\textwidth}
        \centering
        \includegraphics[width=75mm]{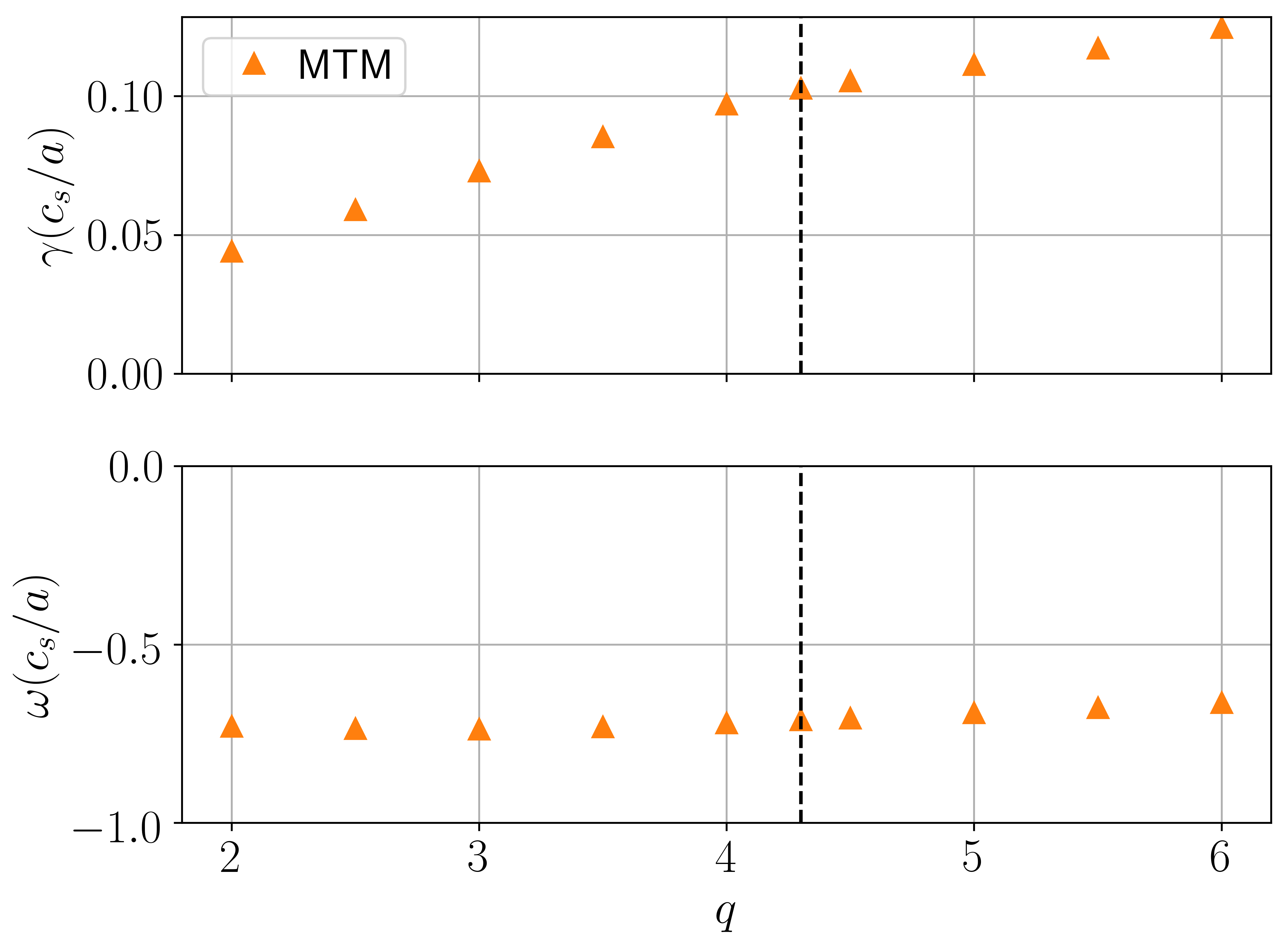}
        \caption{}
        \label{fig:ky_0.35_q_mtm}
    \end{subfigure}
    \begin{subfigure}{0.5\textwidth}
        \centering
        \includegraphics[width=75mm]{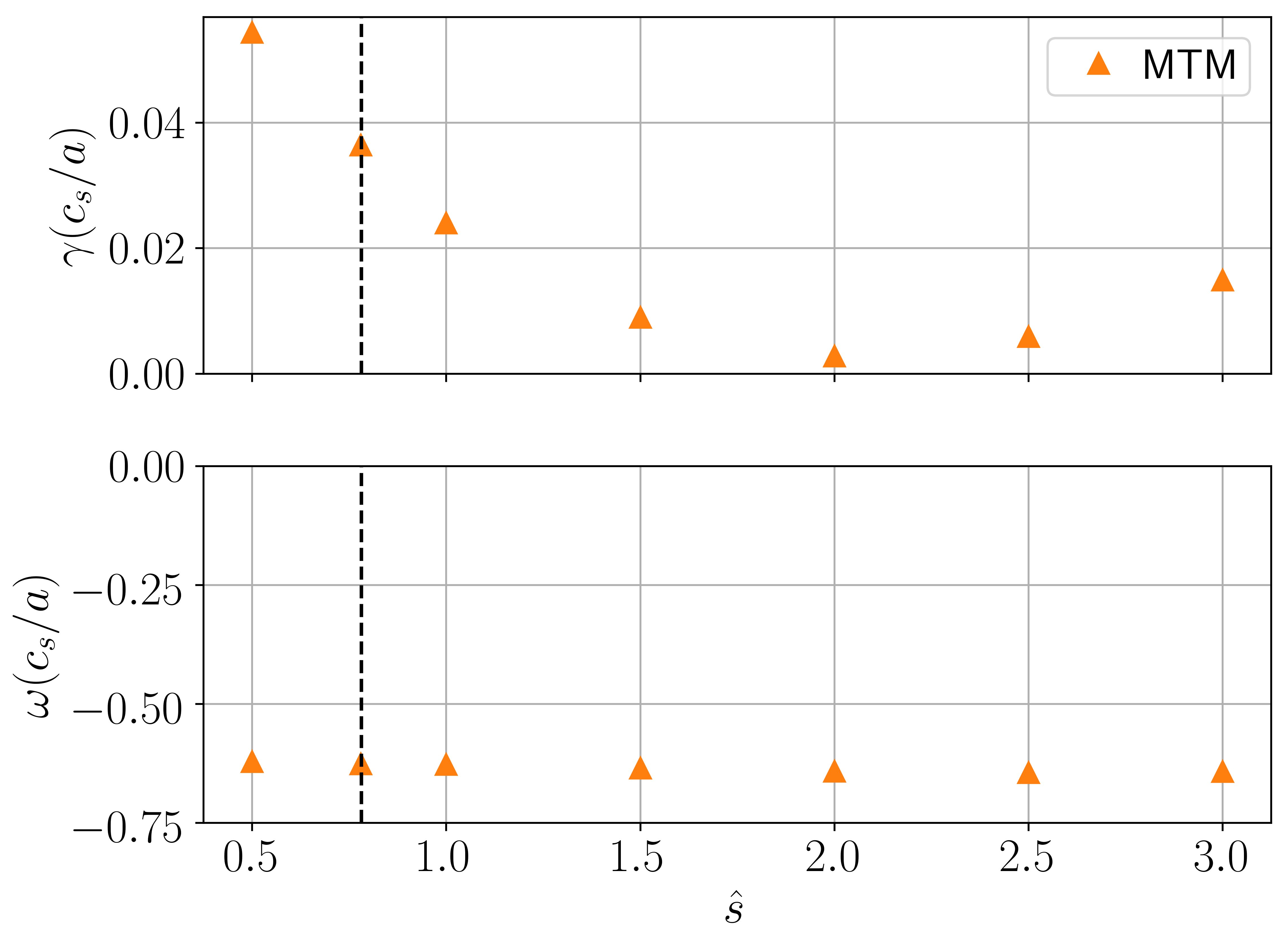}
        \caption{}
        \label{fig:ky_0.35_shear_mtm}
    \end{subfigure}
       \caption{Dominant odd parity mode eigenvalues when changing a) $q$ (at $\nu_{ee}=0.14 c_s/a$) and b) $\hat{s}$ for $k_y\rho_s=0.35$. The vertical dashed black lines shows the equilibrium values.}
    \label{fig:ky_0.35_mtm_magnetic}
\end{figure}

Next, the impact $\hat{s}$ and $q$ have on the low $k_y$ MTMs is explored, and it can be expected that increasing $q$ will be destabilising for this MTM given that $\nu_* \propto q$. To isolate the impact on the MTM and avoid the oKBM being driven unstable, the $q$ scan was run with $\nu_{ee}=0.14 c_s/a$. Figure \ref{fig:ky_0.35_q_mtm} illustrates how there appears to be a linear relationship between $q$ and the MTM growth rate, consistent with a $\nu_*$ scaling. The magnetic shear dependency displays non-monotonic behaviour as shown in Figure \ref{fig:ky_0.35_shear_mtm}.

The impact of $\hat{s}/q$ is often examined as it is related to Landau damping and field line bending. For the MTM seen here it appears that $\hat{s}/q$ has a non-monotonic behaviour, so the impact on the turbulent transport will be difficult to predict. This has been seen before on MAST \cite{dickinson2012kinetic}. However, there are examples where $\hat{s}/q$ is stabilising for MTMs like in DIII-D \cite{jian2019role} and counter-examples where $\hat{s}/q$ tends to be destabilising, such as on NSTX \cite{guttenfelder2012scaling} attributed to higher field line bending. This further highlights the complicated behaviour of MTMs so further work is needed to understand the dependence on the $q$-profile. 

\begin{figure}[!tbh]
    \begin{subfigure}{0.5\textwidth}
        \centering
        \includegraphics[width=75mm]{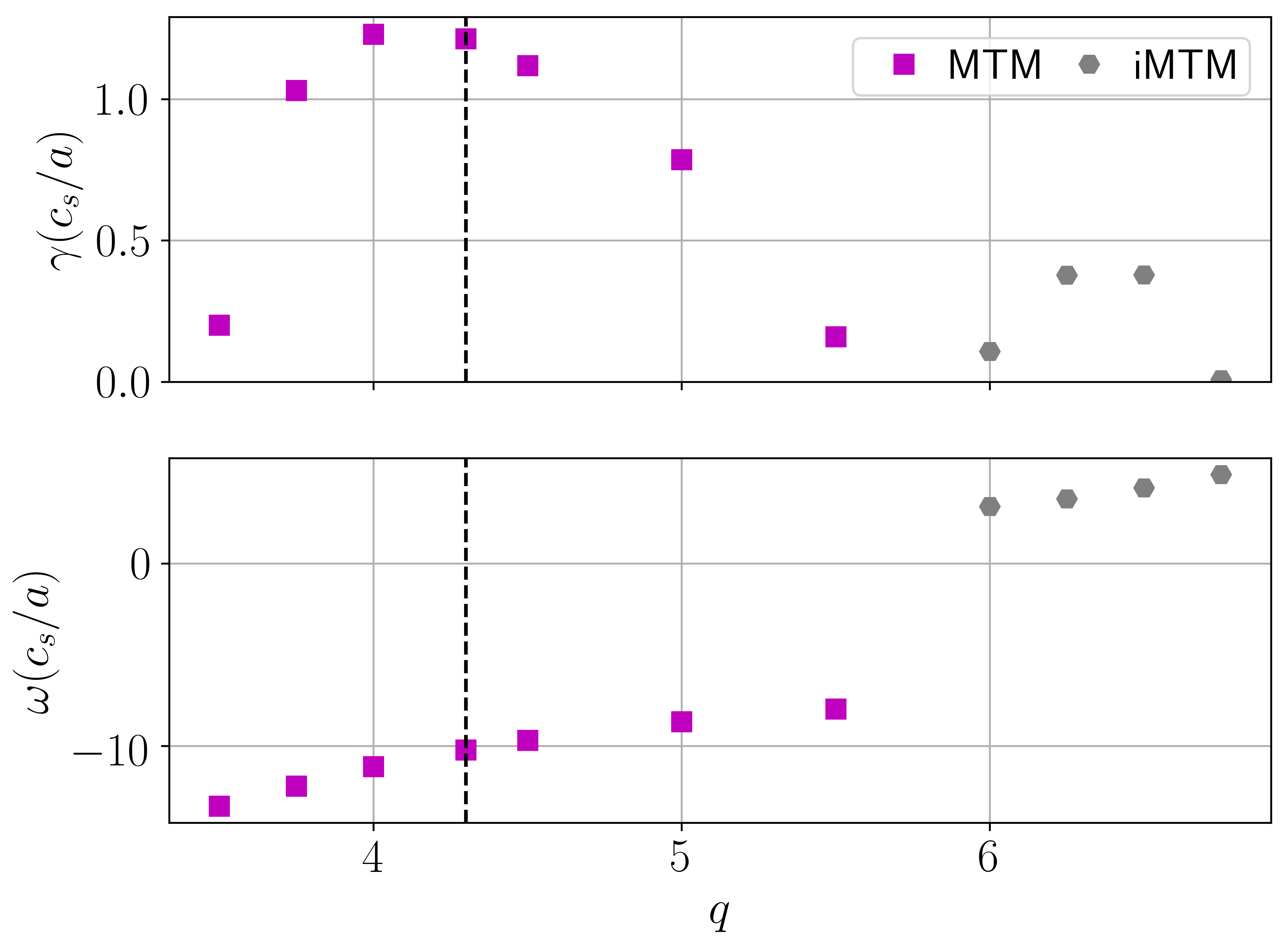}
        \caption{}
        \label{fig:ky_4_mtm_q}
    \end{subfigure}
    \begin{subfigure}{0.5\textwidth}
        \centering
        \includegraphics[width=75mm]{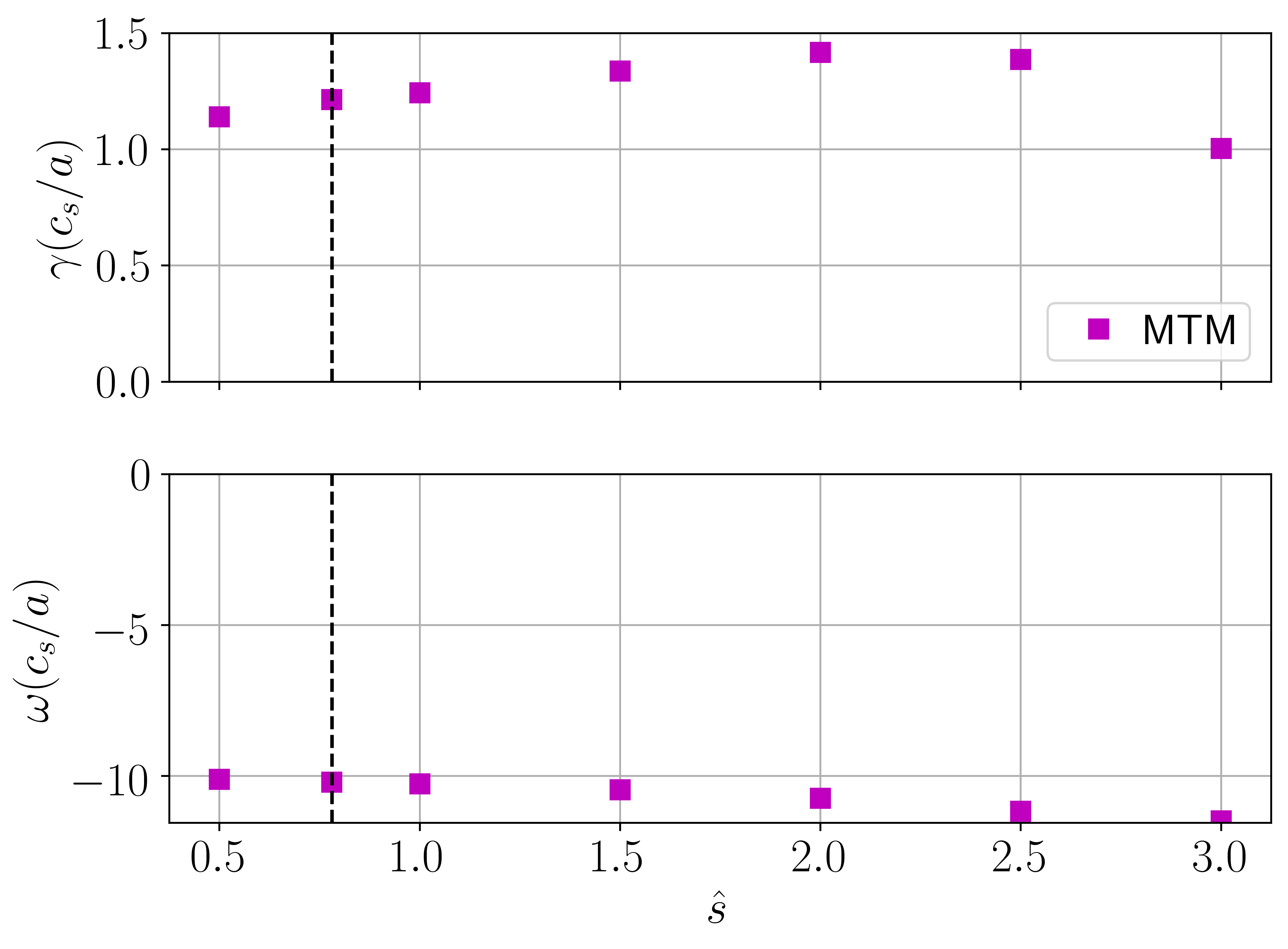}
        \caption{}
        \label{fig:ky_4_mtm_shear}
    \end{subfigure}
       \caption{Impact of a) $q$ and b) $\hat{s}$ on the MTMs at $k_y\rho_s=4.2$. The vertical dashed black lines shows the equilibrium values.}
    \label{fig:ky_4_mtm_magnetic}
\end{figure}

Figure \ref{fig:ky_4_mtm_q} shows a scan in $q$ for the high $k_y$ MTMs, and it can be seen that the growth rate has a peak close to the local equilibrium value of $q$, so increasing or decreasing can reduce the growth rate. A higher $B_\varphi$ device will likely have a higher $q$, so the combination of lower $\beta_{e,\mathrm{unit}}$ and higher $q$ indicates that this mode will be stabilised by a higher $B_\varphi$. At very high $q$ an iMTM mode with $C_{\mathrm{tear}}=0.7$ was found, though it exists for a very narrow window. Looking at the behaviour with $\hat{s}$ in Figure \ref{fig:ky_4_mtm_shear}, there is a non-monotonic dependence similar to the lower $k_y$ MTM. However, it seems to be a weak dependence so will likely not be significant for this MTM.

\subsection{Radial surfaces at $\rho=0.3$ and $\rho=0.85$}
This analysis was repeated for two different flux surfaces at $\rho_\psi=0.3$ and $0.85$ and similar KBMs and MTMs were found, indicating that these modes and the trends reported here exist robustly across the whole equilibrium. These are discussed in more detail in \cite{patel2021confinement}.

\section{Options for stabilising the equilibrium}
\label{sec:options_stable}
Section \ref{sec:mode_dependancy} has indicated potential avenues for stabilising the dominant micro-instabilities based on local equilibrium scans. Here these insights are used to modify the global equilibrium using SCENE with two options being pursued. Firstly, the impact of increased toroidal field is examined. To maintain the same $q$. $I_{\mathrm{p}}$ would need to be increased. However, $I_{\mathrm{p}}$ was fixed to avoid the need for additional auxiliary current drive. Secondly, the impact of a more peaked density profile is considered.

\subsection{Impact of higher toroidal field at constant plasma current}
\label{sec:low_ky_high_field}

\newcommand{\appropto}{\mathrel{\vcenter{
  \offinterlineskip\halign{\hfil$##$\cr
    \propto\cr\noalign{\kern2pt}\sim\cr\noalign{\kern-2pt}}}}}

Increasing $B_\phi$ by going to higher $I_\mathrm{rod}$ at constant $I_{\mathrm{p}}$, will reduce $\beta_e$ and increase $q$. These changes have opposing impacts on the stability of the low $k_y$ MTMs: reducing $\beta_e$ is stabilising, but increasing $q$ is destabilising.

To investigate the impact of higher field, SCENE was used to consistently generate equilibria with a higher toroidal field. The value of $I_{\mathrm{rod}}$ was increased from $30$\si{\mega\ampere} $\rightarrow 50$\si{\mega\ampere} in 5\si{\mega\ampere} increments, whilst the density and temperature profiles were kept fixed. The total plasma current was fixed as well as the shape of the auxiliary current profile. As $I_{\mathrm{rod}}$ increases, there was a small increase in the bootstrap current but this was offset by a small reduction in the diamagnetic current such that the total auxiliary current remained roughly constant. This results in a near linear increase in the safety factor profile with $I_\mathrm{rod}$. Furthermore, most of the local equilibrium parameters stayed within 1\% of the baseline value. The main parameters that did change for $\rho_\psi=0.5$ surface were 

\begin{itemize}
    \item $\beta_{e,\mathrm{unit}}: 0.012 \rightarrow0.004$
    \item $\beta'_{e,\mathrm{unit}} : -0.08 \rightarrow -0.03$
    \item $q : 4.3 \rightarrow 7.3$
    \item $\hat{s} : 0.78 \rightarrow 0.70$
\end{itemize}

\begin{figure}[!hbt]
    \centering
    \begin{subfigure}{0.49\textwidth}
        \centering
        \includegraphics[width=75mm]{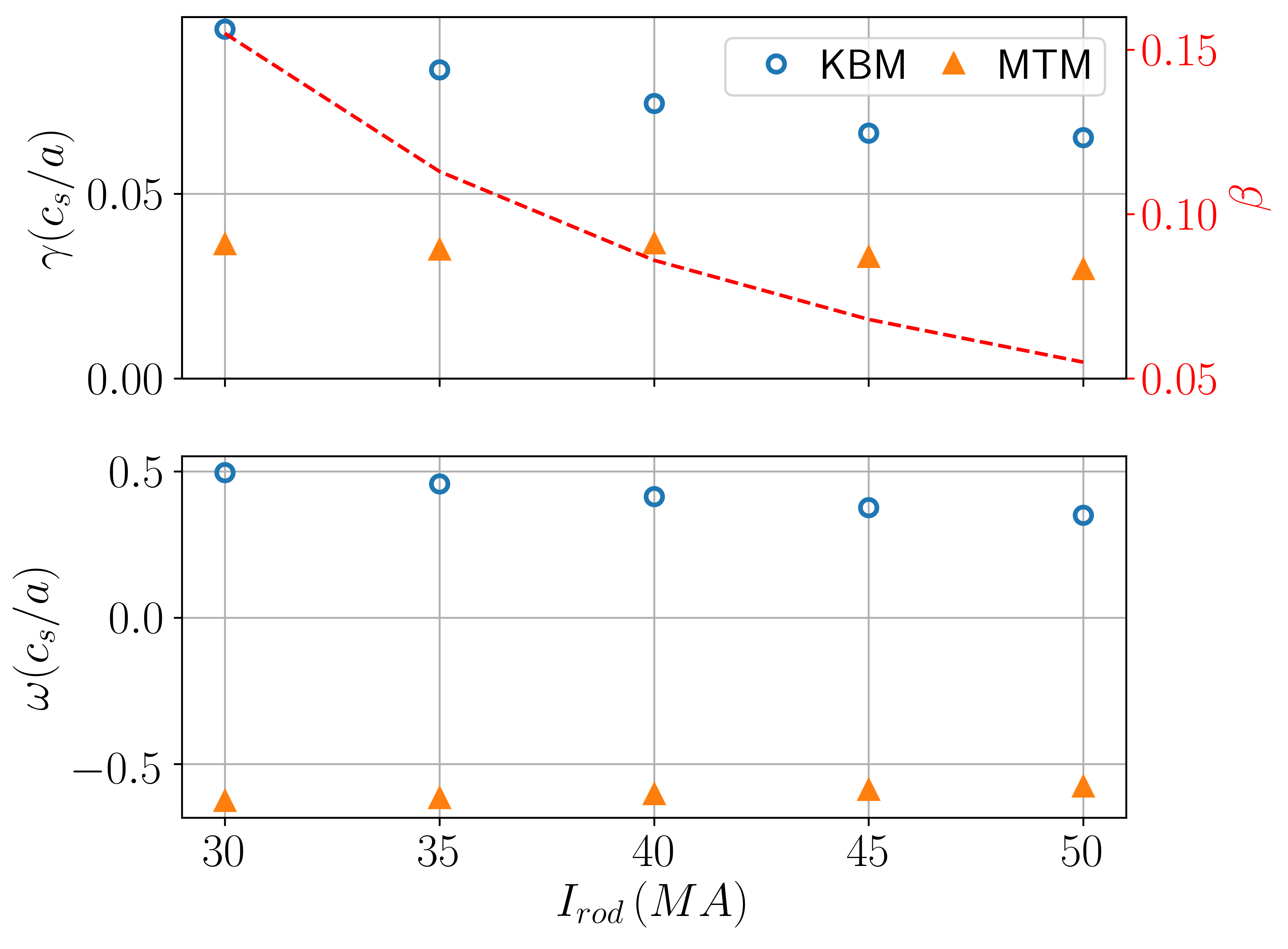}
        \caption{}
        \label{fig:ky_0.35_irod}
    \end{subfigure}
    \begin{subfigure}{0.49\textwidth}
        \centering
        \includegraphics[width=75mm]{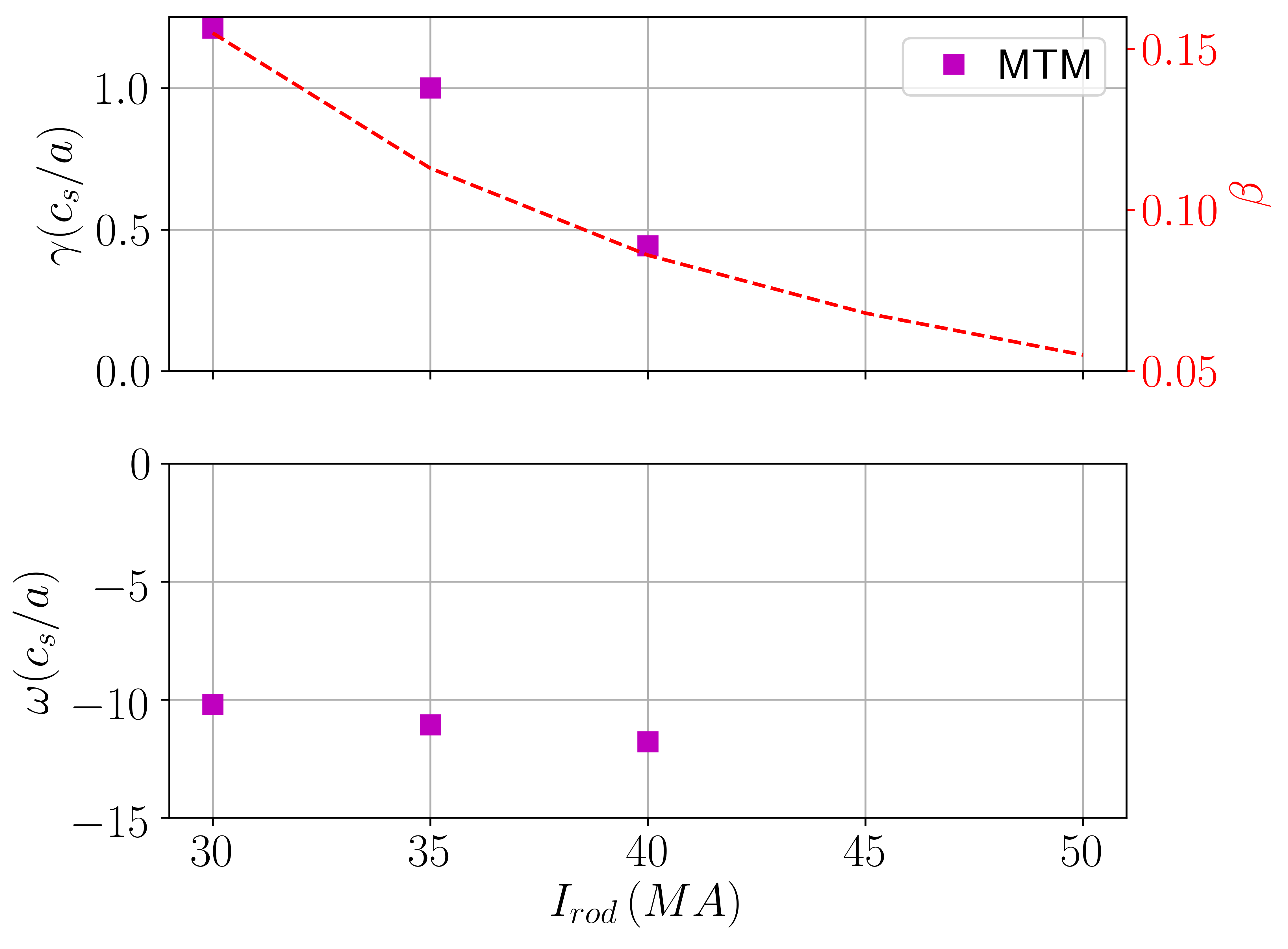}
        \caption{}
        \label{fig:ky_4_irod}
    \end{subfigure}
    \caption{Examining the impact of a higher field device by increasing $I_{\mathrm{rod}}$ for a) $k_y\rho_s=0.35$ and b) $k_y\rho_s=4.2$ at $\rho_\psi=0.5$.}
    \label{fig:gs2_irod}
\end{figure}

Figure \ref{fig:ky_0.35_irod} illustrates how the growth rate and frequency of the $k_y \rho_s=0.35$ mode varies over the scan in $I_\mathrm{rod}$\footnote{Note this is not at fixed toroidal mode number $n$ as $\rho_s$ changes with $B$}, with $\beta_{e,\mathrm{unit}}$ shown by the dashed red line. Firstly it is important to note that the KBM and two different MTMs were found in all of these equilibria, suggesting that these will exist over a range of different high $\beta$ ST equilibria. The KBM is weakly stabilised by the higher rod current. It seems that the stabilisation from the lower $\beta_{e,\mathrm{unit}}$ and higher $q$ is counteracted by the lower $\beta'_{e,\mathrm{unit}}$. For the MTM there is a very weak stabilisation as the increased $q$ counteracts the lower $\beta_{e,\mathrm{unit}}$. If $I_{\mathrm{rod}}$ is sufficiently increased, as $\beta_{e,\mathrm{unit}}\rightarrow0$, these modes should become stable, but $I_{\mathrm{rod}}=50$\si{\mega\ampere}. is pushing the limits of engineering so going further beyond this is unlikely to be feasible.

Examining the impact on the high $k_y$ MTM, Figure \ref{fig:ky_4_irod} shows that operating at higher field stabilises the mode and when $I_{\mathrm{rod}}>40MA$, consistent with the $\beta_{e,\mathrm{unit}}$, $\beta'_{e,\mathrm{unit}}$ scan conducted in Figure \ref{fig:ky_4_mtm_beta_bprime}. However, the ETG seen there is not destabilised in this scan as discussed further in \ref{app:high_ky_etg}.



\subsection{Impact of density peaking}
\label{sec:opt_equ}

The results of the Section \ref{sec:low_ky_kbm_mtm} suggest that the flow shear will wipe out the KBMs and the high $k_y$ MTMs found so far. It was also demonstrated that increasing the density gradient was beneficial in stabilising the low $k_y$ MTMs. If the $a/L_T$ is reduced whilst increasing $a/L_n$ without changing the fusion power, then it may be possible to design an equilibrium in which the majority of the linear instabilities are stable.

\begin{figure}
    \begin{subfigure}{\textwidth}
        \centering
        \includegraphics[width=100mm]{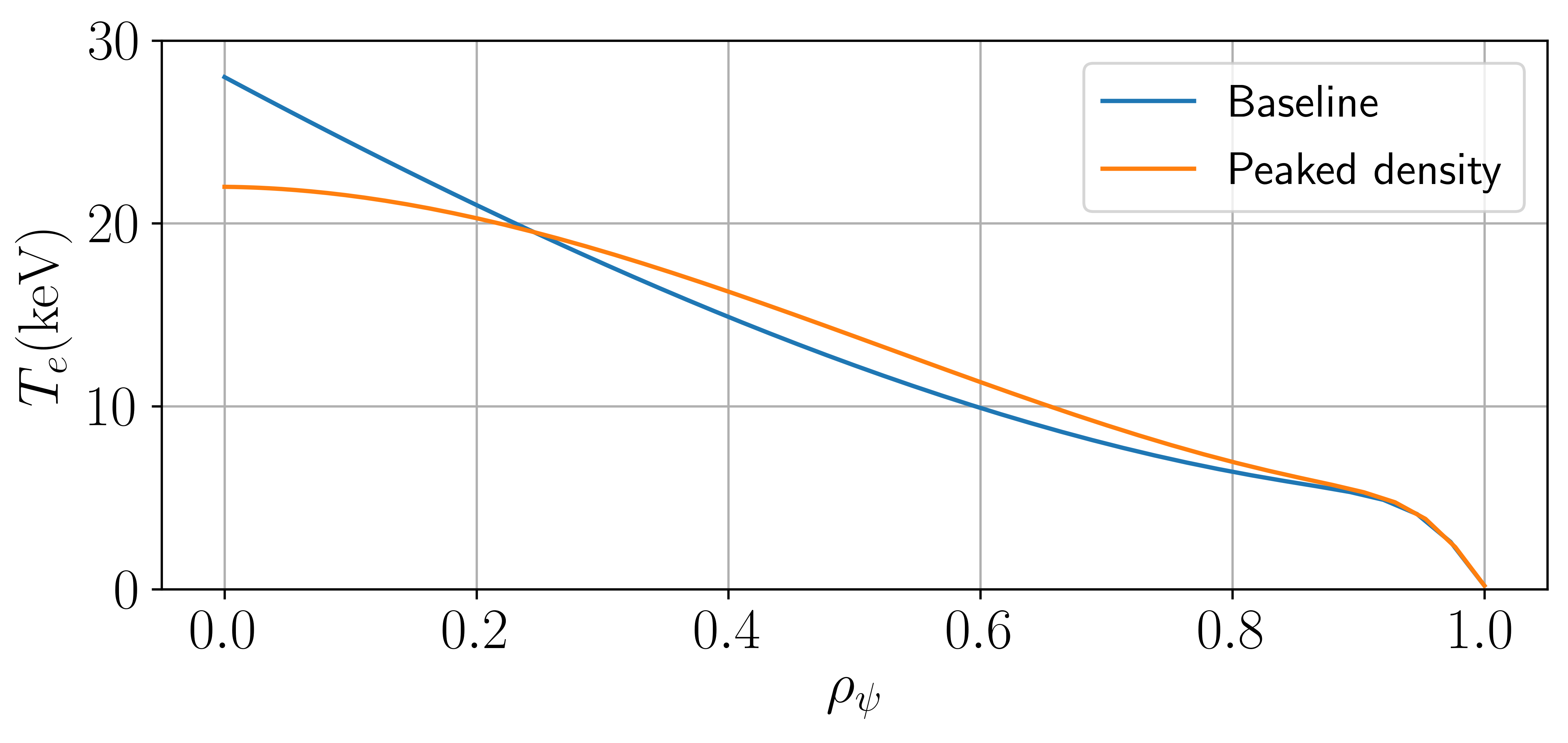}
        \caption{}
        \label{fig:comparison_te}   
    \end{subfigure}
    \begin{subfigure}{\textwidth}
        \centering
        \includegraphics[width=100mm]{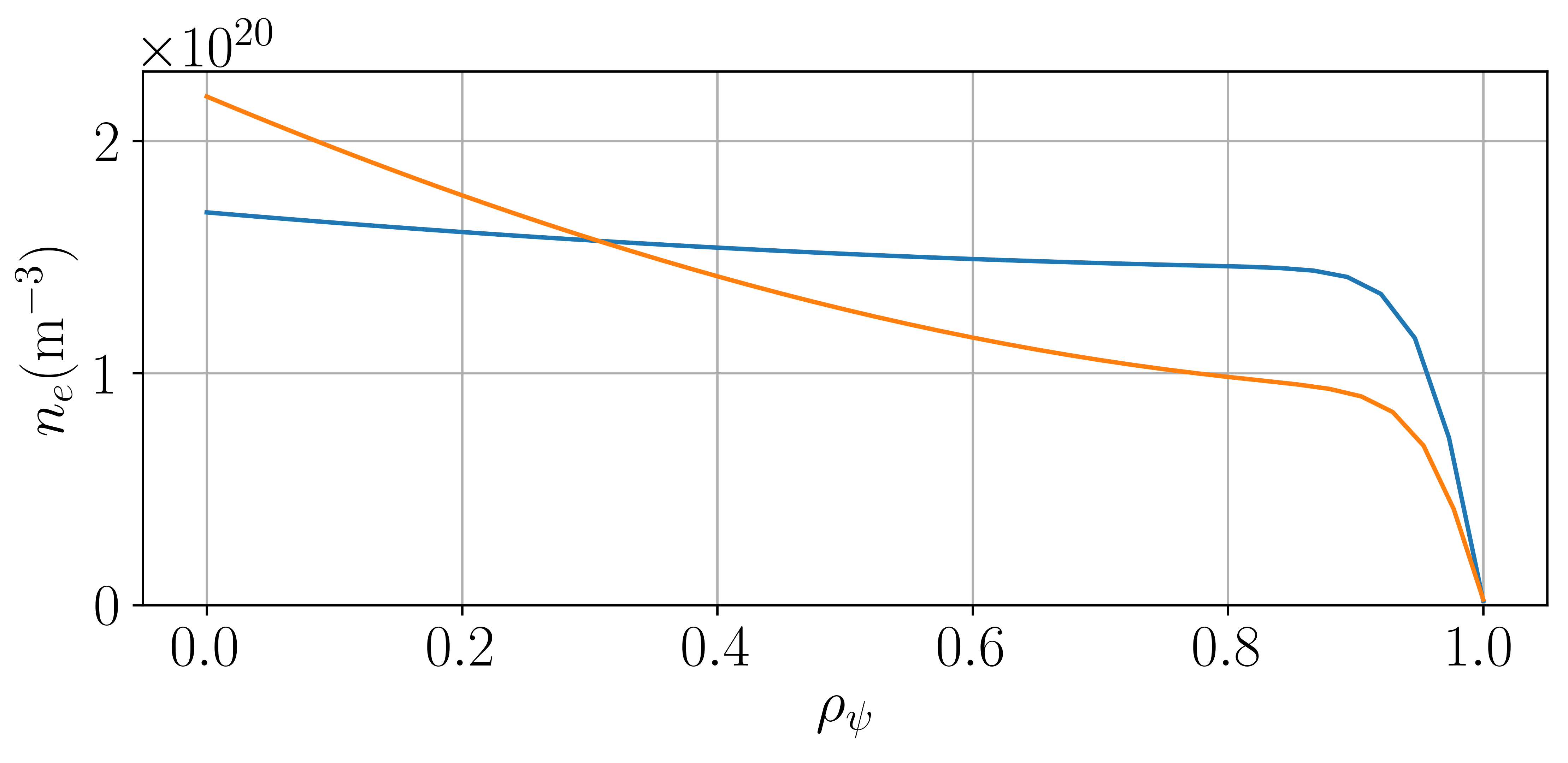}
        \caption{}
        \label{fig:comparison_ne}   
    \end{subfigure}
    \begin{subfigure}{\textwidth}
        \centering
        \includegraphics[width=100mm]{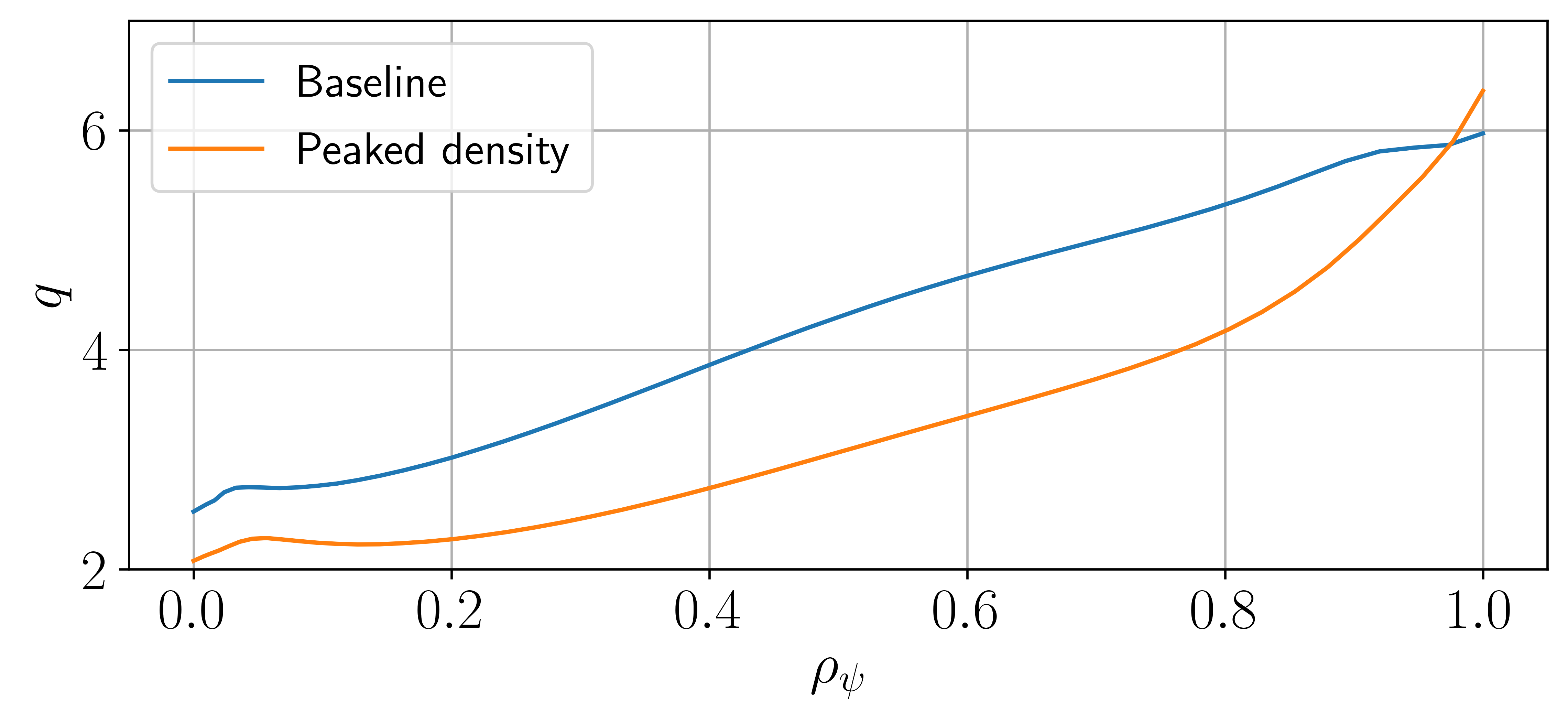}
        \caption{}
        \label{fig:comparison_q}   
    \end{subfigure}
    \caption{Comparison of a) temperature, b) density and c) safety factor profile for the baseline case in Table \ref{tab:baseline_parameters} and peaked density case in Table \ref{tab:optimised_parameters}.}
\end{figure}

\begin{table}[!htb]
    \begin{minipage}{0.45\linewidth}
        \centering
        \begin{tabular}{|c|c|c|}
        \hline
        Parameter & Baseline & Peaked $n$\\
        \hline
        \hline 
        $R_{\mathrm{maj}}$ (m) & 2.5 & 2.5  \\
        $a$ (m) & 1.5 & 1.5  \\
        $R_0$ (m) & 3.15 & 3.05 \\
        $I_{\mathrm{rod}}$ (\si{\mega\ampere}) & 30.0 & 30.0 \\
        $I_{\mathrm{p}}$ (\si{\mega\ampere}) & 21.0 & 21.0 \\
        $I_{\mathrm{aux}}$ (\si{\mega\ampere}) & 8.2 & 7.1 \\
        $P_{\mathrm{fus}}$ (\si{\mega\watt}) & 1100 & 1100 \\
        $P_{\mathrm{aux}}$ (\si{\mega\watt}) & 94 & 83\\
        $\kappa$ & 2.8 & 2.8\\
        $\delta$ & 0.55 & 0.55\\
        $H_{98}$ & 1.35 & 1.26 \\
        $H_{\mathrm{Petty}}$ & 0.94 & 0.83 \\
        $T_{e0} $(\si{\kilo\electronvolt}) & 28.0 & 22.0\\
        $\langle T_e\rangle $(\si{\kilo\electronvolt}) & 14.8 & 14.1\\
        $n_{e0} (10^{20}$\si{\per\cubic\metre}) & 1.72 & 2.19 \\
        $\langle n_e\rangle (10^{20}$\si{\per\cubic\metre}) & 1.54 & 1.38 \\
        $f_{\mathrm{GW}}$ & 0.52 & 0.52 \\
        $l_i$ & 0.27 & 0.38\\
        $\beta_{\mathrm{N}}$ & 5.5 & 5.13 \\
        $q_0$ & 2.51 & 2.07 \\
        \hline
    \end{tabular}
    \caption{Comparing the plasma parameters for the baseline and the peaked density operating points.}
    \label{tab:optimised_parameters}
    \end{minipage}
    \hspace{0.05\linewidth}
    \begin{minipage}{0.5\linewidth}
        \centering
    \begin{tabular}{|c|c|c|}
        \hline
        Parameter & Baseline & Peaked $n$\\
        \hline\hline
        $r/a$ & 0.66 & 0.67 \\
        $R_{\mathrm{maj}}/a$ & 1.79 & 1.83 \\
        $n_{e20}$ $($\si{\per\cubic\metre}) & 1.51 & 1.36\\
        $T_e $(\si{\kilo\electronvolt}) & 12.2 & 12.8\\
        $B_0$ (\si{\tesla}) & 2.16 & 2.18 \\
        $B_{\mathrm{unit}}$ (\si{\tesla}) & 7.52 & 7.34 \\
        $\rho_s$ (\si{\metre}) & 0.0021 & 0.0022 \\
        $a/L_n$ & 0.43 & 1.40 \\
        $a/L_T$ & 2.77 & 2.34\\
        $\Delta$ & -0.57 & -0.48\\
        $q$ & 4.30 & 3.14\\
        $\hat{s}$ & 0.78 & 0.97\\
        $\kappa$ & 3.03 & 2.87\\
        $s_\kappa$ & -0.14 & -0.11\\
        $\delta$ & 0.45 & 0.34\\
        $s_\delta$ & 0.19 & 0.21\\
        $\beta_{e}$ & 0.15 & 0.14 \\
        $\beta_{e,\mathrm{unit}}$ & 0.012 & 0.013 \\
        $\nu_{ee} (c_s/a)$ & 0.017 & 0.013\\
        $\gamma_{\mathrm{dia}} (c_s/a)$ & 0.080 & 0.055 \\
        \hline
    \end{tabular}
    \caption{Local plasma and Miller parameters for the $\rho_\psi = 0.5$ surface of the baseline and peaked density equilibrium.}
    \label{tab:gk_flux_params_opt}
    \end{minipage}
\end{table}

Using this information a new equilibrium was designed in SCENE with a more peaked density profile. An operating point was examined where $n_{e0}/\langle n_e \rangle$, which was equal to 1.12 in the baseline operating point, was increased to $n_{e0}/\langle n_e \rangle =1.58$ at fixed $P_{\mathrm{fus}}=1.1GW$. The pedestal height and width was kept the same. To reduce the temperature gradient drive, the core temperature for both species was dropped from $28keV$ to $20 keV$. The auxiliary current profile was kept fixed. The resulting equilibrium parameters are outlined in Table \ref{tab:optimised_parameters}. The temperature, density and safety factor profiles are illustrated in Figures \ref{fig:comparison_te}, \ref{fig:comparison_ne} and \ref{fig:comparison_q} respectively.


The parameters are broadly similar to the baseline case, with the exception of the density and temperature values. The Miller and plasma parameters for the $\rho_\psi=0.5$ surface are detailed in Table \ref{tab:gk_flux_params_opt}. It can be seen that many of the parameters are very similar to the baseline parameters in Table \ref{tab:gk_flux_params}, with the notable exception of the higher $a/L_n$ and lower $a/L_T$ as expected. However, $a/L_p$ has increased overall from $6.40$ to $7.48$, so it may be expected that the KBMs will be driven more unstable, though as seen in Section \ref{sec:beta_betaprime}, this is counteracted by the increased $\beta'_{e,\mathrm{unit}}$. $\hat{s}/q$ has increased, due to the differences in the bootstrap and diamagnetic current profiles, with this design having a higher bootstrap fraction due to the peaked density profile. Following the trends seen in Figures \ref{fig:ky_0.35_q_kbm} and \ref{fig:ky_0.35_shear_kbm}, this is expected to destabilise the KBMs but its impact on the MTMs is less clear.

\begin{figure}[!htb]
    \centering
    \includegraphics[width=120mm]{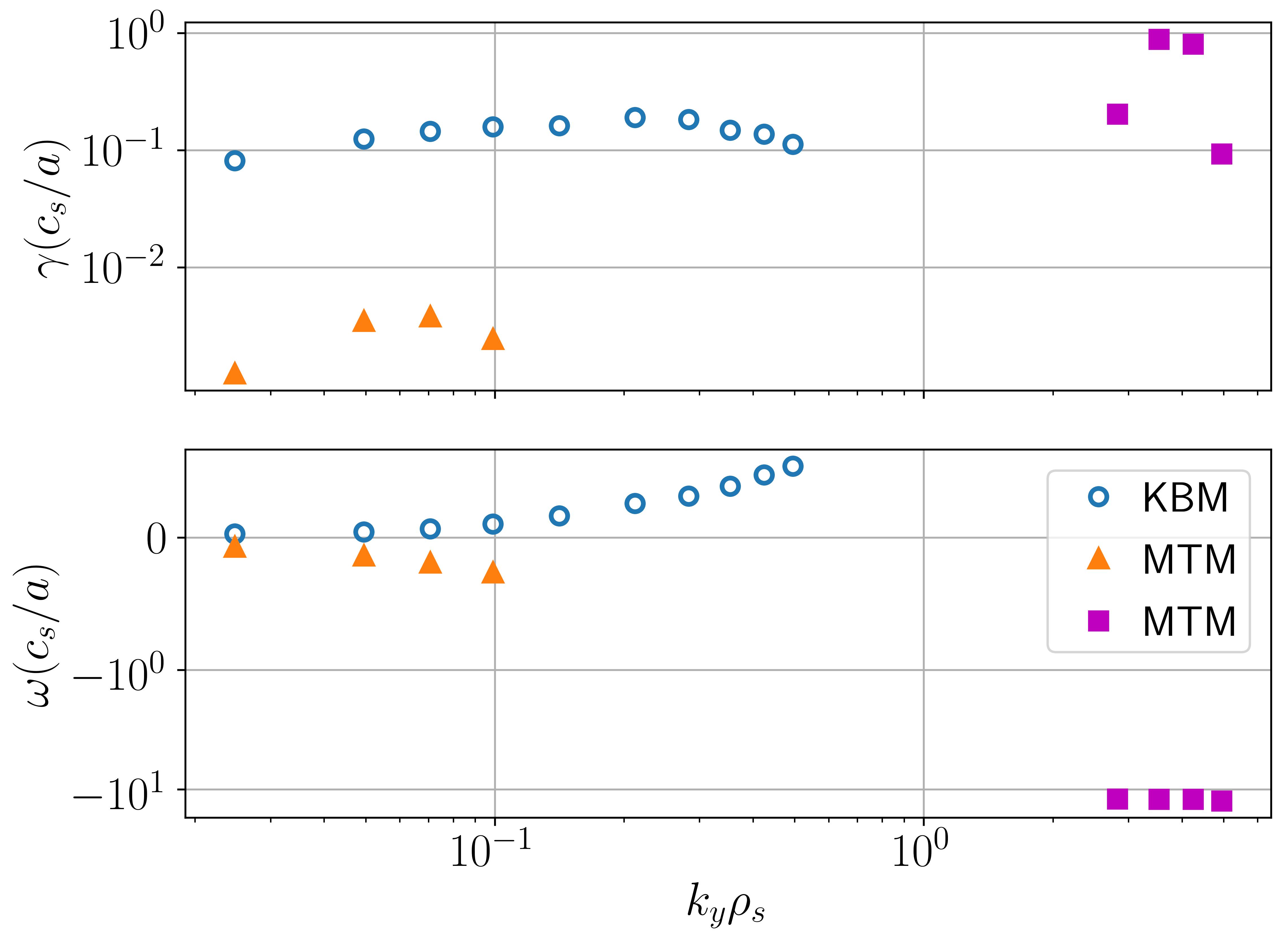}
    \caption{Dominant odd and even mode for the $\rho_\psi=0.5$ surface of the peaked density equilibrium without flow shear outlined in Table \ref{tab:gk_flux_params_opt}.}
    \label{fig:ky_scan_opt}
\end{figure}

The dominant instabilities are presented in Figure \ref{fig:ky_scan_opt}. Similar to the baseline operating point, in the low $k_y$ region KBMs and extended MTMs are found. However, the peak MTM growth rate is $\gamma_{\mathrm{MTM}}=0.004 c_s/a$, significantly lower than the baseline operating point where it was $\gamma_{\mathrm{MTM}}=0.04 c_s/a$. The KBM peak growth rate has increased from $\gamma_{\mathrm{KBM}}=0.09 c_s/a$ to $\gamma_{\mathrm{KBM}} = 0.19 c_s/a$, but these modes are still narrow in $\theta_0$ so it will be shown that again these are wiped out by flow shear. The higher $k_y$ MTMs are also seen and have a lower growth rate reducing from $1.2c_s/a$ to $0.9 c_s/a$, and it is expected that the flow shear will help to stabilise these modes as well. Finally above $k_y\rho_s=6$, the equilibrium was found to be completely stable, similar to the previous case. It should be noted that these modes peak at a lower $k_y$ in Figure \ref{fig:ky_scan_opt} compared to the baseline case in Figure \ref{fig:gs2_dom} and this is expected as $\rho_s$ has not significantly changed but the value of $q$ is lower compared to the baseline case in Table \ref{tab:gk_flux_params}. Specifically, given that $k_y=\frac{nq}{r}$, to find a given toroidal mode number $n$ at lower $q$, a lower $k_y$ is needed. The low $k_y$ MTM here was unstable down to $n=3$.

\begin{figure}[!htb]
    \centering
    \includegraphics[width=100mm]{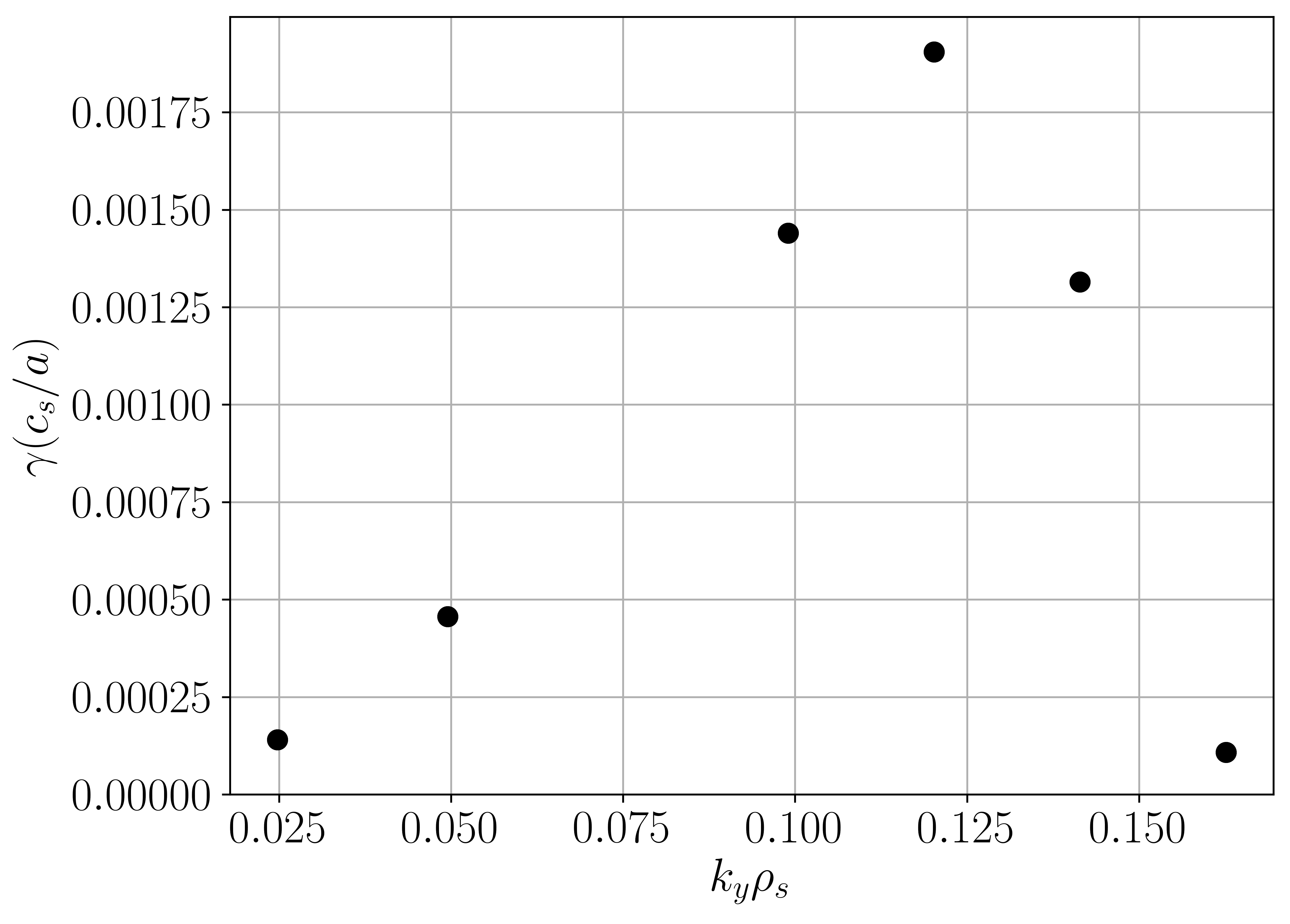}
    \caption{Effective growth rate for the $\rho_\psi=0.5$ surface of the peaked density equilibrium with diamagnetic levels of flow shear.}
    \label{fig:opt_flow}
\end{figure}

A simulation was run for this case with $E\times B$ flow shear at the diamagnetic level of $\gamma_{\mathrm{dia}}=0.055 c_s/a$. The effective growth rate is shown in Figure \ref{fig:opt_flow} and it was found that above $k_y\rho_s=0.2$ the equilibrium was completely stable. Below this value, the MTMs did cause a very slowly growing mode with a growth rate $\mathcal{O}(10^{-3})$. While nonlinear simulations are required to determine the level of driven transport, nevertheless this work hints that density peaking is favourable and the plasma may be able to operate close to neoclassical levels of transport. The impact of impurities and fast ions needs to be examined as they may drive the MTM more unstable, increasing the effective growth rate. In particular, density peaking can lead to impurity accumulation in the core of conventional tokamaks, which would be detrimental to the performance of a reactor. 
While quantifying the level of impurity accumulation and impact on radiative losses is important, this equilibrium suggests a possible pathway to a high performance high $\beta$ ST. The open question is how to generate peaking in the equilibrium density profile.

The level of density peaking will be set by particle transport and fuelling. The transport for each species can be decomposed into a diffusive term $D$ and a convective term $V$, shown in Equation \ref{eqn:pflux}.

\begin{equation}
    \label{eqn:pflux}
    \Gamma = D \frac{1}{n}\frac{\partial n}{\partial r } + V
\end{equation}

\begin{figure}[!htb]
    \begin{subfigure}{0.5\textwidth}
        \centering
        \includegraphics[width=75mm]{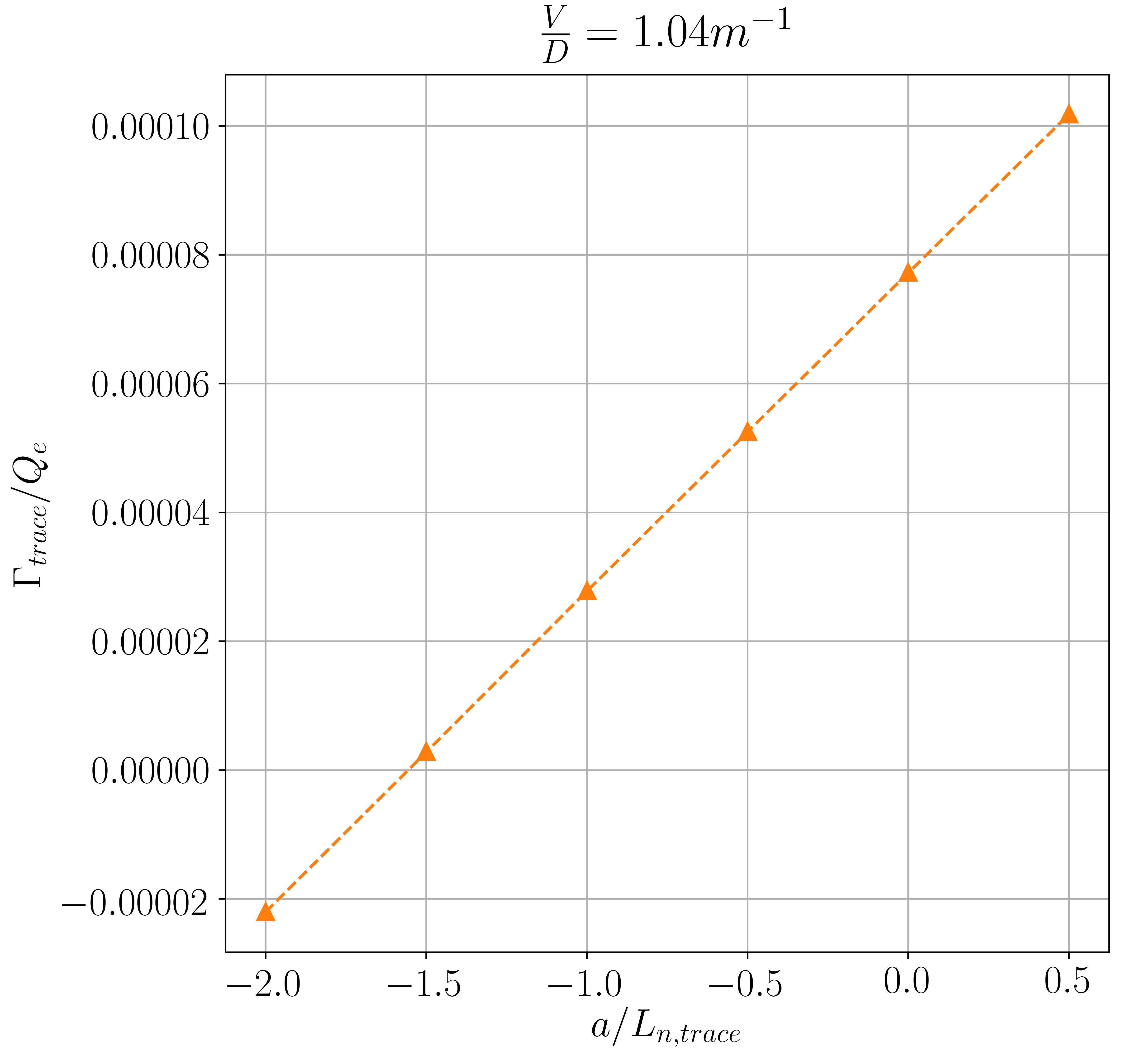}
        \caption{}
        \label{fig:mtm_zero_pflux}
    \end{subfigure}
    \begin{subfigure}{0.5\textwidth}
        \centering
        \includegraphics[width=75mm]{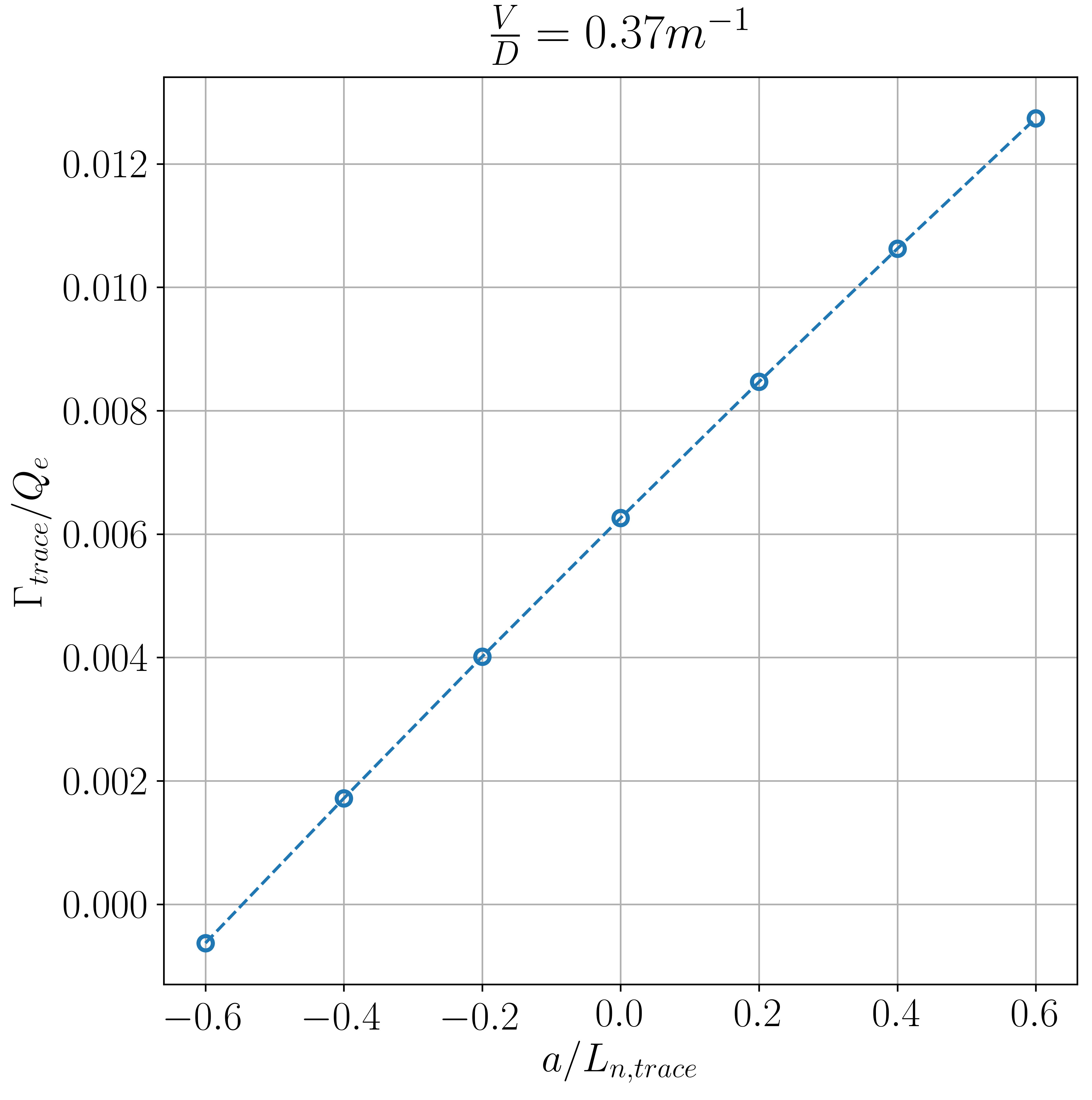}
        \caption{}
        \label{fig:kbm_zero_pflux}
    \end{subfigure}
       \caption{Linear particle flux of a trace deuterium when scanning through the trace density gradients for the a) MTM and b) KBM at $\rho_s=0.35$. The particle flux has been normalised to the electron heat flux.}
    \label{fig:zero_pflux}
\end{figure}

The convective particle pinch term from the low $k_y$ MTMs can be estimated using a quasi-linear argument to determine whether it leads to outward or inward particle transport. If the saturated turbulence has a quasi-linear character, then many attributes of the turbulent transport can be inferred from the properties of the linear instabilities. Assuming the quasi-linear turbulence is dominated by a particular linear mode, it is possible to determine the ratio of the convective pinch and diffusivity terms, $V/D,$ shown in Equation \ref{eqn:pflux}. By introducing a trace Deuterium species and modifying the trace density gradient to find the point where $\Gamma \rightarrow 0$, the ratio of $V/D$ can be determined. Examining $k_y\rho_s=0.35$ when $n_{\mathrm{trace},D}/n_e = 0.05$, Figure \ref{fig:mtm_zero_pflux} illustrates that zero particle flux occurs at $a/L_n=-1.56$ meaning $V/D= -a/L_n / a = 1.56/1.5 = 1.04m^{-1}$. This means the particle convection is outwards making density peaking even more difficult. The particle flux is normalised to the linear electron heat flux and is $\mathcal{O}(10^{-5})$ so the effect may be small but will not help achieve a peaked density profile. A similar scan for the KBMs shows that they also have an outwards convection, shown in Figure \ref{fig:kbm_zero_pflux}. The relative size of the particle flux is $\mathcal{O}(10^{-2})$, larger than the MTMs but still quite small overall. These quasi-linear ideas suggest that peaked density is unlikely to be produced through transport from these KBMs and MTMs. The generation of peaked density profiles will require either other mechanisms contributing to the particle transport, or a core localised particle source.

\section{Conclusions}

An equilibrium has been presented from a potential operating point  of a high $\beta$ spherical tokamak. Local micro-instability studies of the thermal plasma (neglecting fast ions) find that KBMs and MTMs are the dominant micro-instabilities. Furthermore, the equilibria examined here were found to be stable in the electron scale range where $k_y\rho_s>6$. 

Two different types of MTMs were found, a collisional MTM when $k_y\rho_s<0.6$ and a collisionless MTM between $3.0<k_y\rho_s<6.0$. The KBMs and high $k_y$ MTMs were highly ballooning, and unstable only in a very narrow range of $\theta_0\sim0.0$, such that diamagnetic levels of flow shear was sufficient in suppressing these modes. The low $k_y$ MTM, however, was unaffected by flow shear and thus is expected to be the dominant source of transport for this equilibrium. These low $k_y$ MTM were very extended along the field line in ballooning space and require resolving electron scales radially indicating a potential challenge for nonlinear simulations.

The parametric dependency of these modes on different equilibrium parameters was determined and it was found that the low $k_y$ MTMs can be stabilised by the raising density gradient, suggesting that a peaked density profile would be beneficial to confinement. Moreover, the low $k_y$ MTMs were destabilised by increasing $\nu_*$, consistent with previous energy confinements scalings from ST experiments \cite{kaye2007confinement, valovivc2011collisionality, guttenfelder2012scaling}, indicating that a high temperature/low density device will benefit the confinement. A complicated dependency on $\beta_{e,\mathrm{unit}}$, $\beta'_{e,\mathrm{unit}}$ and $q$ led to the interesting result that increasing the toroidal field at constant $I_{\mathrm{p}}$ had little impact on micro-stability of the low $k_y$ modes. Counter-intuitively, this suggests that a higher field device may not improve the confinement when these MTMs are dominant, though there may be a gain if $I_{\mathrm{p}}$ can also  be increased to keep $q$ constant.

An optimised equilibrium was generated with a more peaked density profile and flatter temperature profile. With flow shear, the equilibrium was found to be marginally stable, indicating that operating close to neoclassical levels of transport may be feasible if such an equilibrium can be accessed. The convective pinch term was found to be outwards for both low $k_y$ modes, suggesting that transport from MTMs and KBMs is unlikely to generate the required peaking without a central source of particle fuelling.

To quantify the level of transport driven from each of these different modes requires nonlinear simulations. These are required to test and develop reduced physics models that are capable of describing the transport generated by the various instabilities contributing to the nonlinear turbulence.  Nonlinear calculations are also required to improve our understanding as to how experimental actuators will influence transport in high $\beta$ STs. This will identify the modes that contribute most strongly to transport, and may provide insights in how this turbulence can be controlled.

\section{Acknowledgements}

The author would like to thank Francis Casson, Walter Guttenfelder, Emily Belli and Jeff Candy for useful discussions that assisted this work.

The views and opinions expressed herein do not necessarily reflect those of the European Commission. This work was supported by the Engineering and Physical Sciences Research Council [EP/L01663X/1, EP/R034737/1, EP/T012250/1]. This work used the ARCHER UK National Supercomputing Service (http://www.archer.ac.uk). We acknowledge the CINECA award under the ISCRA initiative, for the availability of high performance computing resources and support. This research used resources of the National Energy Research Scientific Computing Center (NERSC), a U.S. Department of Energy Office of Science User Facility located at Lawrence Berkeley National Laboratory, operated under Contract No. DE-AC02-05CH11231.

\appendix

\section{Cross code validation}
\label{app:cross_code}
\begin{figure}[!htb]
    \centering
    \includegraphics[width=100mm]{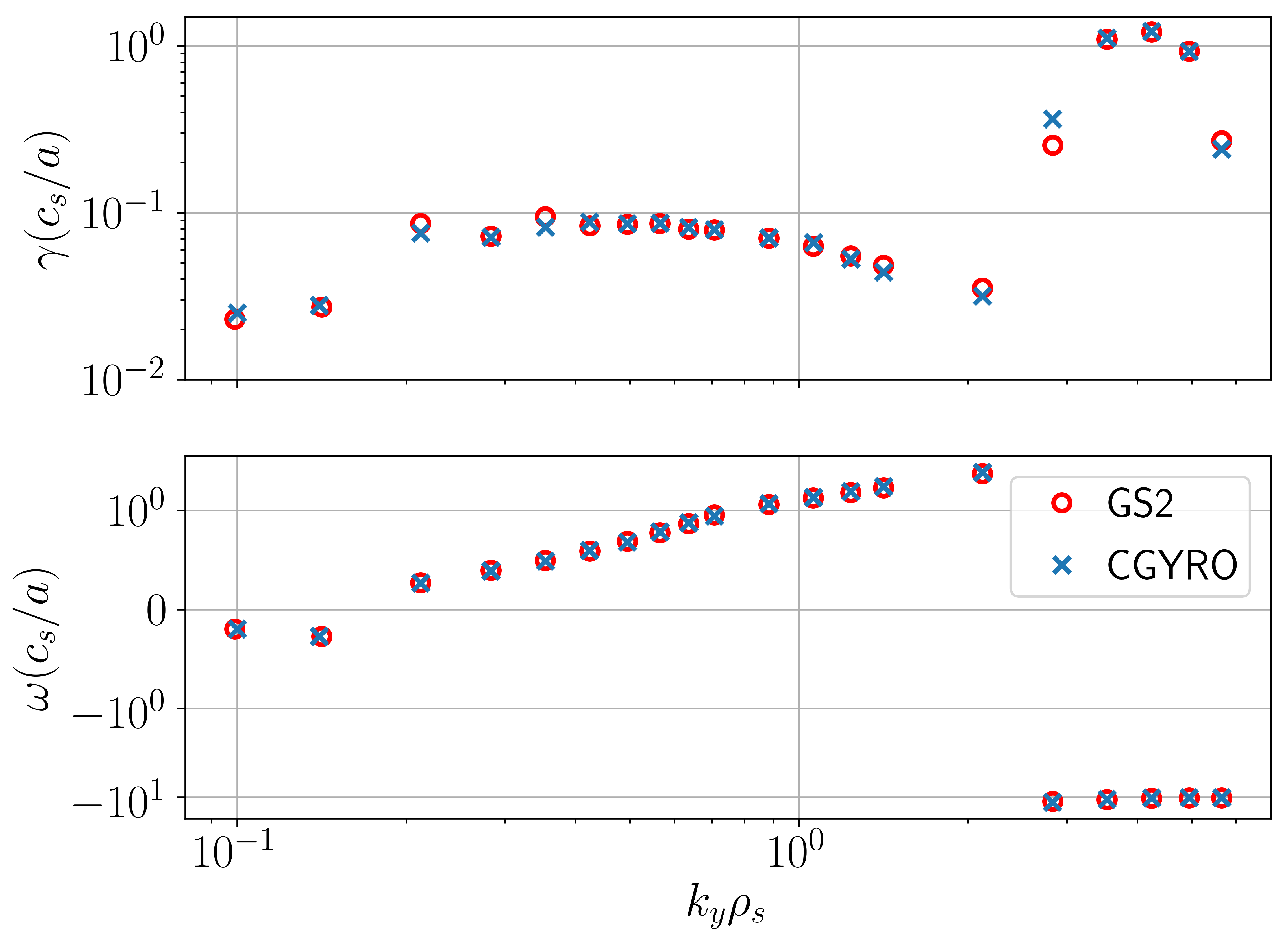}
    \caption{Comparison of the dominant eigenvalue predictions made by GS2 and CGYRO of the equilbrium outlined in Table \ref{tab:gk_flux_params}.}
    \label{fig:cross_code}
\end{figure}

Given the highly exotic nature of this equilibrium, the predictions made by GS2 were compared to CGYRO to ensure that these modes can be consistently found across codes. CGYRO is currently unable to force an odd or even eigenmode so only the dominant instability can be examined. The results for the $\rho_\psi=0.5$ surface outlined in Table \ref{tab:gk_flux_params} are shown in Figure \ref{fig:cross_code}.

It can be seen that both codes are in good agreement for this case, with CGYRO finding these MTMs with the extended mode structure, which indicates that the gyrokinetic equation is being correctly solved and that these modes are likely to appear in high $\beta$ STs like STEP.

\section{Stability of electron scale region}
\label{app:high_ky_etg}

All the equilibria examined so far were found to be stable when $k_y\rho_e\sim \mathcal{O}(1)$. The STPP reactor design was also found to be stable to electron scale modes \cite{wilson2004STPP}. ETG-like instabilities are generally found at the electron scale in other tokamaks. To understand why they are stable here, different plasma parameters were scanned as done in the Section \ref{sec:low_ky_kbm_mtm}. This section examines the un-optimised case in Table \ref{tab:gk_flux_params} for $\rho_\psi=0.5$.

\subsection{Impact of kinetic profiles}
\begin{figure}[!htb]
    \begin{subfigure}{0.5\textwidth}
        \centering
        \includegraphics[width=75mm]{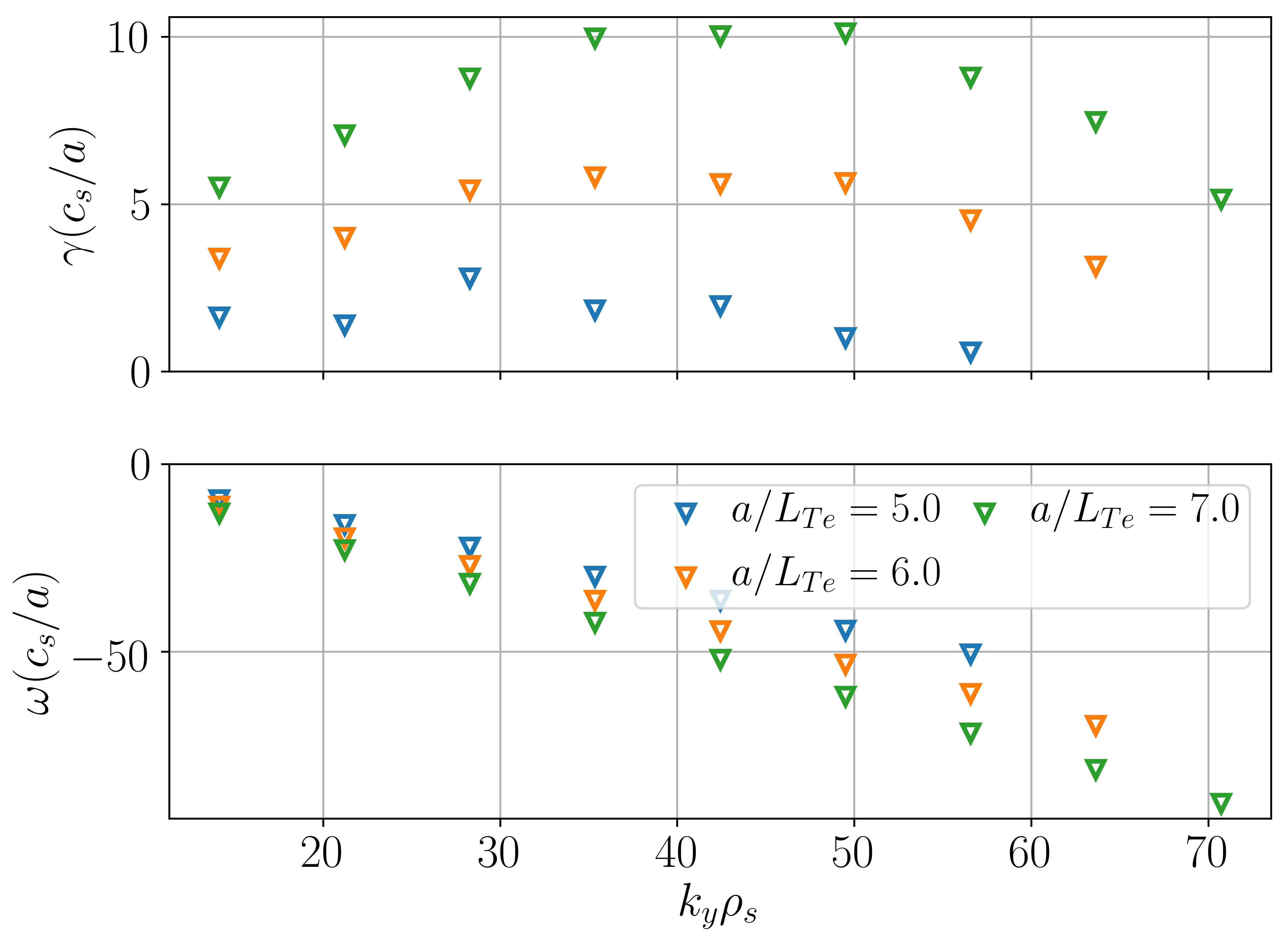}
        \caption{}
        \label{fig:high_ky_lte}
    \end{subfigure}
    \begin{subfigure}{0.5\textwidth}
        \centering
        \includegraphics[width=75mm]{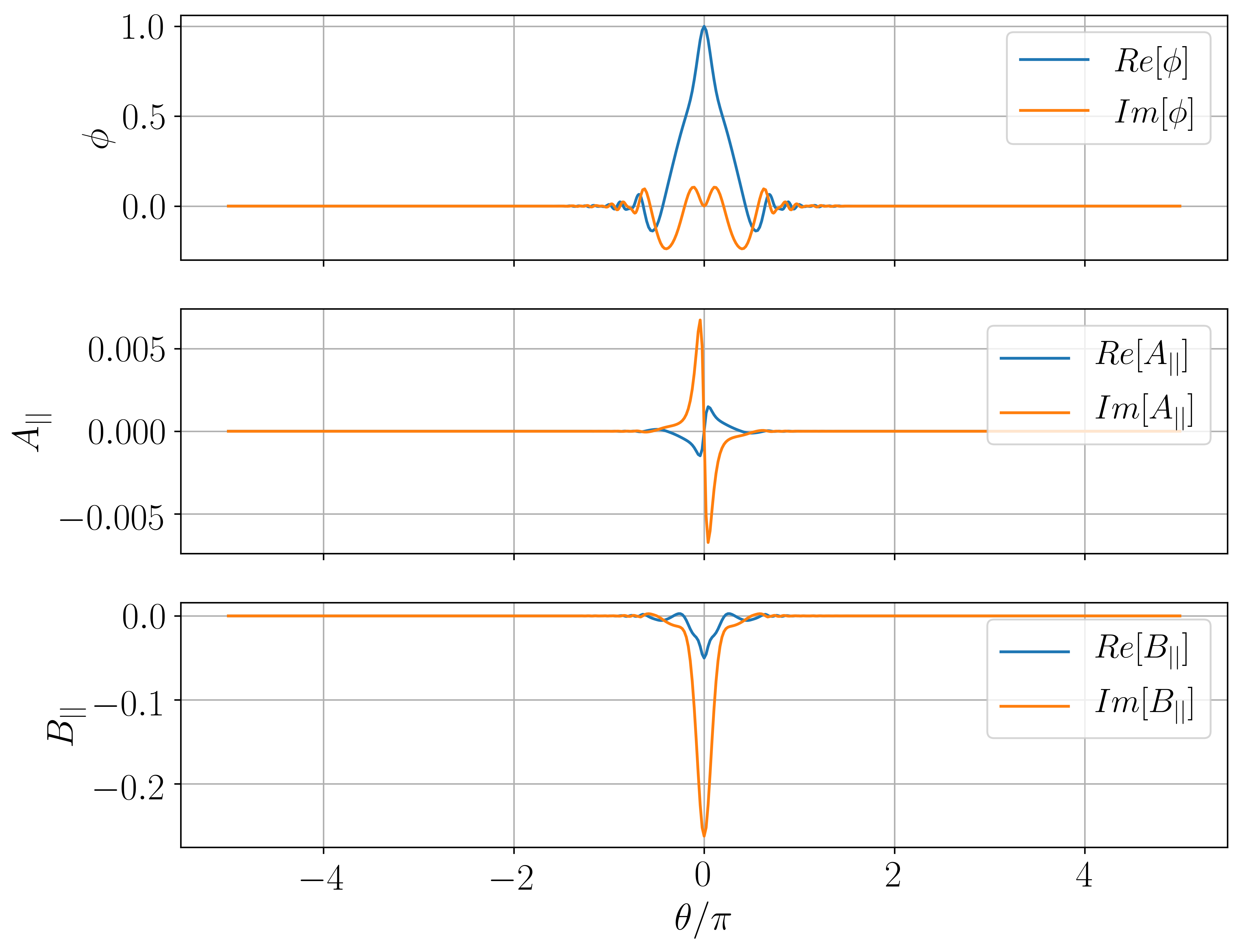}
        \caption{}
        \label{fig:ky_35_eigfunc}
    \end{subfigure}
       \caption{a) Growth rate spectrum for high values of $k_y$ at different $a/L_{Te}$. The equilibrium value was $a/L_{Te}=2.77$. b) Eigenfunctions for $k_y\rho_s=35$ when $a/L_{Te}=7.0$.}
    \label{fig:high_ky}
\end{figure}

Given that ETG is expected, the impact of $a/L_{Te}$ was examined. Scans were for $a/L_{Te}=5.0 , 6.0 \, \& \, 7.0$ for $k_y\rho_s= 15 \rightarrow 70$ . The equilibrium value of $a/L_{Te}=2.77$ is well below this temperature gradient range, illustrating the large increase in drive necessary to find any ETG. Figure \ref{fig:high_ky_lte} shows how an unstable mode can be found if the drive is sufficiently high. It can be seen that increasing $a/L_{Te}$ does drive the mode seen here unstable, consistent with ETG. This shows that a significant increase in temperature gradient would be necessary for this region of $k_y$-space to be unstable. The eigenfunction for $k_y\rho_s=35$ at $a/L_{Te}=7.0$ is shown in Figure \ref{fig:ky_35_eigfunc}. The $A_{||}$ component is very small when normalised to $\phi$ suggesting it is not significant for these ETG modes.

To examine why the critical $a/L_{Te}$ is high for this equilibrium other parameters will be examined. The stability at $k_y\rho_s=35$ will be investigated more carefully.

\subsection{Impact of magnetic equilibrium}

\subsubsection{Impact of $\beta_{e,\mathrm{unit}}$ and $\beta'_{e,\mathrm{unit}}$}
Earlier it was shown that an ETG appeared when reducing $\beta_{e,\mathrm{unit}}$ and $\beta'_{unit}$. As they were both reduced, the high $k_y$ MTM was stabilised and an ETG was destabilised, so it is expected that they will have an impact on the stability.

\begin{figure}[!htb]
    \begin{subfigure}{0.5\textwidth}
        \centering
        \includegraphics[width=75mm]{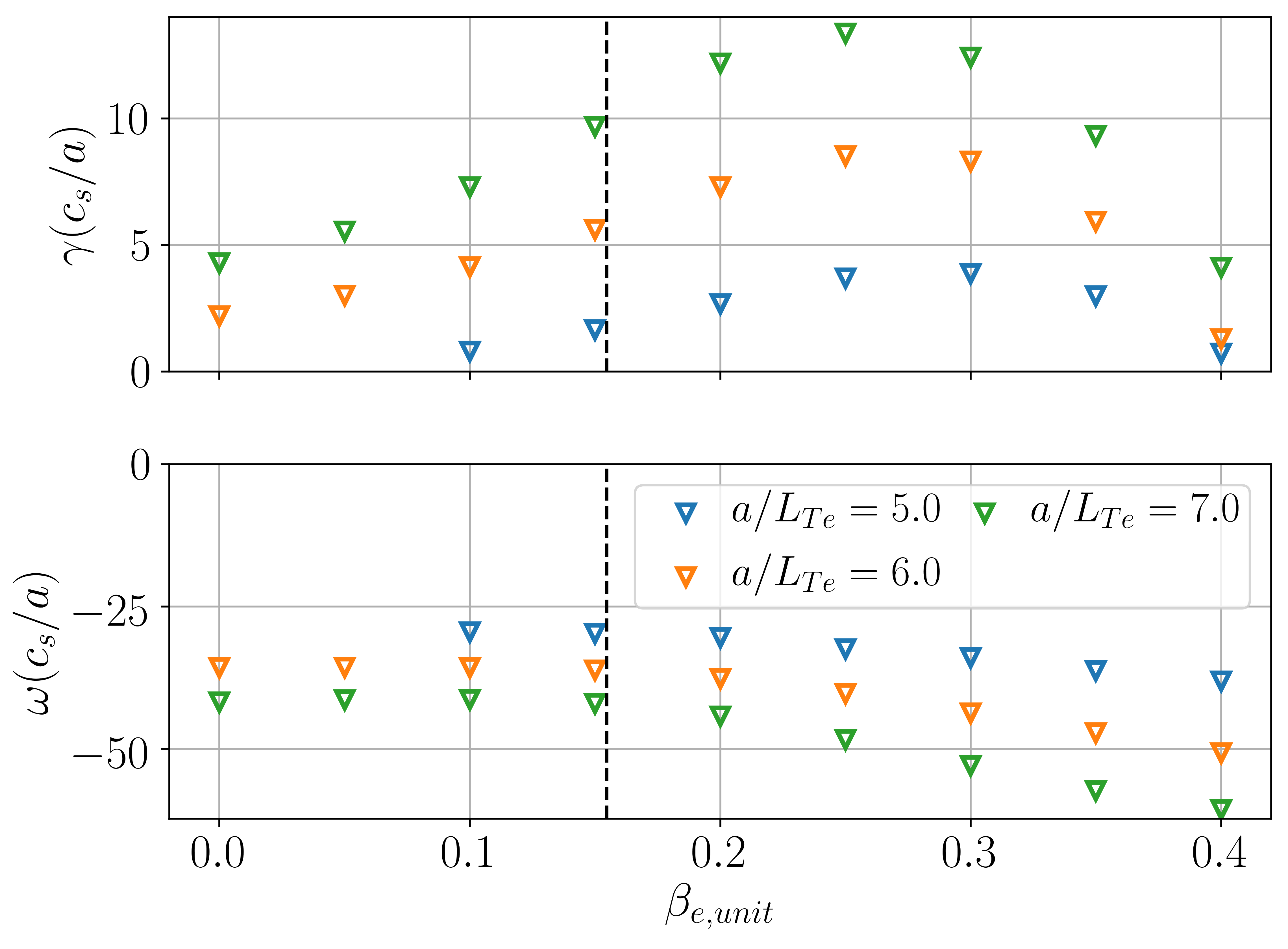}
        \caption{}
        \label{fig:ky_35_beta}
    \end{subfigure}
    \begin{subfigure}{0.5\textwidth}
        \centering
        \includegraphics[width=75mm]{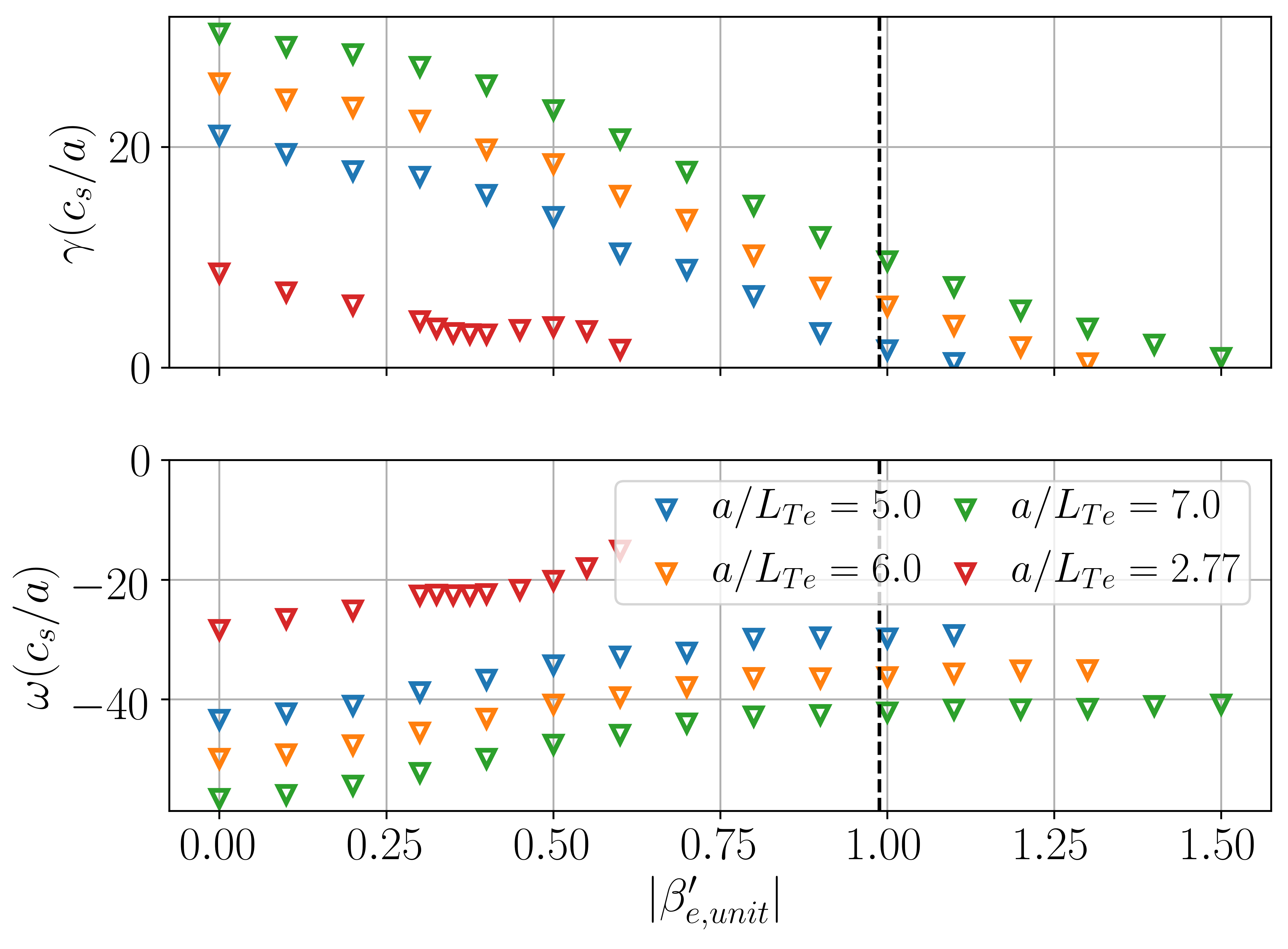}
        \caption{}
        \label{fig:ky_35_bprime}
    \end{subfigure}
       \caption{Impact of a) $\beta_{e,\mathrm{unit}}$ and b) $\beta'_{e,\mathrm{unit}}$ at different $a/L_{Te}$ with $k_y\rho_s=35$.}
    \label{fig:ky_35_em}
\end{figure}

Figure \ref{fig:ky_35_beta} shows how the growth rate changes for the 3 different temperature gradients when changing $\beta_{e,\mathrm{unit}}$. $\beta_{e,\mathrm{unit}}$ is found to be destabilising to these high gradient ETG modes, up to a critical value $\beta_{e,\mathrm{unit}}=0.02$, when there is a rollover and at sufficiently high $\beta_{e,\mathrm{unit}}$ these ETG are stabilised. However, at the equilibrium value of $a/L_{Te}$, the mode was completely stable regardless of the $\beta_{e,\mathrm{unit}}$. A similar scan was conducted with $\beta'_{e,\mathrm{unit}}$ and this appears to be a cause of the stability. Increasing $\beta'_{e,\mathrm{unit}}$ strongly stabilises the ETG mode, such that when $\beta'_{e,\mathrm{unit}}=0$ the ETG mode is unstable even at the equilibrium $a/L_{Te}$. This has been seen before by Roach $et \, al$ \cite{roach1995trapped} and is attributed to high $\beta'$ causing drift reversal at the outboard mid-plane. This all indicates a potential disadvantage with operating at higher field. The resulting lower $\beta'_{e,\mathrm{unit}}$ could destabilise the ETG modes and have a detrimental effect on electron thermal transport.

\subsubsection{Safety factor profile}

A critical gradient formula for ETG-like turbulence was found by Jenko $et \, al$ \cite{jenko2001critical} as follows

\begin{equation}
    \label{eqn:crit_etg}
    \bigg(\frac{a}{L_{Te}}\bigg)^{ETG}_{\mathrm{crit}} \propto \bigg(1.3 + 1.9 \frac{\hat{s}}{q}\bigg) 
\end{equation}

\begin{figure}[!htb]
    \begin{subfigure}{0.5\textwidth}
        \centering
        \includegraphics[width=75mm]{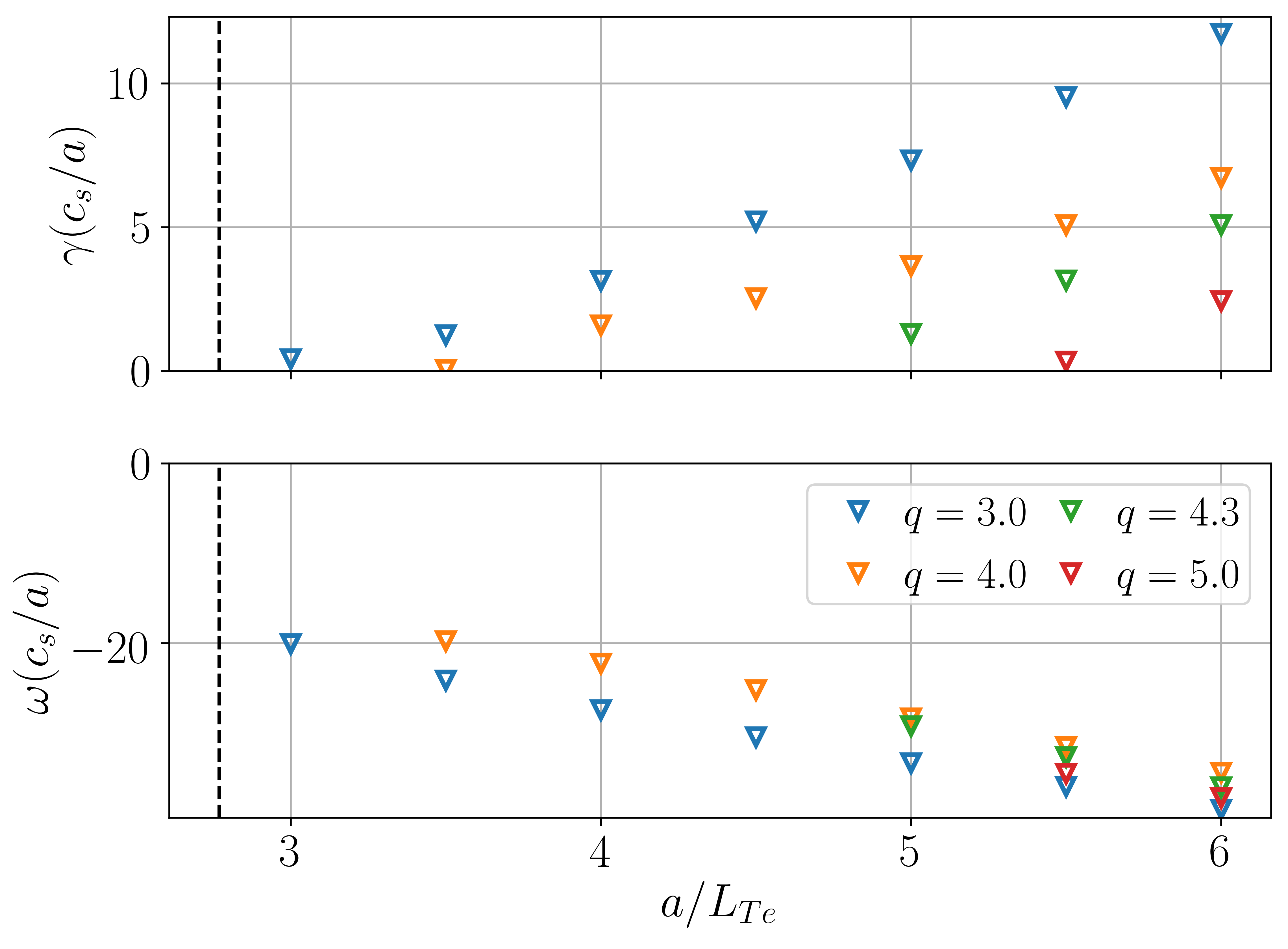}
        \caption{}
        \label{fig:ky_35_q}
    \end{subfigure}
    \begin{subfigure}{0.5\textwidth}
        \centering
        \includegraphics[width=75mm]{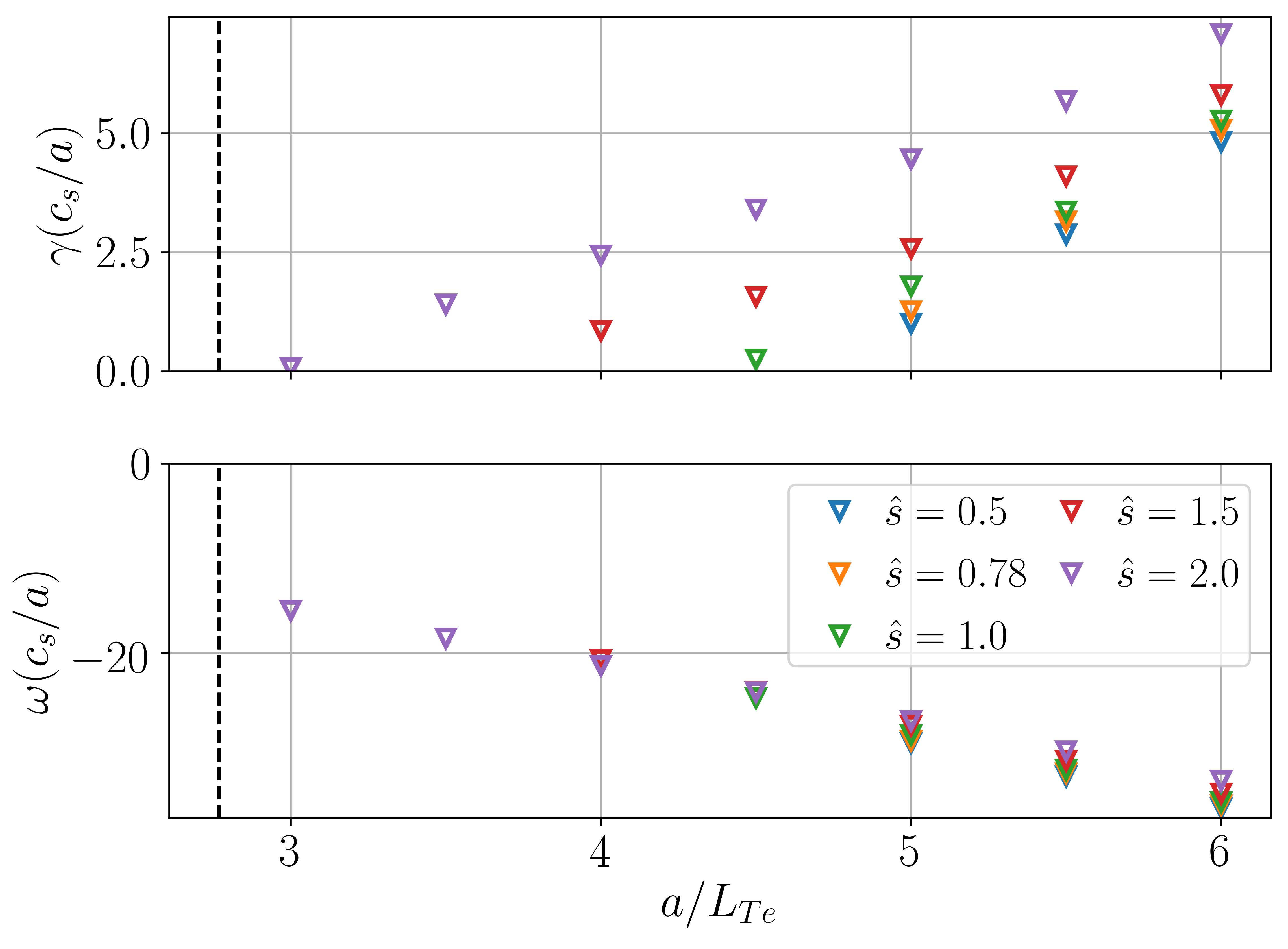}
        \caption{}
        \label{fig:ky_35_shat}
    \end{subfigure}
       \caption{Impact of a) $q$ and b) $\hat{s}$ with different $a/L_{Te}$ at $k_y\rho_s=35$.}
    \label{fig:ky_35_safety}
\end{figure}

\begin{figure}[!htb]
    \begin{subfigure}{0.5\textwidth}
        \centering
        \includegraphics[width=75mm]{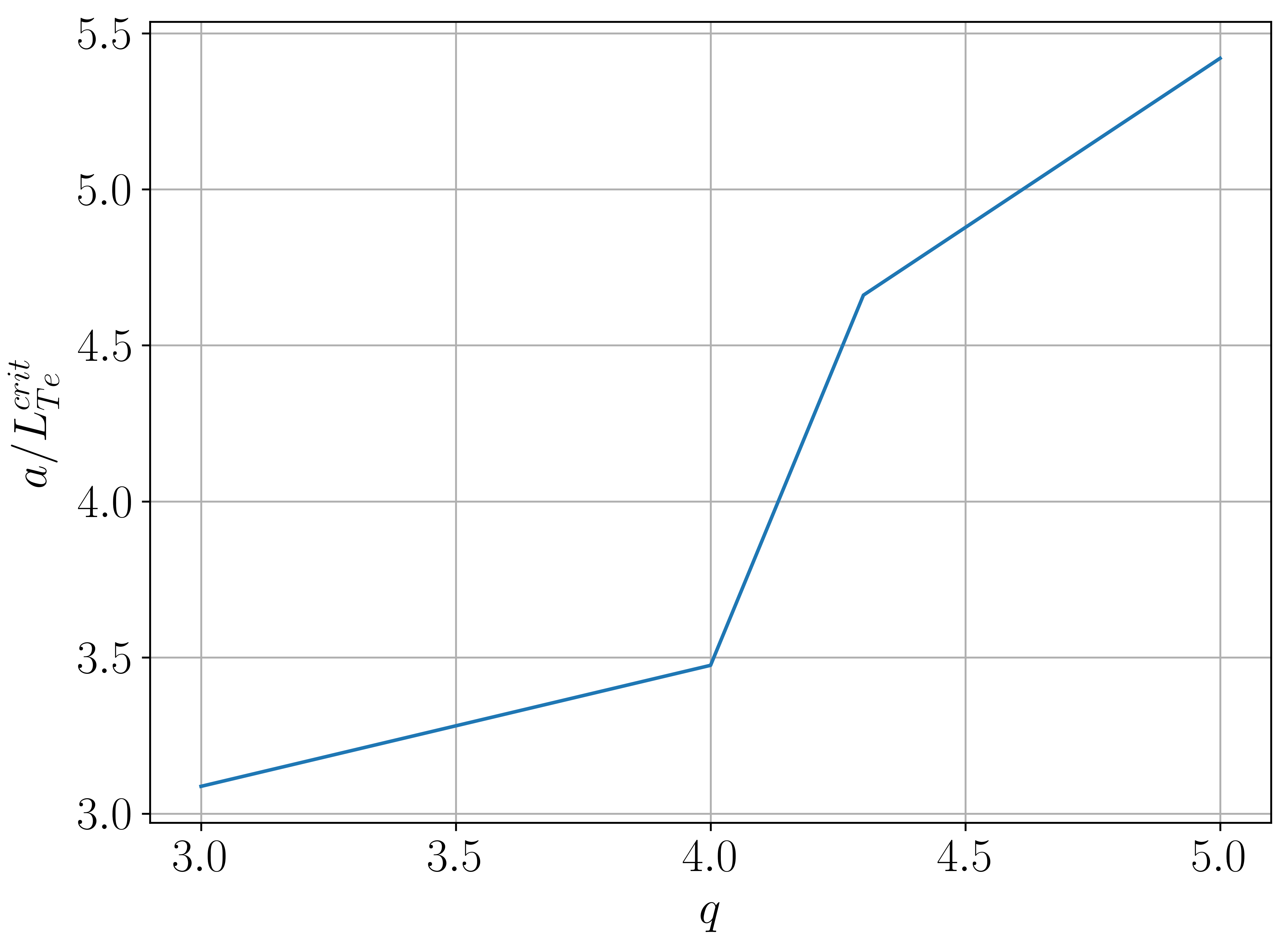}
        \caption{}
        \label{fig:ky_35_q_crit}
    \end{subfigure}
    \begin{subfigure}{0.5\textwidth}
        \centering
        \includegraphics[width=75mm]{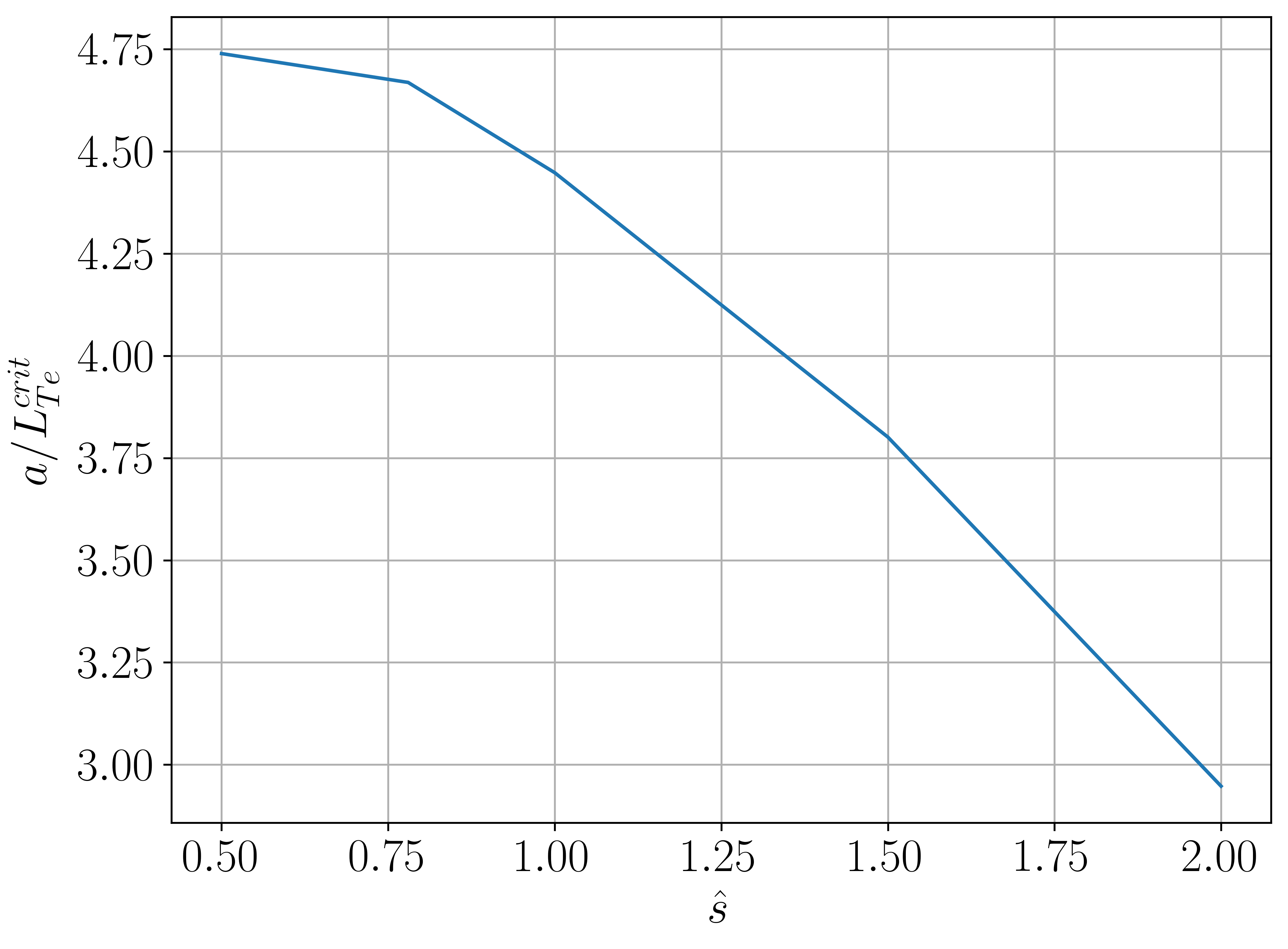}
        \caption{}
        \label{fig:ky_35_crit}
    \end{subfigure}
       \caption{Critical $a/L_{Te}$ threshold for various a) $q$ and b) $\hat{s}$ at $k_y\rho_s=35$.}
    \label{fig:ky_35_safety_crit}
\end{figure}

This was generated from simulations of a low $\beta$, conventional aspect ratio tokamak. To examine its validity for this ST equilibrium, scans were performed in $a/L_{Te}$ at different values of $q$ and $\hat{s}$ to find the critical gradient. Equation \ref{eqn:crit_etg} suggests that the critical gradient will increase with $\hat{s}/q$ but Figures \ref{fig:ky_35_q} and \ref{fig:ky_35_shat} indicate this is not the case here. Performing a linear fit to these growth rates, an estimate of the critical gradient can be made and this is shown in Figure \ref{fig:ky_35_safety_crit}. It is clear that the scaling of Equation \ref{eqn:crit_etg} does not describe this regime. Rather, operating at low $\hat{s}/q$ is actually beneficial to these modes.

This indicates that the scaling laws developed by Jenko are not applicable in this regime such that they cannot be utilised. Though this may not be of a major concern as the ETG was found to be stable for this equilibrium as well as the optimised equilibrium.

\subsection{Higher field device}

\begin{figure}[!htb]
    \centering
    \includegraphics[width=100mm]{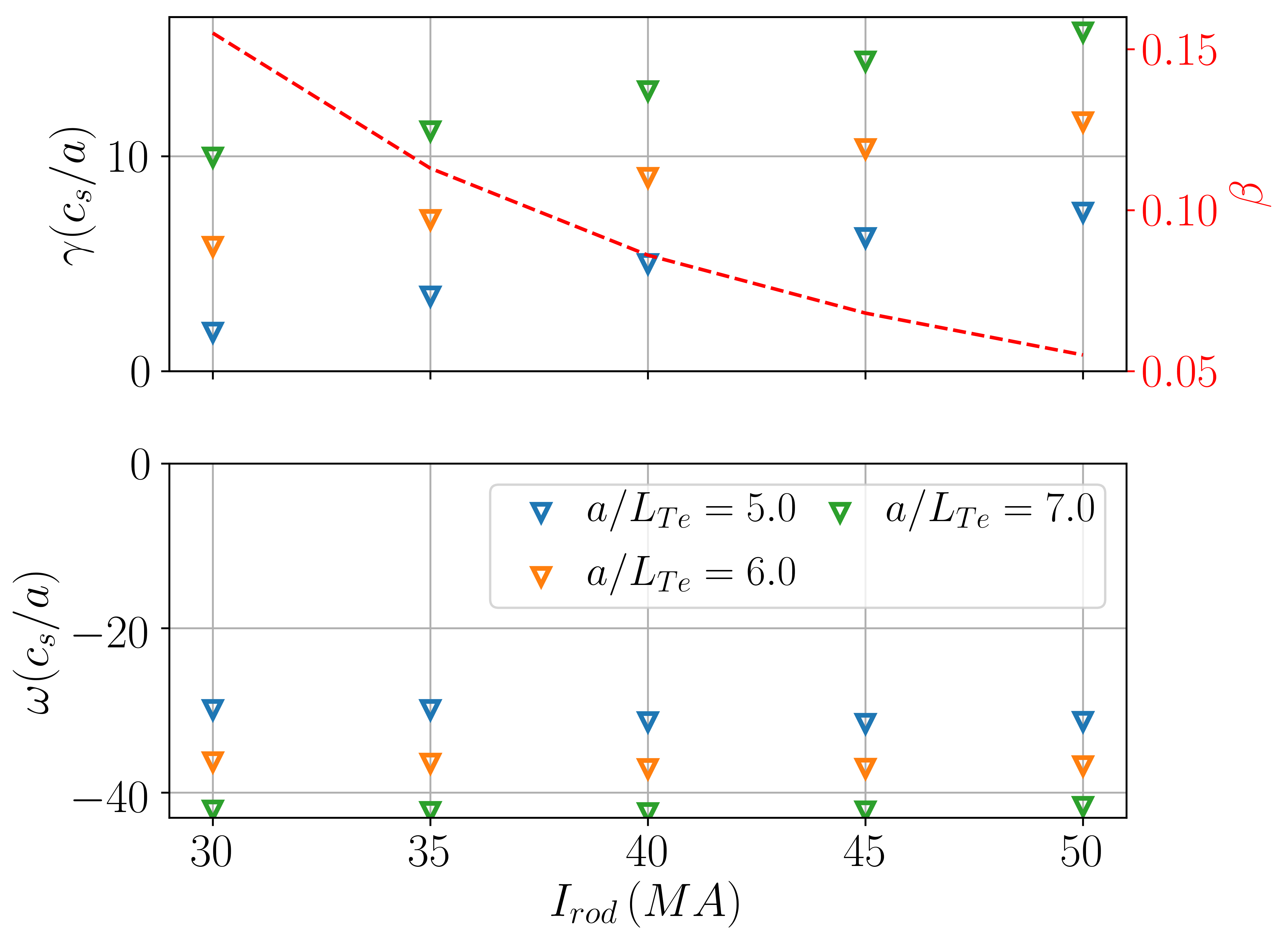}
    \caption{Examining the impact of $I_{\mathrm{rod}}$ on the ETG modes for 3 higher temperature gradients at $k_y\rho_s=35$. Note the reference equilibrium value is $a/L_{Te}=2.77$.}
    \label{fig:ky_35_irod}
\end{figure}

A scan was done looking at the impact of the higher $I_{rod}$ devices on these ETG, as was conducted in Section \ref{sec:low_ky_high_field}. When examining the impact on the higher gradient cases at $k_y\rho_s=35$, Figure \ref{fig:ky_35_irod} shows that operating at higher field does destabilise these ETG modes, indicating the destabilisation from $\beta'_{e,\mathrm{unit}}$ is overcoming the stabilisation from lower $\beta_{e,\mathrm{unit}}$ and higher $q$. Nevertheless, for $a/L_{Te}=2.77$, the modes are stable, so for this case it is not a concern. But it does highlight that operating at higher field may destabilise ETG modes.

\printbibliography[]
\end{document}